\documentclass[12pt]{article}

\usepackage{anyfontsize}
\usepackage[utf8]{inputenc}
\usepackage{amsmath,amssymb}
\usepackage{amsthm}
\usepackage{graphicx}
\usepackage{amsfonts}
\usepackage{caption}
\usepackage{changepage}
\usepackage{subcaption}
\usepackage{color}
\usepackage{float}
\usepackage{bm}
\usepackage{bbm}
\usepackage{comment}
\usepackage{natbib}
\usepackage{xr}
\usepackage{hyperref}
\hypersetup{
    colorlinks=true,
    linkcolor=blue,
    filecolor=magenta,      
    urlcolor=cyan,
    citecolor=blue
}
\usepackage{pdfpages}

\usepackage[ruled, titlenumbered, algosection]{algorithm2e}
\usepackage{hanging}

\newcommand{\xx}{\mathbf{x}}

\newcommand{\XX}{\mathbf{X}}

\newcommand{\LL}{\mathbf{L}}
\newcommand{\UU}{\mathbf{U}}

\newcommand{\HH}{\mathbf{H}}
\newcommand{\II}{\mathbf{I}}

\newcommand{\R}[2]{\mathbb{R}^{#1 \times #2}}

\newcommand{\iRNR}{\in \mathbb{R}^{n \times r}}
\newcommand{\iRPP}{\in \mathbb{R}^{p \times p}}

\newcommand{\BTh}{\mathbf{\Theta}}
\newcommand{\BSig}{\mathbf{\Sigma}}

\definecolor{purple}{rgb}{0.73, 0.33, 0.83}

\newcommand{\vertiii}[1]{{\left\vert\kern-0.25ex\left\vert\kern-0.25ex\left\vert #1 \right\vert\kern-0.25ex\right\vert\kern-0.25ex\right\vert}}
\DeclareMathOperator*{\argmin}{arg\,min}
\DeclareMathOperator{\Tr}{Tr}

\newtheorem{thm}{Theorem}
\newtheorem{prop}{Proposition}

\newtheorem{cor}{Corollary}

\newtheorem{remark}{Remark}

\usepackage{pdfpages}
\makeatletter
\newcommand*{\addFileDependency}[1]{
\typeout{(#1)}
\@addtofilelist{#1}
\IfFileExists{#1}{}{\typeout{No file #1.}}
}\makeatother

\newcommand*{\myexternaldocument}[1]{%
\externaldocument{#1}%
\addFileDependency{#1.tex}%
\addFileDependency{#1.aux}%
}

\myexternaldocument{SupportingInformation}

\begin{document}

\def\spacingset#1{\renewcommand{\baselinestretch}%
{#1}\small\normalsize} \spacingset{1}

\newcommand{\blind}{1}

\if1\blind
{
    \title{Low-Rank Covariance Completion for Graph Quilting with Applications to Functional Connectivity}
    \author{Andersen Chang$^1$ \and Lili Zheng$^2$ \and Genevera I. Allen$^{2, 3, 4, 5, 6}$}

    \date{\small%
        $^1$Department of Neuroscience, Baylor College of Medicine\\%
        $^2$Department of Electrical and Computer Engineering, Rice University\\%
        $^3$Department of Computer Science, Rice University\\%
        $^4$Department of Statistics, Rice University\\%
        $^5$Department of Pediatrics-Neurology, Baylor College of Medicine\\%
        $^6$Jan and Dan Duncan Neurological Research Institute, Texas Children’s Hospital
    }

  \maketitle
} \fi

\if0\blind
{
  \bigskip
  \bigskip
  \bigskip
  \begin{center}
    {\LARGE\bf Low-Rank Covariance Completion for Graph Quilting with Applications to Functional Connectivity}
\end{center}
  \medskip
} \fi

\bigskip
\begin{abstract}

As a tool for estimating high-dimensional networks, graphical models are commonly applied to calcium imaging data to estimate functional neuronal connectivity. However, in many calcium imaging data sets, the full population of neurons is not recorded simultaneously, but instead in partially overlapping blocks. This leads to the Graph Quilting problem, as first introduced by \cite{vinci2019}, which attempts to infer the structure of the full graph when only subsets of features are jointly observed. In this paper, we study a two-step approach to Graph Quilting, which first imputes the complete covariance matrix using low-rank covariance completion techniques before estimating the graph structure. While prior works have studied low-rank matrix completion, we are the first to address the challenges brought by block-wise missingness and to investigate this problem in the context of graph learning. We study three approaches to this problem, block singular value decomposition, nuclear norm penalization, and non-convex low-rank factorization from both theoretical and applied perspectives. From our empirical studies, we observe that the functional connectivity networks estimated from these methods more closely replicate the structure of functional connectivity graphs derived from having simultaneous observations of all neurons compared to those estimated via other Graph Quilting procedures.

\end{abstract}

\noindent%

\vfill

\newpage

\spacingset{1.9} 

\section{Introduction}\label{sec:intro}

Graphical models are a commonly used unsupervised learning technique for estimating sparse conditional dependency structures in multivariate data. Various graphical modeling approaches have been used in many different fields, including neuroscience \citep{yatsenko2015}, genomics \citep{allenliu2013}, network biology \citep{wang2016}, and finance \citep{talih2005} to analyze conditional relationships in high-dimensional settings. There exists a wide array of literature on the theoretical and empirical performance of different classes of graphical models, including Gaussian an exponential family graphical models \citep{lauritzen1996, yang2015efgm}. Additionally, many methods have been developed to account for different external effects, such as latent variables \citep{ravikumar2010, pfau2013} and covariates \citep{cai2013, chen2016}. 

One particular context in which graphical models are used is in the analysis of data from calcium imaging, which is used to record in vivo firing activity of individual neurons in the brain of an experimental subject under controlled or natural stimulus conditions \citep{stosiek2003}. In particular, calcium imaging data can be used for the study of functional connectivity, defined as the statistical relationships between the activity of neurons in the brain \citep{horwitz2003}. Intrinsic functional neuronal connectivity is of interest in the field of neuroscience as a potential way to better understand how neuronal circuits in the brain are organized and to find patterns that underlie how neurons pass information to one another \citep{feldt2011}, which is of particular interest in the development of brain-computer interfaces \citep{Daly2012, Leeuwis2021}. Functional neuronal connectivity may also serve as a tool for estimating synaptic connectivity between individual neurons in the brain \citep{honey2009predicting}, as well provide insights to how structure and function in the brain are related under different stimuli and conditions \citep{deco2014}.  

Modern calcium imaging technology allows for the recording of the activity of up to thousands of individual neurons simultaneously in vivo; because of this high-dimensional setting, graphical models are a natural choice for studying biological neuronal networks. However, in many calcium imaging experiments, multiple scans are used to record the firing activity of neurons within a full brain volume of interest \citep{greinberger2012}; these scans are often taken in sequential layers of the brain volume, which leads to partially overlapping blocks of observed neurons between consecutive scans \citep{berens2017}. Because of this data collection scheme, the activity traces between many pairs of neurons in a full calcium imaging data set are never simultaneously observed. Thus, in order to obtain a graphical model estimate for the full set of observed neurons, the network structure of the unobserved portion of the set of joint pairwise observations must be inferred from the existing data. 

This leads to the Graph Quilting problem, which seeks to estimate a full graphical model when measurements exist only for partially overlapping patches of the full covariance matrix. Graph Quilting has been studied previously in the contexts of RNA sequencing \citep{ozsolak2011, gan2020} and neuroscience \citep{vinci2019}. The latter work outlined the challenges of the Graph Quilting problem, and showed that it is possible to not only recover graph edges associated with observed elements of the covariance, but also to recover a superset of edges associated with completely missing entries of the covariance. Their approach solves the Maximum Determinant graph quilting (MAD\textsubscript{GQ}) problem by fitting an $\ell_1$-regularized MLE of the observed covariance with the constraint that no edges are affiliated with unobserved elements. Thresholding and Schur complements are then used to identify graph edges and a minimal superset of edges. While this approach comes with strong theoretical guarantees, it makes several assumptions that may be uncheckable in practice and contains multiple steps with tuning parameters that may lead to suboptimal performance in practice. Furthermore, for the many node pairs that are not jointly observed, the MAD\textsubscript{gq} algorithm method recovers a super set of true edges that can be much denser than the true graph.

\begin{figure}[t]
    \centering
    \includegraphics[width=0.35\textwidth]{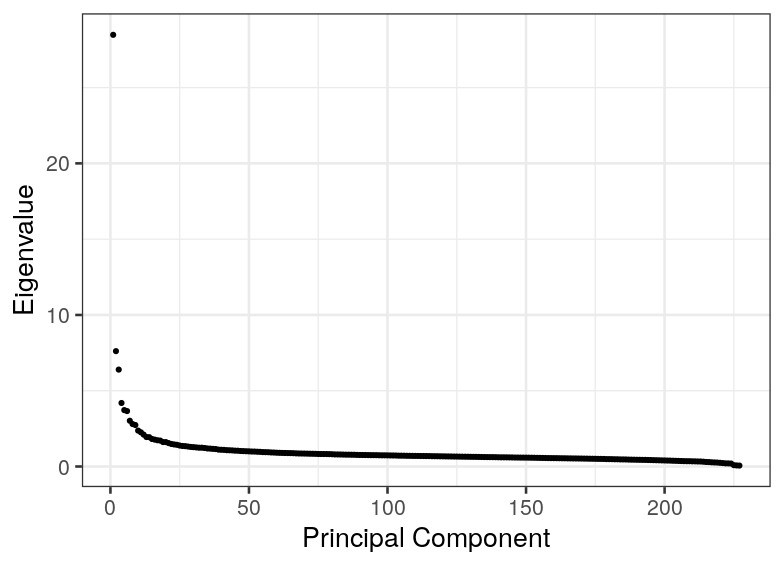}
    \includegraphics[width=0.35\textwidth]{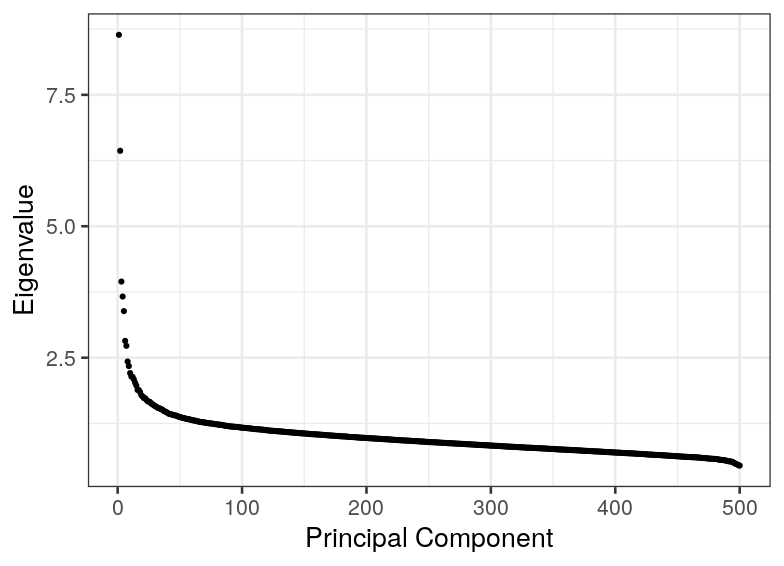}
    \caption{Eigenvalues of correlation matrices of calcium imaging data sets from Allen Institute (left) and Janelia Research Campus (right) from single recording sessions.}
    \label{fig:abalow}
\end{figure}

One characteristic we find in high-dimensional calcium imaging data sets is the low-rankness in the functional activity recording data, which we may be able to exploit to help better solve the Graph Quilting problem. We show two examples of this behavior in Figure \ref{fig:abalow}, in which we plot the eigenvalues from the decomposition of the correlation matrices for the fluorescence traces of single recording sessions from two separate calcium imaging data sets, one from the Allen Institute \citep{lein2007} and another from the Janelia Research Campus \citep{stringer2018}. Here, we tend to observe that the first few principal components appear to explain a substantial proportion of total variance, with rapidly decaying eigenvalues, which suggests an approximate spiked eigenvalue structure \citep{johnstone2001} in the empirical covariance matrix. Because of this common characteristic found in calcium imaging data, we choose to study in this paper an alternative approach that \citep{vinci2019} briefly suggested but chose not to explore: a two-step approach of low-rank covariance completion followed by graph learning.

Many methods specifically for low-rank matrix completion have been extensively studied in previous works \citep{candes2010power,gross2011recovering,recht2011simpler,candes2011tight,chen2020noisy}. However, these methods all assume each entry in the covariance matrix to be missing at random; on the other hand, the Graph Quilting problem assumes a block-wise measurement pattern, in which missingness is systematic and common. Thus, the low-rank matrix completion procedure used for Graph Quilting must be robust with regards to these attributes. Furthermore, the prior works on matrix completion with block-wise missingness all consider different settings from ours, and their methods or theory falls short for our graphical model learning purposes. For example, \cite{cai2016structured} assume that certain rows and columns are fully observed without noise, and that the missing entries form one submatrix. While \cite{zhou2021multi} consider observing multiple blocks of a symmetric PSD matrix and propose a method with solid theory, they assume each block is sampled randomly which leads to overlaps between any two blocks with high probability, making it inappropriate for our calcium imaging application as our blocks are taken sequentially and only consecutive blocks have overlaps. The most closely related work to our Graph Quilting setting is \cite{bishop2014deterministic}. However, they only provide a Frobenius norm error bound, which cannot rule out the situation where the imputed covariance matrix has large errors for a small fraction of its entries; even a small number of badly estimated pairwise covariance can still lead to many false positives the graph estimate. Instead, a sufficiently small $\ell_{\infty}$-norm error bound for the covariance completion step is required and poses a significant challenge for our graph learning purposes.

In this paper, we consider several potential approaches to the Graph Quilting problem in the case where the full sample covariance matrix is assumed to be either exactly or approximately low-rank, and we study the potential application of these methods for estimating functional connectivity networks from calcium imaging data. All of the methods discussed in this work follow the two-step covariance completion graph quilting framework discussed above. Specifically, we incorporate several different low-rank covariance completion methods currently used in the literature and apply them as part of the covariance imputation step of our proposed low-rank graph quilting methods. While the performance of these methods have been studied for the general problem of imputing missing values of a covariance matrix, we consider their potential applicability in a novel context, specifically with respect to the Graph Quilting problem for calcium imaging data. We first study whether our proposed general two-step approach is appropriate for the Graph Quilting problem from a theoretical perspective. To do this, we show that an entry-wise error bound for the imputed covariance matrix is required for graph selection consistency; this type of error bound has not been previously proven in the literature on matrix completion for block-wise missingness patterns. Furthermore, we show the entry-wise error bound requirements hold for one of the imputation methods, leading to graph selection consistency guarantees of the corresponding low-rank graph quilting method. We then compare the empirical performance of the different low-rank Graph Quilting methods, along with the MAD\textsubscript{gq} method of \cite{vinci2019}, through a simulation study as well as through two examples using real-world calcium imaging data sets; these empirical studies suggest that our low-rank Graph Quilting approaches have superior performance compared to previous mthods.

\begin{figure}[t]
\begin{center}
\minipage{0.24\textwidth}%
  \includegraphics[width=\linewidth]{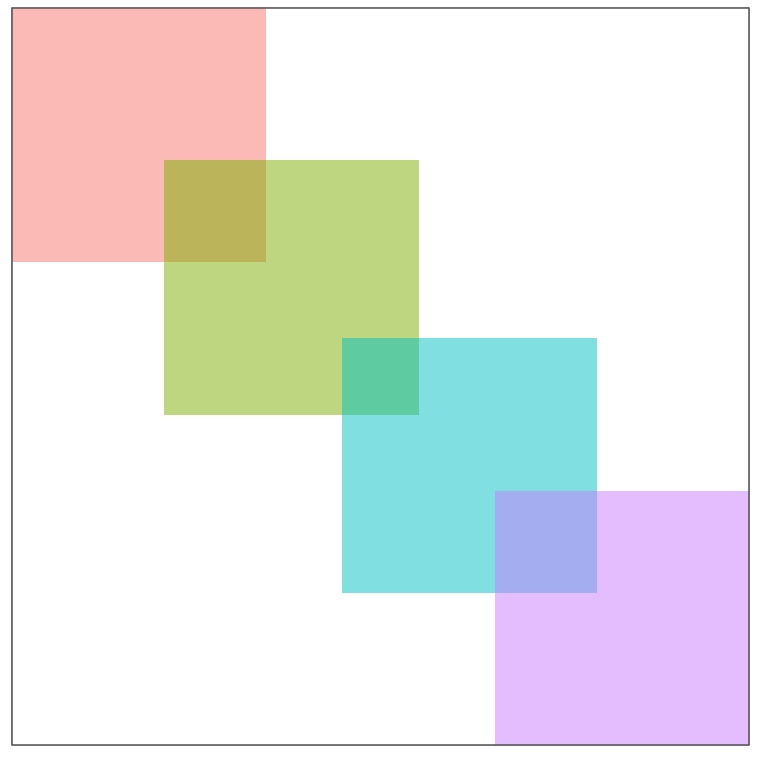}
\endminipage \Huge{ = }%
\minipage{0.12\textwidth}%
  \includegraphics[width=\linewidth]{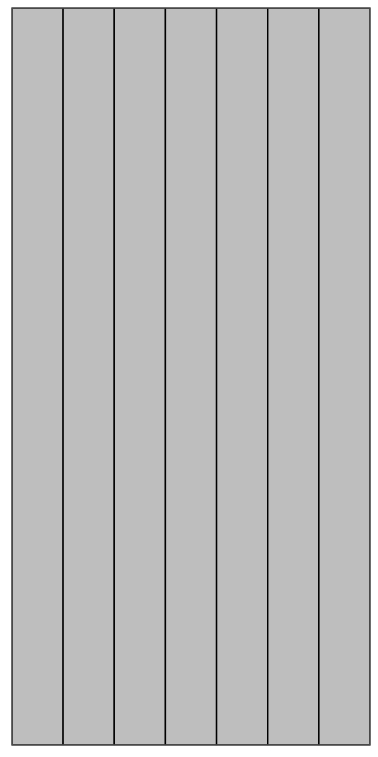}
\endminipage \Huge{$\times$}%
\minipage{0.12\textwidth}%
  \includegraphics[width=\linewidth]{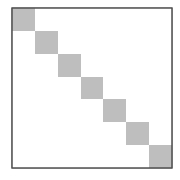}
\endminipage \Huge{$\times$}%
\minipage{0.24\textwidth}%
  \includegraphics[width=\linewidth]{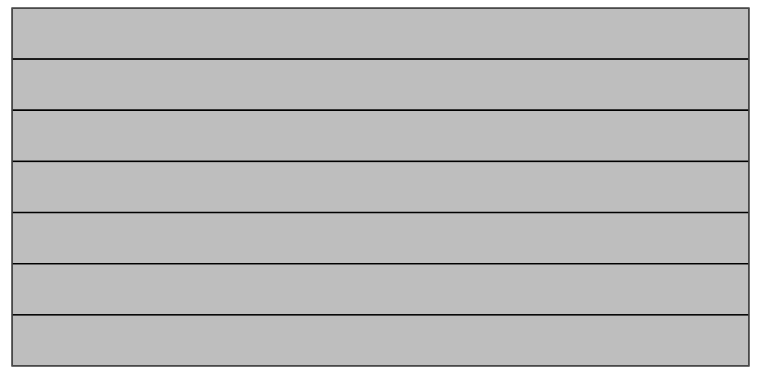}
\endminipage
\end{center}
\caption{An example of an incomplete empirical covariance matrix for four partially overlapping patches of nodes; each square represents the nodes in a particular patch, while the parts of the covariance matrix in not covered are never jointly observed.}
\end{figure}

The paper is organized as follows. In Section \ref{sec:methods}, we formally introduce the general two-step graph quilting algorithm and three specific models for low-rank Graph Quilting, and discuss the theoretical justification for these methods as well as practical model selection procedures. We study the empirical performance of low-rank graph quilting on simulation studies in Section \ref{sec:sim}. Lastly, in Section \ref{sec:cal}, we investigate the efficacy of the low-rank Graph Quilting methods for estimating functional neuronal connectivity from calcium imaging.


\section{Low-Rank Graph Quilting} \label{sec:methods}

We start by defining the notations that will used throughout the paper. For any matrix $\mathbf{A}\in \mathbb{R}^{p_1\times p_2}$, we let $\|\mathbf{A}\|_{\max} = \max_{1\leq j\leq p_1,1\leq k\leq p_2}|\mathbf{A}_{j,k}|$ be the maximum absolute value of its entries; let $\|\mathbf{A}\|_{F} = \left(\sum_{j,k}\mathbf{A}_{j,k}^2\right)^{\frac{1}{2}}$ be its Frobenius norm; and let $\|\mathbf{A}\|_* = \sum_{j=1}^{\min\{p_1,p_2\}}s_j(\mathbf{A})$ be the nuclear norm, where $s_j(\mathbf{A})$ is the $j$th singular value of $\mathbf{A}$. If $\mathbf{A}$ is a square matrix with $p_1=p_2$, we let $\|\mathbf{A}\|_{1, \, off} = \sum_{1\leq j\neq k\leq p_1}\mathbf{A}_{j,k}$ denote its off-diagonal $\ell_1$ norm.

We first define the general Graph Quilting problem, following the prior work of \cite{vinci2019}.
Consider the Gaussian graphical model, where each sample vector $\xx_i\in \mathbb{R}^p$ follows Gaussian distribution $\mathcal{N}(0, \BSig^*)$, with mean zero and covariance $\BSig^*\in \mathbb{R}^{p\times p}$. The primary objective is to recover a sparse inverse covariance matrix, denoted as $\BTh^* = \BSig^{*-1}$, whose non-zero pattern encodes the conditional dependency structure between the $p$ features \citep{ravikumar2010}: we want to recover the edge set $E = \{(j,k): \BTh^*_{j,k}\neq 0\}$. However, instead of having simultaneous observations for the full set of joint feature pairs, we only have joint observations for $K$ partially overlapping subsets of features, denoted as $k \in \{1, \hdots, K\}$. We denote the set of features observed in each subset $k$ as $V_k$ of size $p_k<p$, and the corresponding observed data matrix as $\XX^{(k)}\in \R{n_k}{p_k}.$ From this, we define the full pairwise observation set as $O = \bigcup_{k = 1}^K V_k \times V_k.$ Our goal is to obtain a graphical model estimate of the full feature set from the incomplete measurements $\{\XX^{(k)}\}_{k=1}^K.$ Without making any additional assumptions, this is an extremely challenging task; as discussed in \cite{vinci2019}, even when the number of samples $n_k$ for each block $k$ approaches infinity and the sub-covariance of each block is perfectly estimated, the whole graph structure is still non-identifiable, particularly for the edges between nodes that are never jointly observed. Therefore, instead of pursuing full graph recovery, \cite{vinci2019} proposes a method to estimate a super set of the edges in $O^c$. However, as has been noted in Figure \ref{fig:abalow}, the real calcium imaging data sets exhibits an approximately low-rank structure, motivating us to leverage such structures to develop a new approach suited to the neuroscience applications.

More specifically, let us define the observed incomplete sample covariance matrix $\widehat{\BSig}_O = \{\widehat{\BSig}_{ij}: (i, j) \in O\},$ computed empirically from the available joint observation pairs in $\{\XX^{(k)}\}_{k=1}^K$. In particular, we compute the sample covariance for each pair of nodes based on their joint observations:
\begin{equation*}
\begin{split}
    \widehat{\BSig}_{ij}=\widehat{m}_{ij}-\widehat{m}_i\widehat{m}_j,\quad \widehat{m}_{ij}=\frac{1}{\sum_{k:\,i,\,j\in V_k}n_k}\sum_{k:
    \,i,\,j\in V_k}\XX^{(k)\top}_{:,i}\XX^{(k)}_{:,j},\quad \widehat{m}_{i}=\frac{1}{\sum_{k:\,i\in V_k}n_k}\sum_{k:\,i\in V_k}\XX^{(k)\top}_{:,i}1_{n_{k}},
\end{split}
\end{equation*}
where we note that the summation is over the observational blocks indexed by $k$.
One possible framework that can be applied for the Graph Quilting problem is a two-step process where we first apply covariance completion methods to obtain a full covariance estimate, denoted by $\widetilde{\BSig}$, before the graphical Lasso \citep{originalglasso, friedman2008} is applied to get an estimated inverse covariance, denoted by $\widehat{\BTh}_G$, encoding the estimated graph structure; we outline this approach in Algorithm \ref{alg:lrgq}. For the first step, we impose a low-rank or approximately low-rank structure on the imputed covariance matrix $\widetilde{\BSig}$; we call this the low-rank Graph Quilting problem.
\spacingset{1.1}
{\LinesNumberedHidden
\begin{algorithm}[t] 
  \caption{Two-Step Low-Rank Graph Quilting} \label{alg:lrgq}
  \begin{small}
  \SetAlgoNoLine
  \SetKwFunction{Union}{Union}\SetKwFunction{FindCompress}{FindCompress}
    \SetKwInOut{Input}{Input}\SetKwInOut{Output}{Output}
   \textbf{Input:} Incomplete observed covariance matrix $\widehat{\BSig}_O$, sparsity tuning parameter $\lambda$, rank of full covariance matrix $r$.\\
  \vspace{5pt}
    (1) Obtain imputed covariance matrix $\widetilde{\BSig}$ using low-rank covariance completion methods.\\
    \vspace{5pt}
    (2) Apply the graphical Lasso to the imputed full covariance matrix $\widetilde{\BSig}$ in order to obtain the estimated graph $\widehat{\BTh}_G$: $$\widehat{\BTh}_G = \argmin_{\BTh \iRPP, \BTh \succ 0}  \, \Tr(\widetilde{\BSig}\BTh) - \log \det(\BTh) + \lambda \|\BTh\|_{1, \mathrm{off}} $$\\
\vspace{5pt}
    \textbf{Output:} $\widetilde{\BSig}$, $\widehat{\BTh}_G$
    \end{small}
\vspace{2pt}
\end{algorithm}}
\spacingset{1.9}

For the covariance completion step, we consider two ways for imposing the low-rankness on the imputed covariance. One natural idea might be to consider a spiked eigenvalue structure for the population covariance matrix, i.e. 
\begin{equation}\label{eq:apprlr_cov_model}
    \BSig^*  = \LL^* + \sigma^{*2}\II, \, \LL \iRPP, \, \sigma^{*2} \in \mathbb{R}^{+},
\end{equation} where $\LL^*$ is a low-rank positive definite matrix of rank $r^*$; this formulation gives an approximate low-rank structure for small values of $\sigma^{*2}$ \citep{johnstone2001}. We will further discuss \eqref{eq:apprlr_cov_model} in Section \ref{sec:low-rank-graph_disc} in the context of graphical models. Motivated by \eqref{eq:apprlr_cov_model}, we will propose methods that encourage a spiked eigenvalue (approximately low-rank) structure for $\widetilde{\BSig}$. Another possibility for the covariance imputation step is to constrain $\widetilde{\BSig}$ to have an exactly low-rank structure with rank $r$. 
Although the population covariance is invertible, the sample covariance computed from fully observed data, if available, would not be of full-rank under the high-dimensional setting. Imposing an exactly low-rank structure on the $\widetilde{\BSig}$ can also sometimes be desirable for the purpose of imputing the sample covariance, a task that may be important in different applications. In Sections \ref{sec:bsvdgq} and \ref{sec:NNgq_LRFgq}, we will introduce three different potential low-rank covariance completion schemes for the first step of the Graph Quilting estimation process to derive a full covariance matrix $\widetilde{\BSig}$ from the incomplete observed covariance matrix $\widehat{\BSig}_O$; detailed computational procedures for each of the methods are shown in Sections A and B of the Supporting Information.

\subsection{Low-Rankness in Graph Structures}\label{sec:low-rank-graph_disc}

Before introducing the covariance completion methods, we first discuss the rationale behind the low-rank assumption and the two-step procedure in the context of graphical models. In the context of graphical models, assuming approximate low-rankness of the covariance matrix is not a straightforward idea. One intriguing question is how the low-rank assumption we made constrains the class of graphical models. In fact, low-rank graphs are the most typical graphical models that can have approximately low-rank covariances; in particular, if the weighted adjacency matrix is low-rank ($\LL_0$ in Proposition \ref{prop:lrgraph}) with appropriate spectrum, the true covariance matrix will satisfy the low-rank + diagonal decomposition and hence justifies our low-rank covariance completion approaches under model \eqref{eq:apprlr_cov_model}.
\begin{prop}\label{prop:lrgraph}
    Consider graph $\mathcal{G}=(V,E)$ with precision matrix $\BTh^*=c\II - \LL_0\succ 0$, where $\LL_0$ is a rank-$r$ positive semi-definite matrix and $\lambda_r(\LL_0)>\frac{c}{2}$. Then there exists another rank-$r$ positive semi-definite matrix $\LL$, such that $\BSig^* = \BTh^{*-1} = \LL + \frac{1}{c} \II$, with $\lambda_r(\LL)>\frac{1}{c}$. 
\end{prop}
\begin{figure}[t]
    \centering
    \includegraphics[width = 0.75\textwidth]{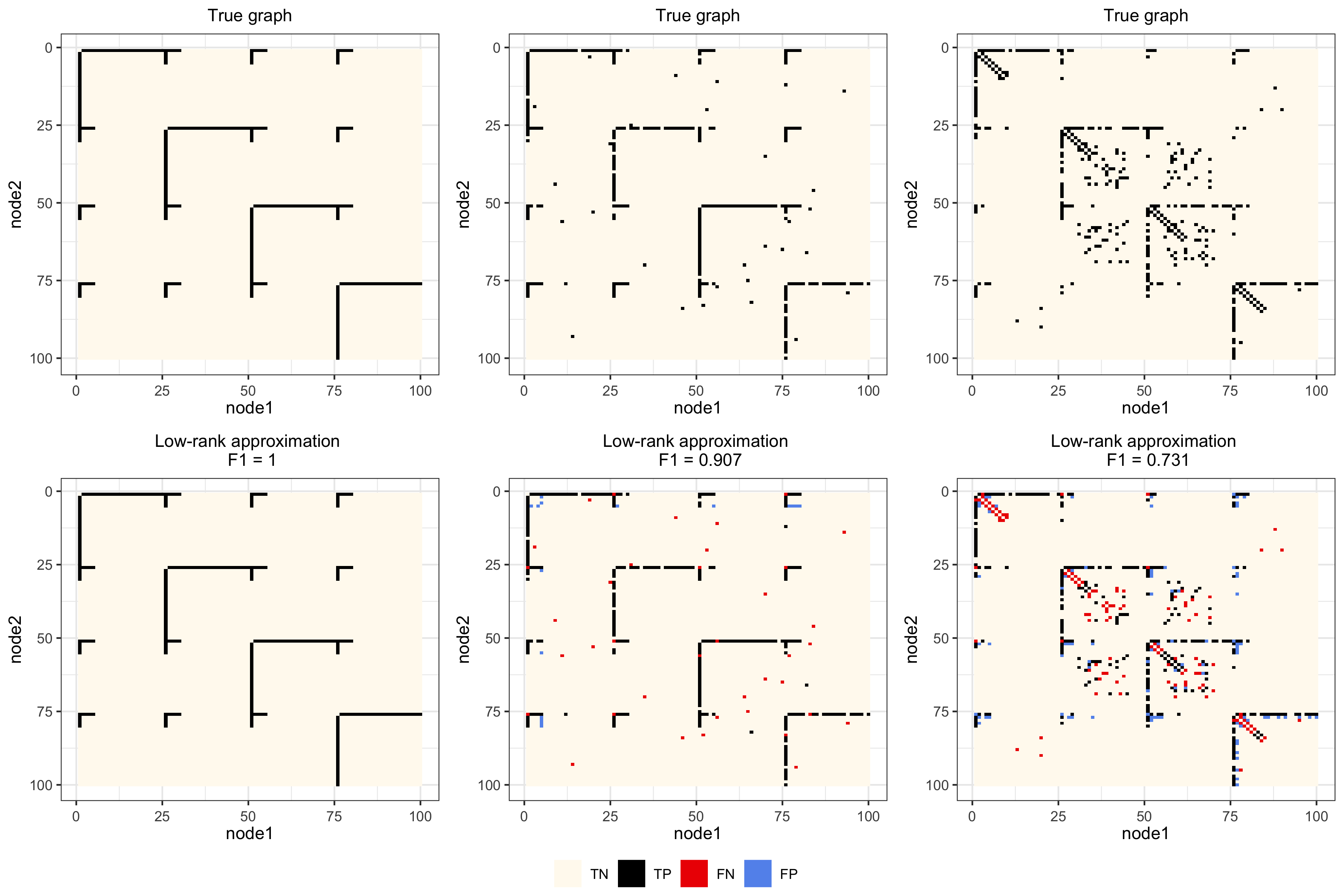}
    \caption{\small Three true graphs (top), and best approximation graph when constraining the nuclear norm of the covariance ($\|\Sigma^*\|_*\leq 2\|\Sigma^*\|$) (bottom). From left to right, we consider a four-star graphs with interconnected stars, a star graph with some edges randomly changed, and a combination of a star, a chain, an Erd\"{o}s R\'{e}nyi, and a small-world graph.}\label{fig:pop_illustration}
\end{figure}
Although low-rankness and sparsity often do not appear together, there are indeed a class of graphs satisfying both and commonly studied in the literature, e.g. multi-star graphs and block graphs, as well as graphs with repeated low-rank motifs. As depicted by many existing graph theory \citep{newman2018networks}, the top singular subspace of graphs often reflects valuable connectivity properties, such as {\em centrality, community, and hubs}. Thus, {\em when making a low-rank approximation for a non-low-rank covariance in a graphical model, we may still preserve important graph structures such as hubs and communities.} To make this intuition more concrete, we conduct a toy simulation on three graph examples, presented in Figure \ref{fig:pop_illustration} (top row). We then project each population covariance matrix onto a small nuclear norm ball and apply the neighborhood Lasso to obtain a new graph (bottom row). The left column is a low-rank four-star graph with interconnected stars; here imposing the low-rank constraint does not alter the graph structure at all. When the graph deviates from the exact low-rank structure, we see that the edges affiliated with the non-low-rank component tend to be missed in the bottom graphs, while the hubs and communities can still be captured, hence supporting our intuition above.

\subsection{Theoretical Motivation for the Two-Step Procedure}
Under the (approximately) low-rank covariance assumption, our choice of using a two-step procedure is motivated by prior theoretical results in \cite{liu2012high}, in which the success of the graphical Lasso is shown to solely depend on the entry-wise estimation error of the covariance matrix. Hence, any imputation method can be applied in the first step, as long as the the imputed covariance is sufficiently close to the true covariance. This result is summarized as an informal meta-theorem as follows: 
\begin{thm}[Graph Selection Consistency of Algorithm 2.1]\label{prop:main}
Consider Algorithm 2.1 and its output $\widehat{\BTh}_{G}$.  
If Assumption 2.1 (incoherence condition) in the Supplement holds, $\|\widetilde{\BSig}-\BSig^*\|_{\max}$ is sufficiently small, and $\lambda$ is appropriately chosen,
then
$\{(i,j):i\neq j, (\widehat{\BTh}_G)_{ij}\neq 0\}=\{(i,j):i\neq j, \BTh^*_{ij}\neq 0\}$.
\end{thm}
\noindent The detailed version of Theorem \ref{prop:main} is included in Section C of the Supporting Information. The previous theorem implies that graph selection consistency of our low-rank Graph Quilting methods can be shown via the infinity-norm bounds of the covariance imputation procedure used in the first step of low-rank Graph Quilting algorithm, an error bound that has not been well-studied under the block-wise missingness assumption. 

\subsection{Block Singular Value Decomposition (BSVDgq)}\label{sec:bsvdgq}
The first method we study utilizes the procedure proposed by \cite{bishop2014deterministic} for covariance completion, which applies sequential blockwise singular value decompositions on overlapping principal submatrices of the full covariance matrix in order to impute the missing values. Specifically, the algorithm finds the singular value decomposition for each principal submatrix sequentially while also performing orthonormal transformations of the overlapping parts of the principal submatrices in between iterations in order to align the submatrices. While this is not the most common method for low-rank covariance completion, we consider it because it also assumes a block-wise structure on the pairwise observation subsets. In particular, for the Graph Quilting problem, we can use each of the partially overlapping subsets of observed feature pairs as the principal submatrices that are used for imputation with this method. This can be done with the $\widehat{\BSig}_O$ matrix in order to achieve an exact low-rank solution, or with $(\widehat{\BSig}_O - \widehat{\sigma}^2\II)$ as the input for an approximately low-rank covariance matrix, where $\widehat{\sigma}^2$ is an estimate of $\sigma^{*2}$. To estimate the latter quantity, we use $\widehat{\sigma}^2 = \text{median}(\widehat{\BSig}_{ii}), \, i \in \{1, \hdots, p\},$ as has been proposed previously for spiked models in \citep{johnstone2009consistency, cai2015optimal}.

Since this covariance completion approach has not been applied before in the context of graphical model estimation, we present some preliminary theoretical results to demonstrate the validity of the BSVDgq method for solving the Graph Quilting problem. Inspired by the condition on the imputation error $\|\widetilde{\BSig}-\BSig^*\|_{\max}$ in Theorem \ref{prop:main}, we show that this error term can indeed be bounded appropriately for the BSVDgq covariance imputation algorithm with sufficiently large sample sizes, leading to a consistent estimate of the graph.
Consider the spiked covariance model \eqref{eq:apprlr_cov_model}, where $\boldsymbol{L}^*$ is positive definite and of rank $r^*$. For $1\leq k\leq K$, 
define the effective rank of each block by $\tau_k=\frac{\mathrm{tr}(\BSig^*_{V_k,V_k})}{\lambda_1(\BSig^*_{V_k,V_k})}$.
\noindent A theoretical guarantee for the BSVDgq method is as follows: 
\begin{thm}[Guarantees for BSVDgq]\label{thm:bsvdgq}
Under Assumptions C.2-C.4 in Appendix C, with probability at least $1-C\sum_{k=1}^Kp_k^{-c}$,
the output $\widetilde{\BSig}$ of the BSVDgq algorithm with the input $\widehat{\BSig}_O-\sigma^{*2}\II$ and $r=r^*$ satisfies
\begin{equation}\label{eq:BSVD_err}
    \|\widetilde{\BSig}-\BSig^*\|_{\max}\leq C^*\max_k\sqrt{\frac{(r+\tau_k)(\tau_k \vee \log p_k)}{n_k}}, 
\end{equation}
where $c>0$ is a universal constant, and $C^*>0$ is a constant depending on the true covariance, whose specific form is included in Section C of the Supplement.
\end{thm}
\noindent The detailed version of Theorem \ref{thm:bsvdgq}, its full proof, more discussion on its implication can be found in Section C and D of the Supplement. Theorem \ref{thm:bsvdgq} shows that the imputed covariance matrix based on the block singular value decomposition method can be entry-wise close to the true covariance if the sample size for each block sufficiently large compared to the rank and effective ranks of the true covariance. Here, Theorem \ref{thm:bsvdgq} assumes $\sigma^{*2}$ to be known only for simplicity, while we expect similar results would still hold qualitatively when $\sigma^{*2}$ is estimated well by $\widehat{\sigma}^2$. 
The error bounds in Theorem \ref{thm:bsvdgq} and Theorem \ref{prop:main} immediately implies the following graph selection consistency of the BSVDgq algorithm:
\begin{cor}\label{cor:bsvdgq_consistency}
Suppose we apply the block SVD algorithm with input $\widehat{\BSig}_O-\sigma^{*2}\II$ and $r=r^*$ as the first step of Algorithm~\ref{alg:lrgq}. If Assumptions C.1-C.4 in Supporting Information C hold, $\lambda$ is appropriately chosen, and for $1\leq k\leq K$,
$n_k\geq C^*\left(r+\tau_k\right)(\tau_k\vee \log p_k)$,
where $C^*$ depends on the model parameters, then with probability at least $1-C\sum_{k=1}^Kp_k^{-c}$, we achieve exact edge recovery of the graph:
$\{(i,j):i\neq j, (\widehat{\BTh}_G)_{ij}\neq 0\}=\{(i,j):i\neq j, \BTh^*_{ij}\neq 0\}$.
\end{cor}
 Under comparable sample size conditions to the literature in graphical models and spiked covariance estimation \citep{ravikumar2011high,koltchinskii2017concentration}, Corollary \ref{cor:bsvdgq_consistency} establishes that the graph consistency can be achieved with high probability using our BSVDgq method. Since in calcium imaging data sets, each session of the neuronal recordings usually spans hours with around 5-10 Hz frequencies ($n_k \approx 10,000$) and covers thousands of neurons ($p_k\approx 5000$) \citep{microns2021functional}, Corollary \ref{cor:bsvdgq_consistency} is an encouraging result for us to apply this method on our motivating neuroscience applications.
 
 
\subsection{Nuclear Norm Penalization (NNgq) and Low-Rank Covariance Factorization (LRFgq)}\label{sec:NNgq_LRFgq}

Here, we propose two other approaches for low-rank graph quilting which solve two squared loss minimization problems to infer the full covariance matrix in the first step of the graph quilting procedure. The first of these, which we call nuclear norm penalization or NNgq for short, uses the objective of minimizing the sum of a Frobenius norm penalty on the difference between the imputed and observed covariance matrix and a nuclear norm penalty on the imputed covariance matrix \citep{mazumder2010, kolt2011}. In the case of an exact low-rank assumption on the completed covariance matrix, this gives us the objective function
\begin{equation}\label{eq:nnmgq}
    \widetilde{\BSig} = \argmin_{\BSig \iRPP} \frac{1}{2} \|\BSig_O - \widehat{\BSig}_O\|_F^2 + \nu \|\BSig\|_{*},
\end{equation} while for the approximate low-rank assumption we get
\begin{equation}\label{eq:NNM_comp_spiked}
\begin{split}(\widehat{\LL}, \widehat{\sigma}^2) =& \argmin_{\LL \iRPP, \, \sigma^2 \in \mathbb{R}^{+}} \frac{1}{2} \|(\LL + \sigma^2\II)_O - \widehat{\BSig}_O\|_F^2 + \nu \|\LL\|_{*}; \; \widetilde{\BSig} = \widehat{\LL} + \widehat{\sigma}^2\II.
\end{split}
\end{equation} A proximal gradient descent algorithm can be used in order to derive estimates from the likelihood, as by the symmetry and convexity of the loss function $\frac{1}{2} \|\BSig_O - \widehat{\BSig}_O\|_F^2 + \nu \|\BSig\|_{*}$, $\widetilde{\BSig}$ is guaranteed to be symmetric. In the case where $\widehat{\BSig}_O$ is not positive semi-definite, we project $\widehat{\BSig}_O$ to the positive semi-definite cone using a weighted $\ell_{\infty}$ norm loss. 

One other common method for low-rank matrix completion utilizes the low-rank factorization and solves an optimization problem with respect to the low-rank factors 
\citep{kesh2009, wen2012}. We consider a low-rank approximation of the full unobserved covariance matrix such that it 
can be factorized as $$\BSig = \UU\UU^{\top},\, \UU \iRNR,\, r \ll p.$$ Following this, we minimize of the Frobenius norm between the observed portion of the covariance matrix and the corresponding entries of the imputed full covariance matrix for the first step of the graph quilting procedure described above. For the exact low-rank covariance completion, this gives us 
\begin{equation}
\begin{split}\widehat{\UU} =& \argmin_{\UU \in \mathbb{R}^{p \times r}} \frac{1}{2} \|(\UU\UU^{\top})_O - \widehat{\BSig}_O\|_F^2; \; \widetilde{\BSig} = \widehat{\UU}\widehat{\UU}^{\top},
\end{split}
\end{equation} while for an approximate low-rank covariance matrix we can use the objective 
\begin{equation}\label{eq:LRF_spiked_loss}
     (\widehat{\HH}, \widehat{\sigma}^2) = \argmin_{\HH \in \mathbb{R}^{p \times r}, \, \sigma^2 \in \mathbb{R}^{+}} \frac{1}{2} \|(\HH\HH^{\top} + \sigma^2\II)_O - \widehat{\BSig}_O\|_F^2.
\end{equation} This method can utilize the resulting imputed covariance matrix from either the BSVDgq or NNgq described above for initialization; estimates from the likelihood are then found using gradient descent. 

Both the covariance completion techniques used in the NNgq and LRFgq procedures have been previously studied for the imputation of low-rank matrices from both a theoretical and empirical perspective \citep{candes2010power,candes2011tight,ma2018implicit}. However, these works have primarily been focused on the case when the missing elements are random. For the Graph Quilting problem, we instead assume that the non-missing entries are arranged as semi-overlapping blocks, meaning that the missingness is highly patterned. Because of this, it is unclear whether existing empirical studies and theoretical guarantees for these covariance completion methods will apply for the graph quilting problem. We investigate the former in Sections \ref{sec:sim} and \ref{sec:cal}, and we leave the development of theoretical properties as future work.

\subsection{Practical Model Selection}\label{sub:pms}

The low-rank Graph Quilting methods require two hyperparameters to be selected: the rank or nuclear norm penalty of the imputed covariance matrix, as well as the sparsity penalty on the subsequent graphical model estimation. To select the hyperparameters that control the rank of the full covariance matrix in the first step of low-rank Graph Quilting, different techniques will be required depending on method. For the low-rank covariance factorization and block singular value decomposition methods, the rank can be selected using a penalized maximum likelihood criteria such as BIC \citep{burnham2004}; for this particular problem either the raw rank or the nuclear norm of the resulting imputed covariance matrix can be penalized. For the nuclear norm penalization approach, we can apply a cross-validation procedure in which scattered feature pairs are randomly selected to be removed from the observed set and used to compare the imputation estimates along a chosen regularization path \citep{mazumder2010}. Then, to select the sparsity of the estimated graphical model, a stability selection approach such as the one proposed in \citep{liu2010} can be used.

\section{Simulation Studies}\label{sec:sim}

\subsection{Illustrative Example}

\begin{figure}[t]
    \centering
    \includegraphics[width = 0.75\textwidth]{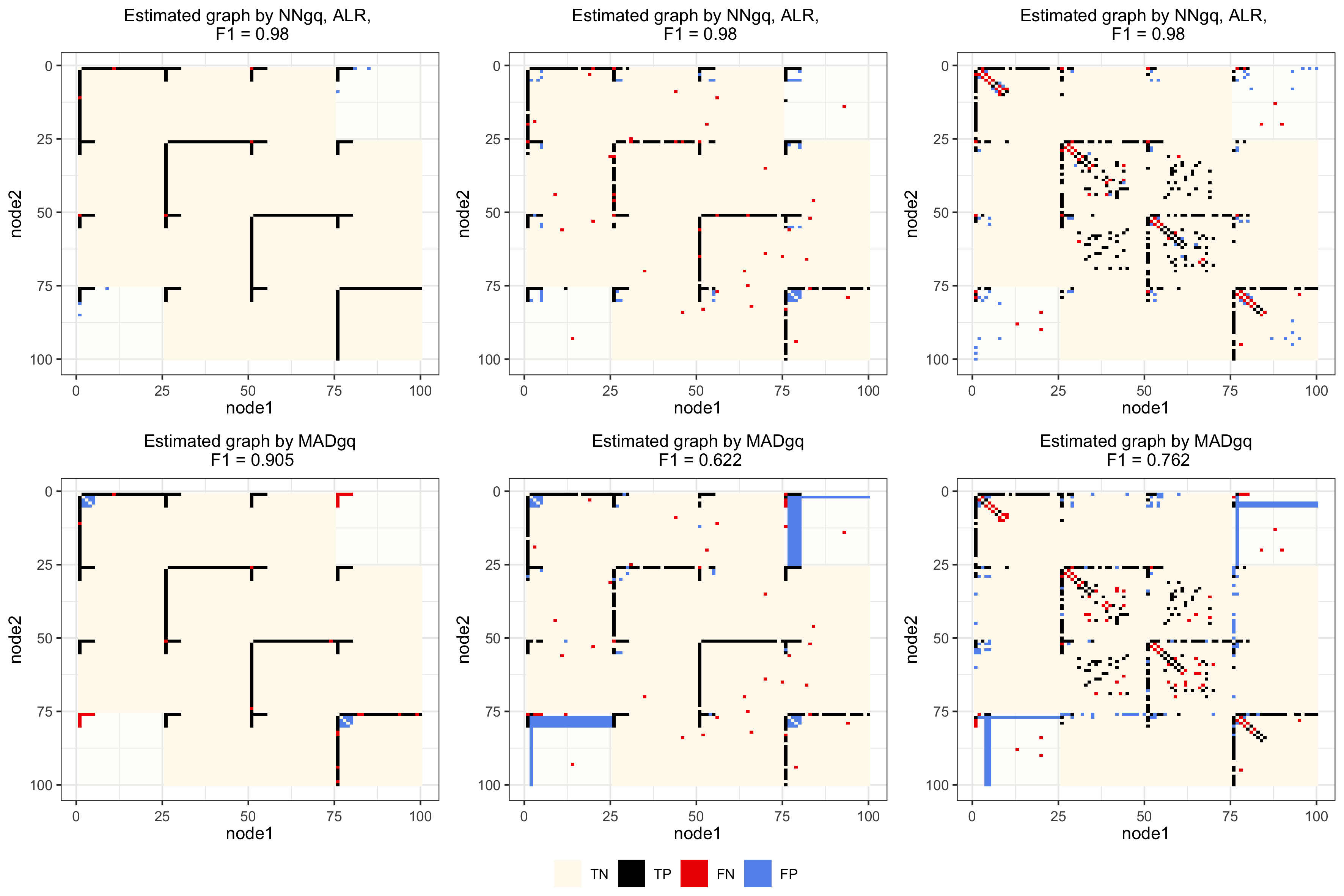}
    \caption{\small Illustration of low-rank graph quilting (NNgq) vs. MAD\textsubscript{gq} \citep{vinci2019} on the graph examples shown in Figure \ref{fig:pop_illustration}, under graph quilting observational pattern. NNgq identifies the edges in $O^c$ better if they are affiliated with the low-rank component.}
    \label{fig:FiniteSample_illustration}
\end{figure}

To further aid the intuition and understand our methods, we first return to the toy graph examples presented in Figure \ref{fig:pop_illustration} in Section \ref{sec:methods} in order to illustrate how our method works for exactly low-rank and non-low-rank graphs given quilting observations. We consider two observational blocks, each of size $75$, highlighted by the solid background in Figure \ref{fig:FiniteSample_illustration}. This shows the graph estimation results using our NNgq method and the MAD\textsubscript{gq} method \citep{vinci2019}. MAD\textsubscript{gq} tends to either not identify any edge or select too many false positives in $O^c$ (as depicted by the theory in \cite{vinci2019}); while our NNgq method can recover the edges in $O^c$ much better if they are affiliated with the low-rank component, making hub and community detection easier. More details on the simulation setup and algorithm implementations are included in the Supporting Information.

\subsection{Systematic Simulation Studies}

We now investigate the performance of the low-rank graph quilting methods on two systematic simulation studies, one using data generated a Gaussian graphical model and one using data from a real-world calcium imaging data set. In both simulation studies, we compare each of the low-rank Graph Quilting methods described in Section \ref{sec:methods} with both the exact low-rank and spiked covariance model assumptions, along with the MAD\textsubscript{gq} algorithm proposed by \citep{vinci2019} and a zero imputation approach. We evaluate the Graph Quilting approaches based on three different criteria: the Frobenius norm of the difference between the imputed and true correlation matrices, i.e. $\|\widetilde{\BSig} - \BSig\|_F,$ the infinity norm of the difference between the imputed and true covariance matrices, i.e. $\|\widetilde{\BSig} - \BSig\|_{\infty}$, and the F1 scores of the graphical Lasso estimates derived from each of the imputed covariance matrices with respect to recovering the set of non-zero entries in the true underlying graph. Sparsity for the graph is selected using oracle tuning with respect to the full covariance matrix in order to fairly compare all methods; for the MAD\textsubscript{gq} algorithm, we additionally set the minimum threshold hyperparameter to be 0. For each simulation setting, we run 50 replications and report the mean and standard deviation of the Frobenius norm, infinity norm, and F1 scores metrics. Code for simulating data, covariance imputation, and graph estimation are available at \url{https://github.com/DataSlingers/LowRankGraphQuilting}.

\subsubsection{Gaussian Graphical Model}

Below, we study the proposed Graph Quilting methods on data simulated from Gaussian graphical models. For each simulation trial, we generate an inverse covariance matrix with a pre-specified structure, which is then used to produce an $n \times p$ data matrix from a multivariate Gaussian distribution and a fully observed empirical covariance matrix. After centering and scaling the columns of the data, we create a partially observed empirical covariance matrix, structured as $K$ patches of $o$ features each, with the features shuffled so that the patch assignment of each node is independent of its neighborhood set. We use these as the input to each of the low-rank and zero-imputation Graph Quilting methods in order to obtain imputed covariance matrices and estimated graphs. We separately apply the MAD\textsubscript{gq} algorithm directly from the partially observed empirical covariance matrix. 

We investigate the performance of the LRGQ methods on three different graph structures: a stochastic block diagonal graph of 5 communities with edge probabilities of 0.8 within each block group and with no edges outside of the blocks, a multistar graph with 4 hubs in which each non-hub node is connected to exactly one hub node, and an Erd\"{o}s R\'{e}nyi graph with edge probability 0.02. Non-zero entries for each graph are generated from a uniform distribution of range 0 to 2, and diagonal entries are initially generated from a uniform distribution of range 1 to 2. For each graph, we also ensure positive definiteness by subsequently adding a constant to each diagonal entry. Below, we study the case with data matrices with $n = 2000$ observations of $p = 100$ features for patch sizes $o = 55, 60$ and $65$ with $K = 2$ patches, and for $K = 2, 3$ and $4$ patches with patch sizes $o = 60, 40$ and $30$, respectively. To select the rank of each procedure for the graph types that induce low-rankness, we use the optimal rank, i.e. 5 for the stochastic block model and 4 for the multistar graph. For the Erd\"{o}s R\'{e}nyi graph, we limit the estimated rank to 20 in order to study the effect of imposing a low-rank assumption to a non low-rank graph structure.

\begin{figure}[t]
\begin{center}
    \begin{subfigure}[t]{0.3\linewidth}
    \centering
        \includegraphics[width=\linewidth]{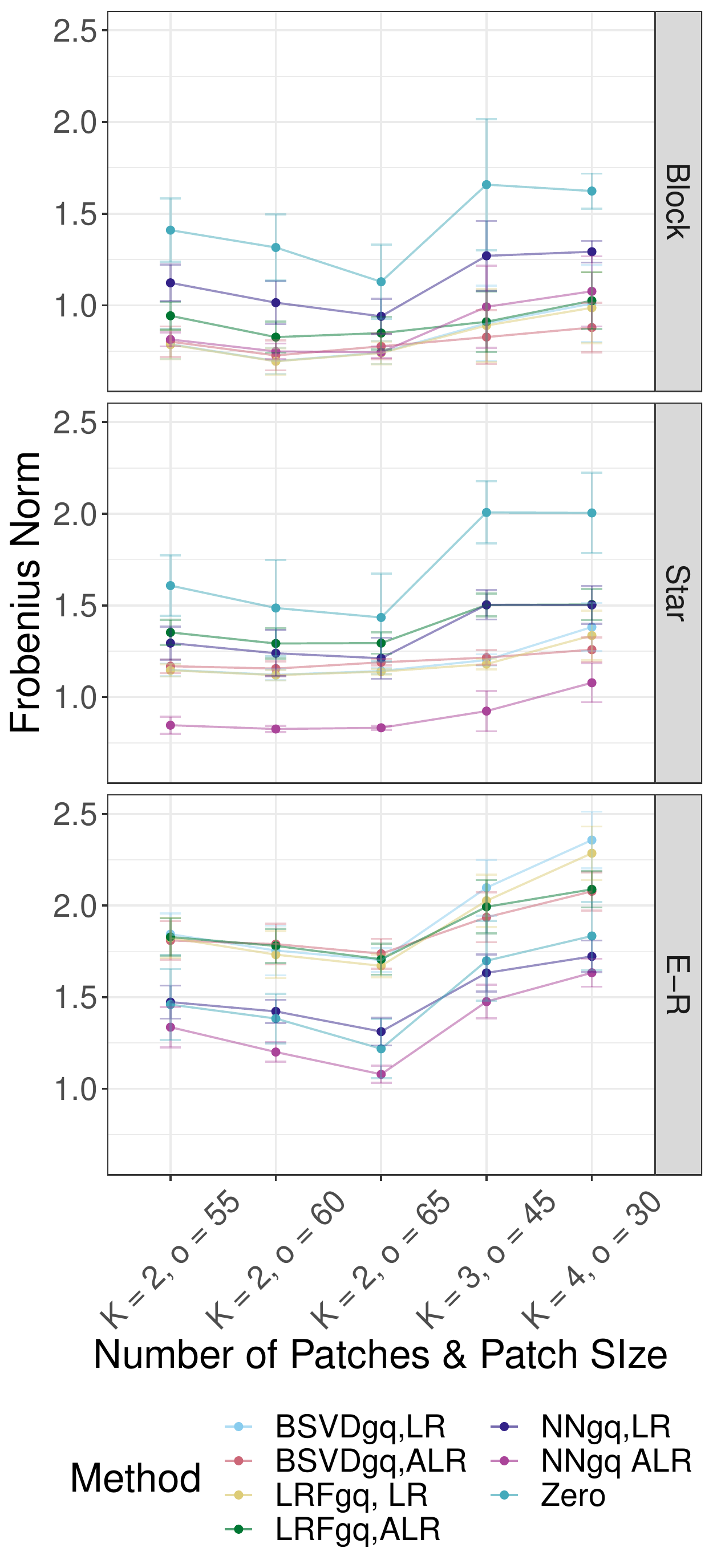}
        \caption{Frobenius norm.}
    \label{fig:simfrob}
    \end{subfigure}
    \begin{subfigure}[t]{0.3\linewidth}
    \centering
        \includegraphics[width=\linewidth]{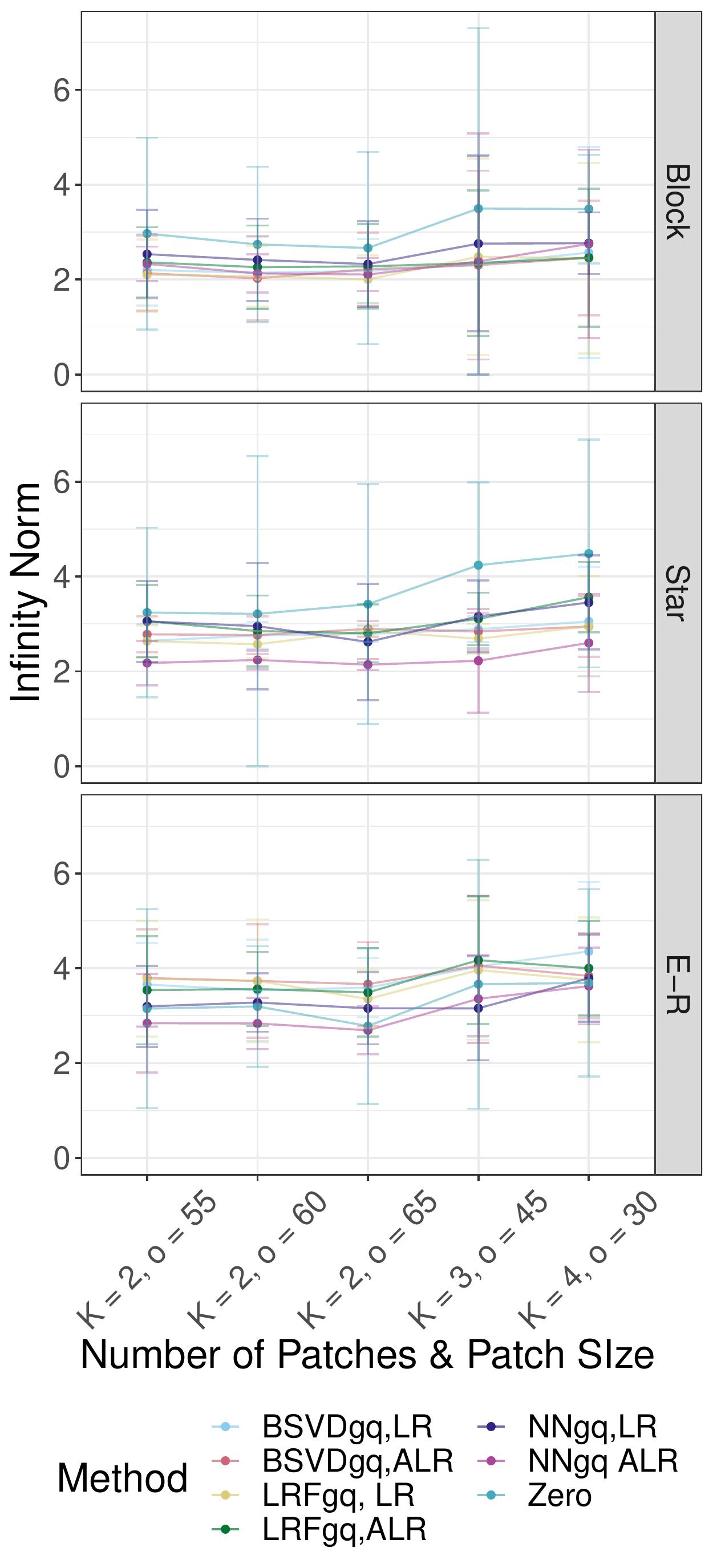}
        \caption{Infinity norm.}
    \label{fig:siminf}
    \end{subfigure}
    \vspace{-0.5cm}
    \begin{subfigure}[t]{0.3\linewidth}
    \centering
        \includegraphics[width=\linewidth]{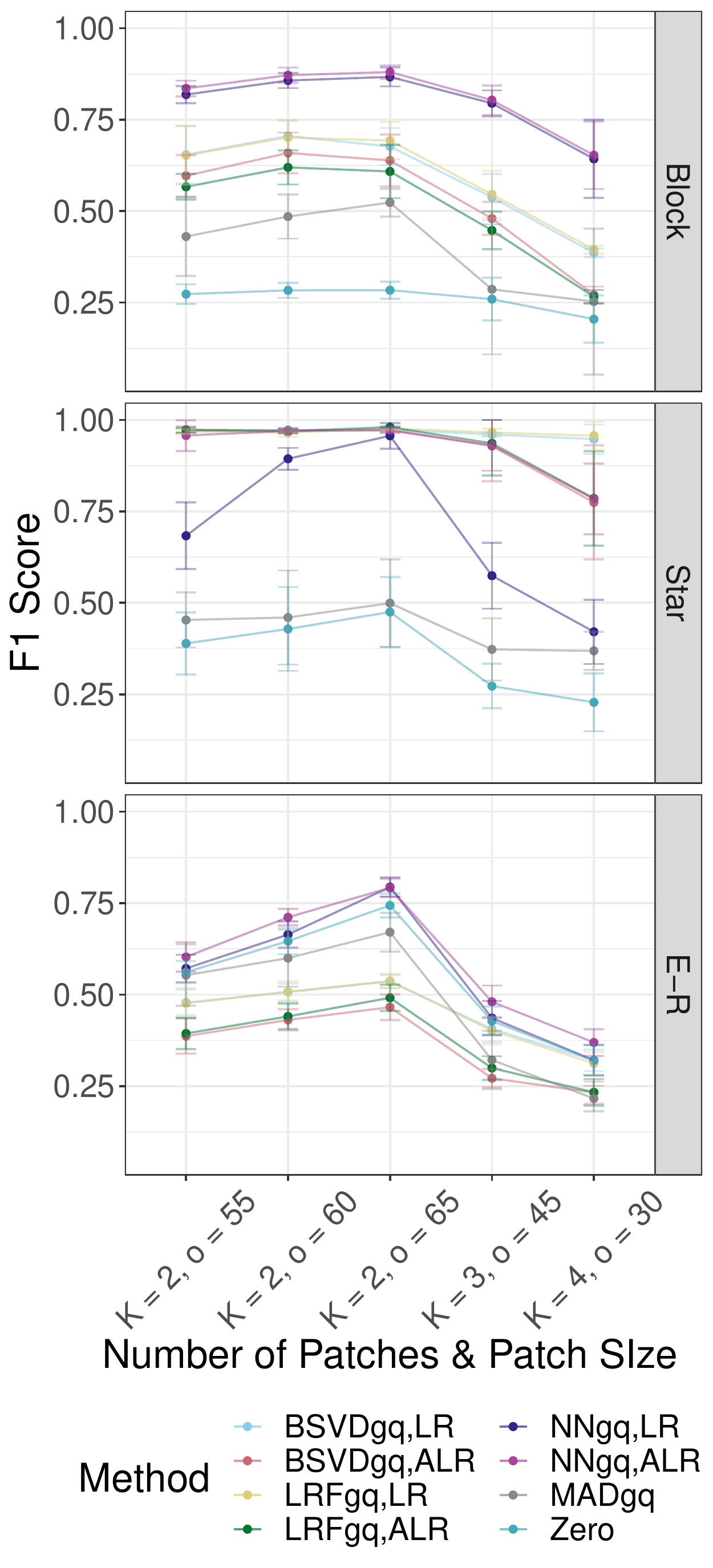}
        \caption{F1 score.}
    \label{fig:simf1}
    \end{subfigure}
\end{center}
\vspace{-0.5cm}
    \caption{Performance of LRGQ, MAD\textsubscript{gq}, and zero imputation methods for covariance imputation and graph recovery on Gaussian graphical model data simulation studies.}
    \label{fig:sim_stuff}
\end{figure}

The results of the Gaussian graphical model simulation study are shown in Figure \ref{fig:sim_stuff}. In the case of the stochastic block diagonal graph, we find that the block singular value methods generally perform the best in terms of recovering the original covariance matrix, but the resulting graph estimate from the covariance matrices imputed with the nuclear norm methods tend to capture the graph structure the best. For the multistar graph, the nuclear norm method with the spiked covariance model tends to return substantially more accurate imputed covariance matrices, but the block singular value and low rank factorization methods give better graph estimates. For the Erd\"{o}s R\'{e}nyi graph, we find that the zero-imputation method outperforms the majority of the low-rank Graph Quilting methods; this aligns with what we would expect from estimating a graph that does not follow a low-rank structure. On the other hand, the MAD\textsubscript{gq} method tends to have a relatively low Fq score compared to the LRGQ methods, which is likely due to the fact that the former was designed to select a strict superset of possible edges in the unobserved portion of the graph rather than the best estimate. Across all graph types, imputation of the covariance matrix and edge recovery are more accurate when the size of each patch is increased and when there are fewer patches, which is what we expect from our theoretical results. Overall, our results show that the low-rank Graph Quilting methods are broadly applicable for graph recovery in the quilting setting when the structure of the graph is low rank. In the Supporting Information, we repeat this simulation study using data-driven tuning of all hyperparameters using the procedure outlined in Section \ref{sub:pms}.

\subsubsection{Real-World Inspired Simulation}

We study the low-rank Graph Quilting methods using simulations based on real-world calcium imaging data. The data set we analyze comes from the Janelia Research Campus \citep{stringer2018} and contains fluorescence traces from a single recording session for approximately 2000 simultaneously recorded neurons measured over 15000 time points from a mouse V1 visual cortex. Before applying our analysis, we first detrend the raw fluorescence traces by first differencing, then centering and scaling each column individually. From this, we calculate the empirical covariance matrix of the full data set and estimate a graph using graphical Lasso; we consider these to be the true underlying covariance matrix and graph to which we compare the estimates from LRGQ. We then divide the empirical covariance matrix into synthetic observation blocks and mask the entries of the covariance matrix outside of the observed set and use it as the input to the LRGQ, MAD\textsubscript{gq}, and zero imputation methods. For this simulation study, we study the impact of changing the size of patches $o$ and the number of patches $K$ while keeping the total overlap $o \times K$ constant on the performance of the low-rank Graph Quilting methods. For graph estimation on the full data set, we use stability selection to select the number of edges. The sparsity of the graph estimates from each of the Graph Quilting methods is then chosen to be the same as that of the graph estimated on the full data. The ranks for the LRGQ method are selected to be 5, following from the scree plot in Figure \ref{fig:abalow}.

\begin{figure}[t]
\begin{center}
    \begin{subfigure}[t]{0.3\linewidth}
    \centering
        \includegraphics[width=\linewidth]{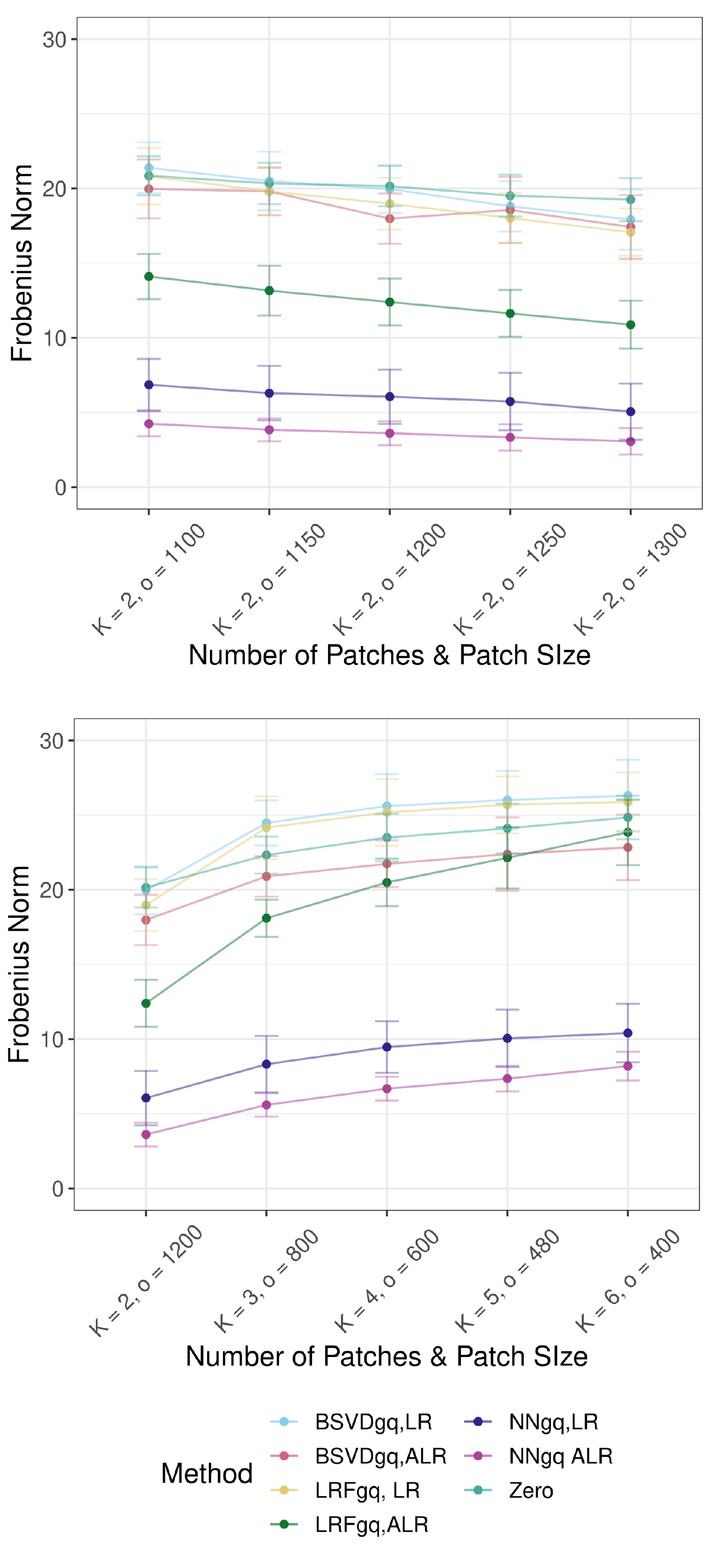}
        \caption{Frobenius norm.}
    \label{fig:janfrob}
    \end{subfigure}
    \begin{subfigure}[t]{0.3\linewidth}
    \centering
        \includegraphics[width=\linewidth]{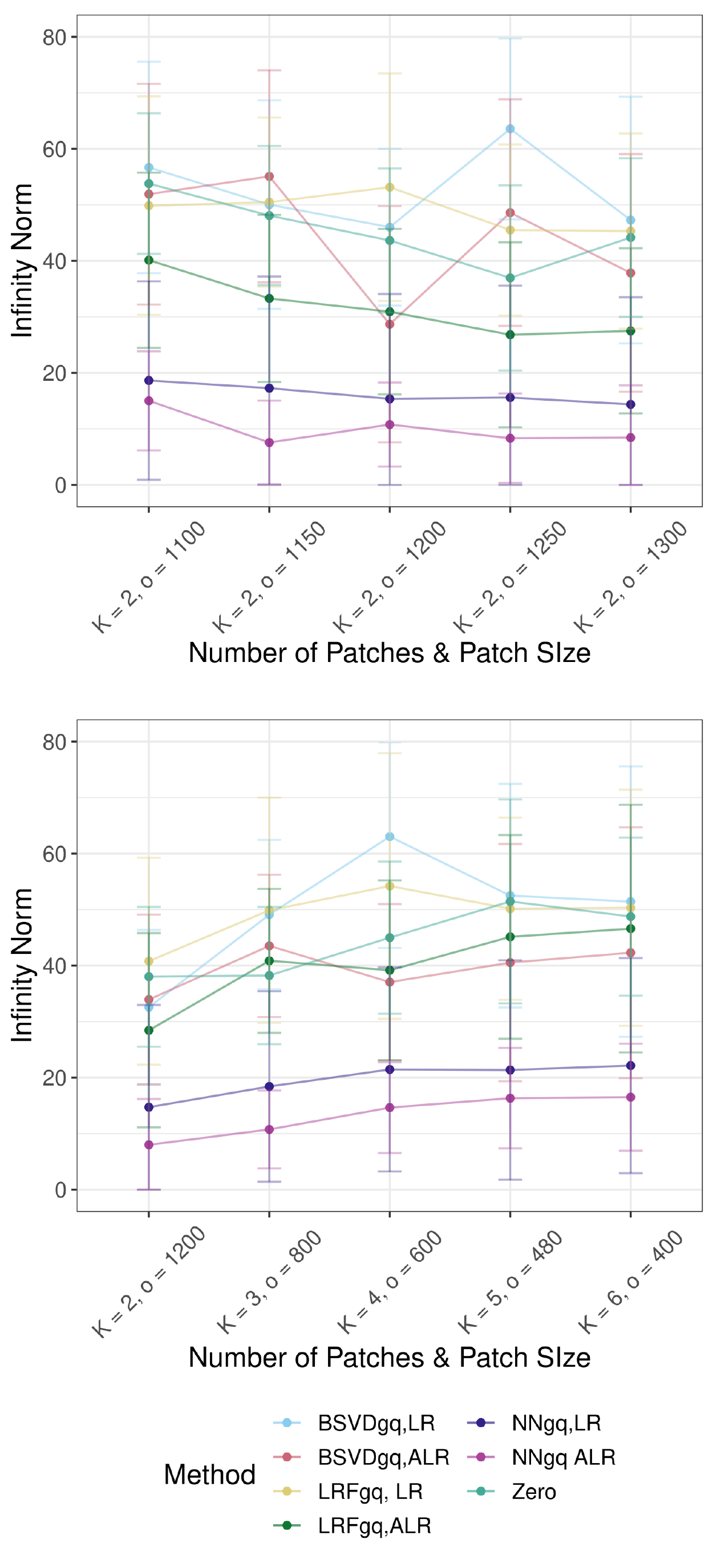}
        \caption{Infinity norm.}
    \label{fig:janinf}
    \end{subfigure}
    \vspace{-0.5cm}
    \begin{subfigure}[t]{0.3\linewidth}
    \centering
        \includegraphics[width=\linewidth]{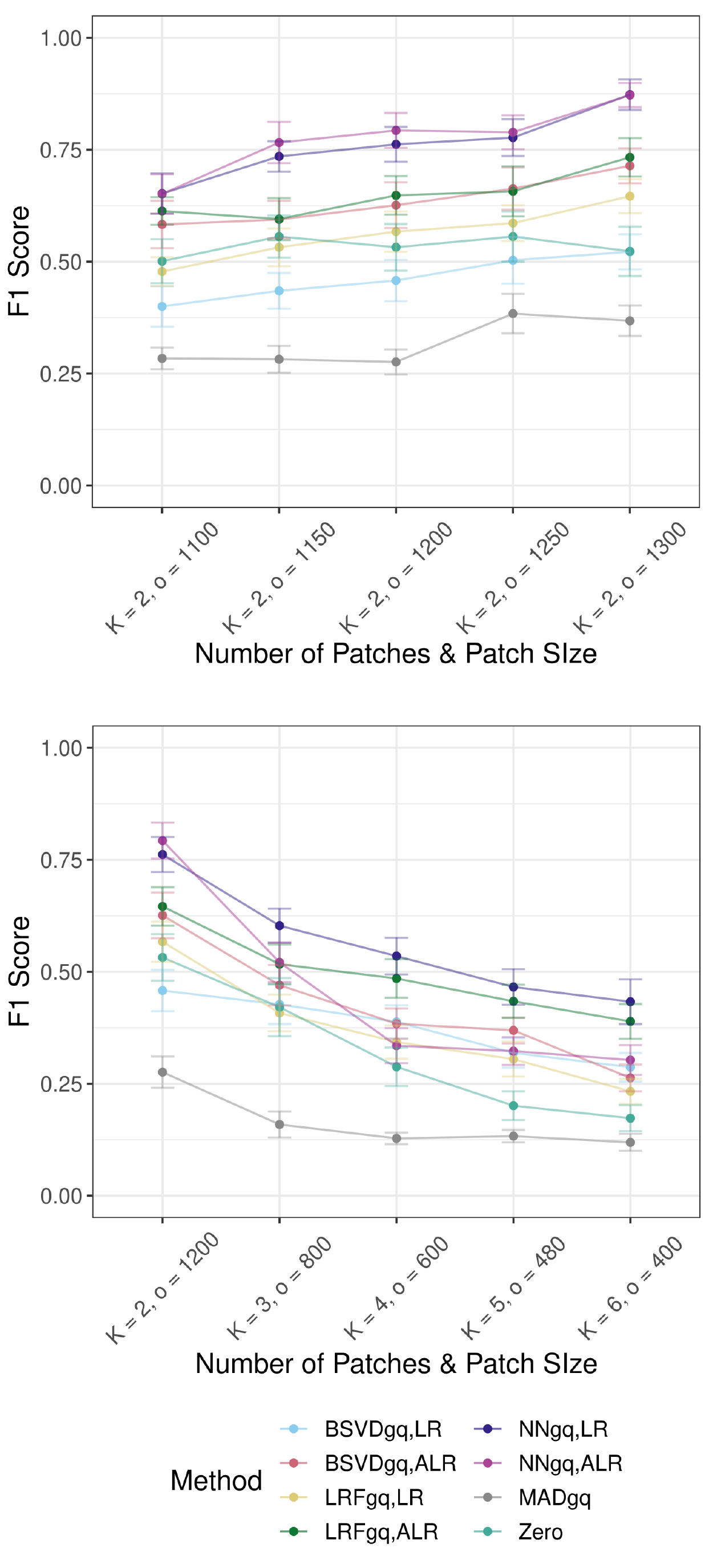}
        \caption{F1 score.}
    \label{fig:janf1}
    \end{subfigure}
\end{center}
\vspace{-0.5cm}
    \caption{Performance of low-rank Graph Quilting, MAD\textsubscript{gq}, and zero imputation methods for covariance imputation and graph recovery on calcium imaging simulation studies.}
    \label{fig:jan_stuff}
\end{figure}

\begin{figure}[t]
\begin{center}
    \begin{subfigure}[t]{0.4\linewidth}
    \centering
        \includegraphics[width=\linewidth]{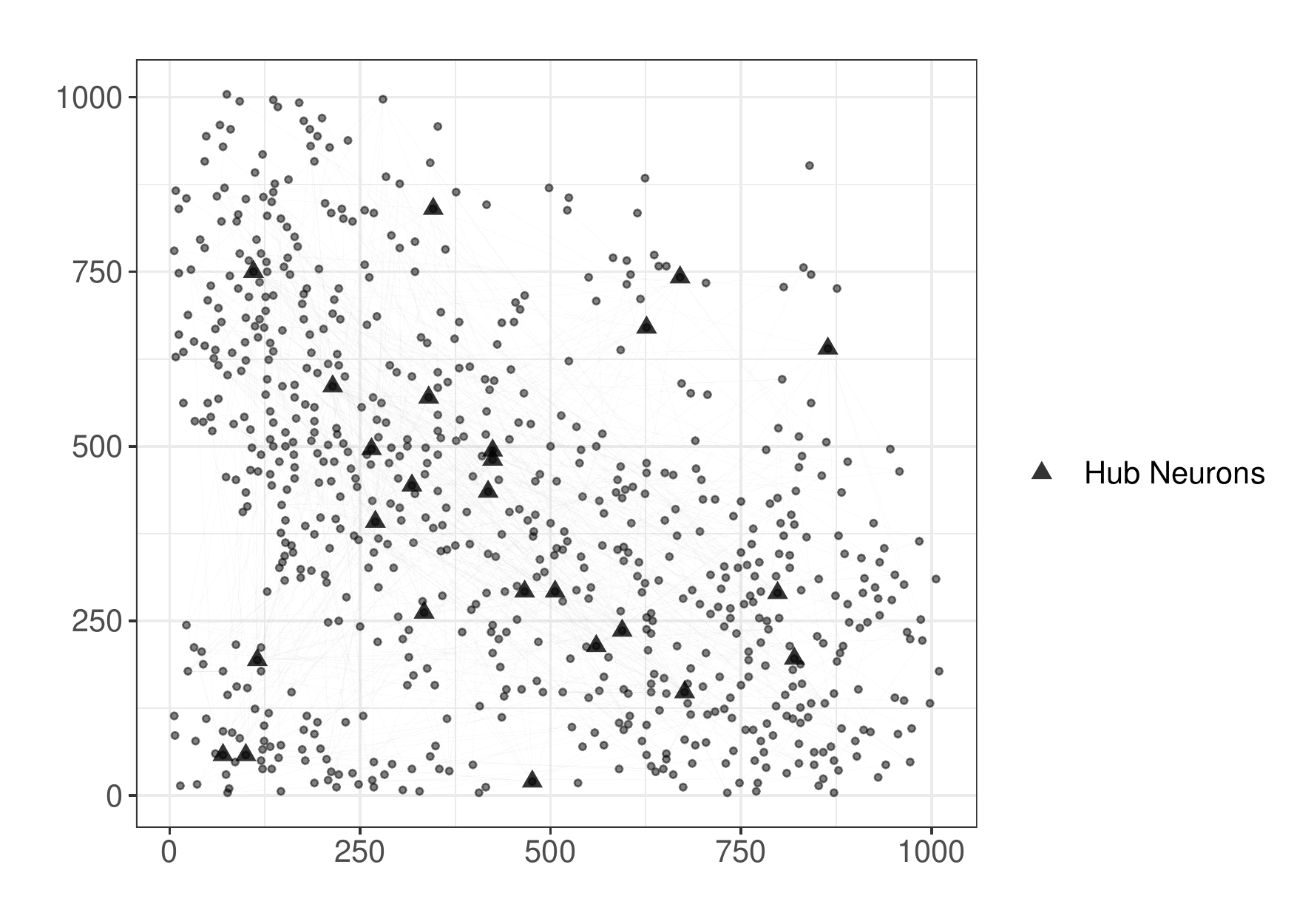}
        \caption{Full joint observations.}
    \label{fig:jsg}
    \end{subfigure}
    \begin{subfigure}[t]{0.4\linewidth}
    \centering
        \includegraphics[width=\linewidth]{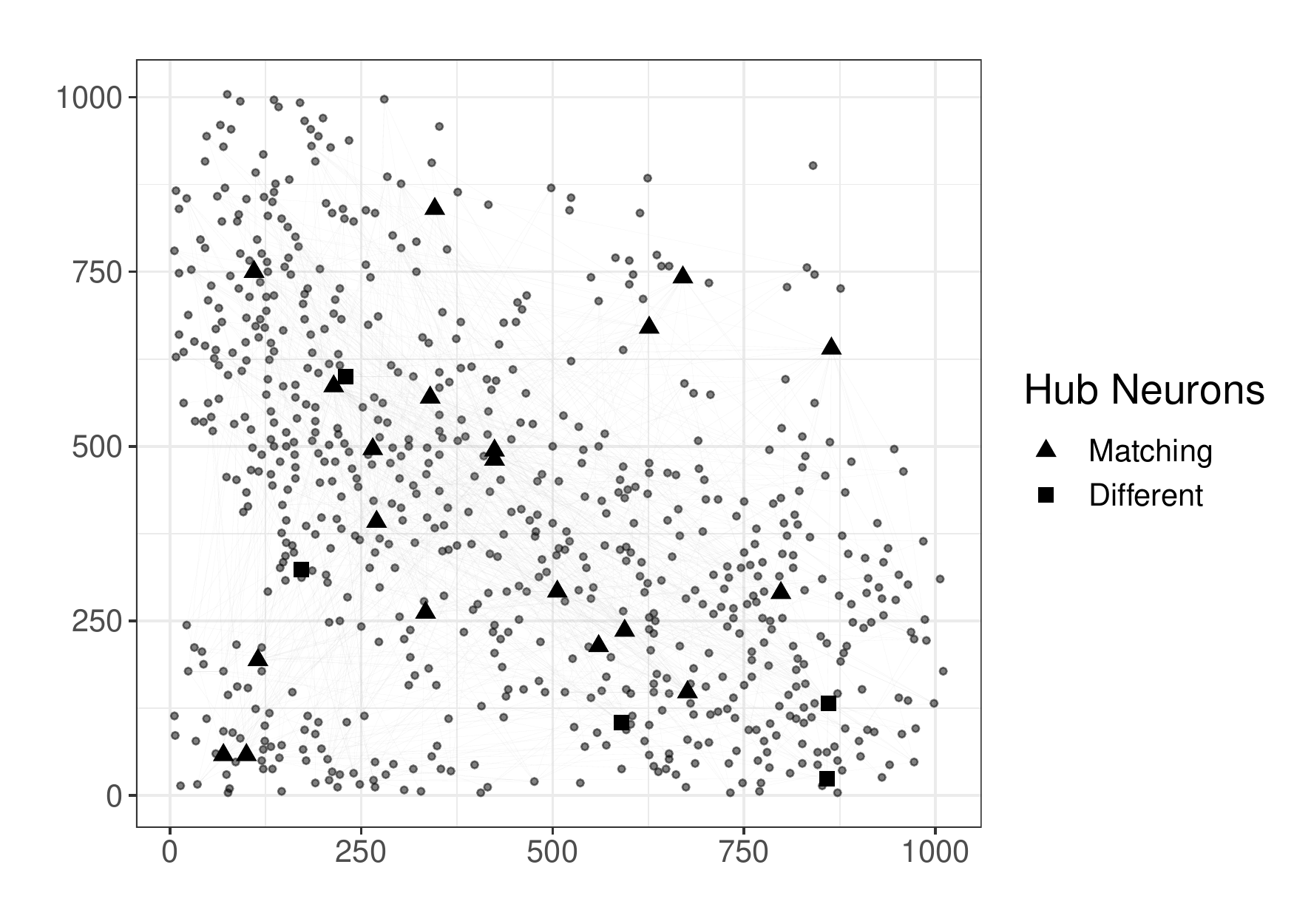}
        \caption{Approximate BSVDgq.}
    \label{fig:jsg2}
    \end{subfigure}
    \vspace{-0.5cm}
    \begin{subfigure}[t]{0.4\linewidth}
    \centering
        \includegraphics[width=\linewidth]{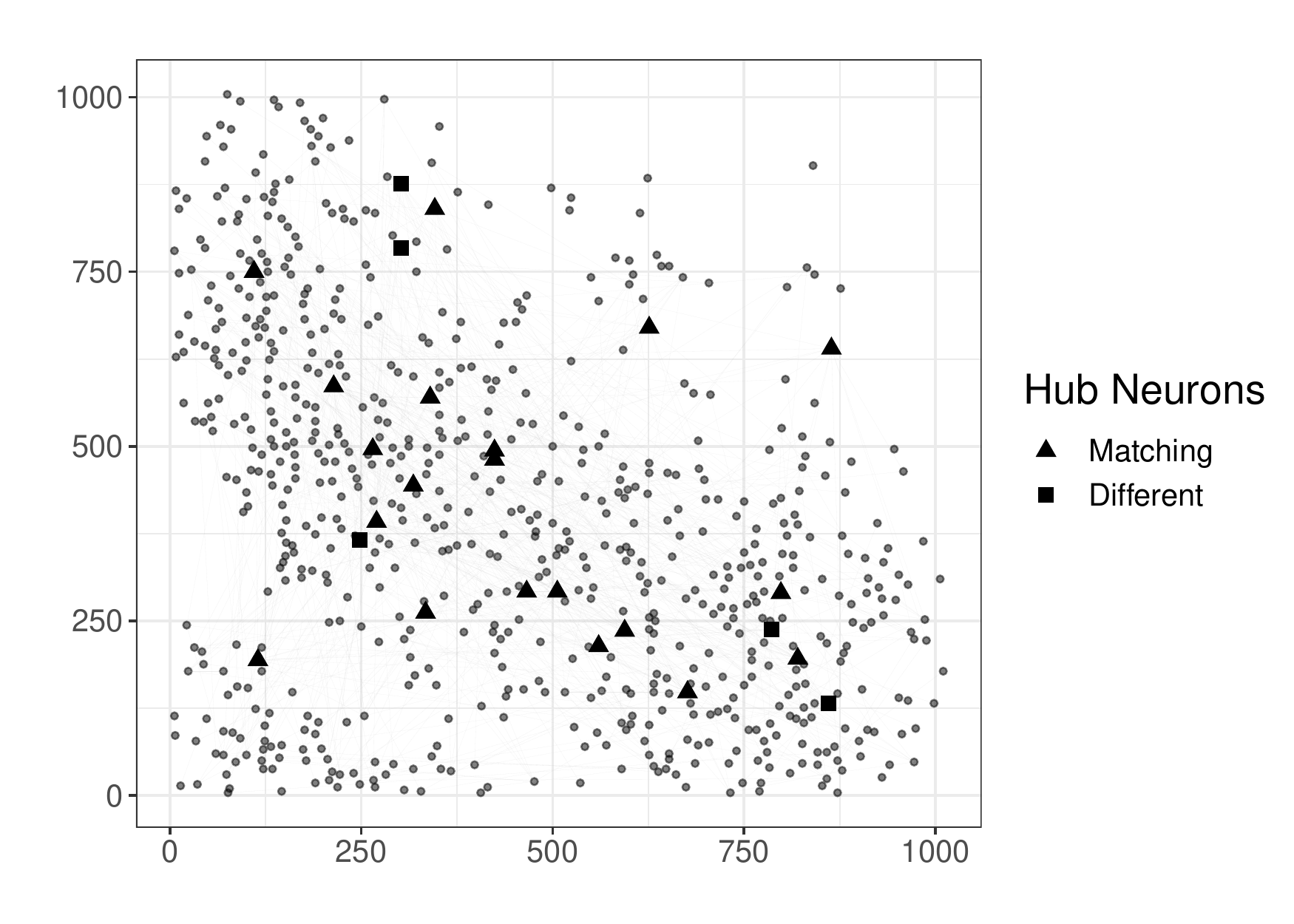}
        \caption{Approximate NNgq.}
    \label{fig:jsg4}
    \end{subfigure}
    \begin{subfigure}[t]{0.4\linewidth}
    \centering
        \includegraphics[width=\linewidth]{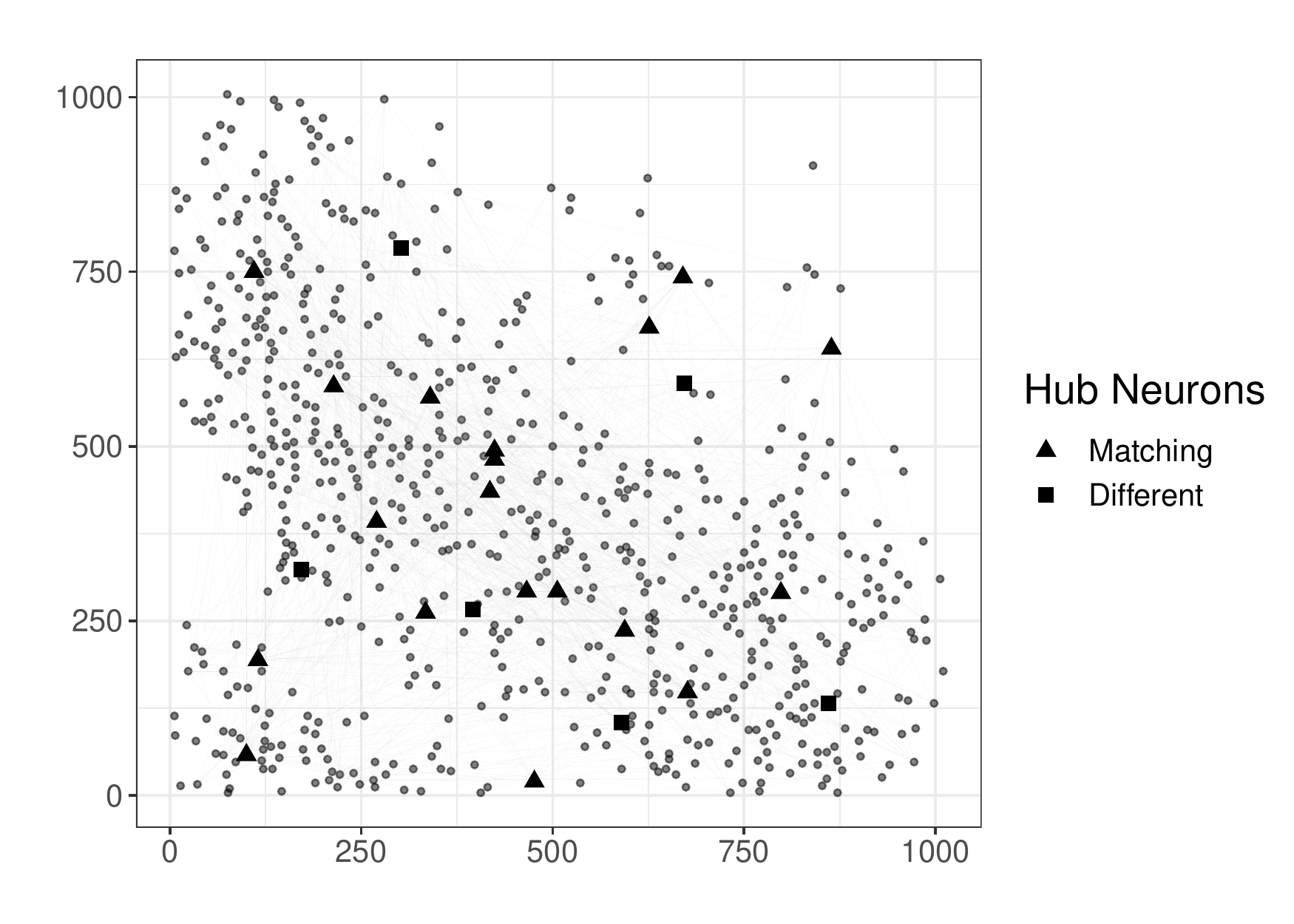}
        \caption{Approximate LRFgq.}
    \label{fig:jsg3}
    \end{subfigure}
\end{center}
\vspace{-0.5cm}
    \caption{Functional connectivity estimates on Janelia calcium imaging data with $K = 2, o = 1200$, visualized for one z-plane, with 25 most highest degree neurons specially denoted. Each functional connectivity graph contains the same number of edges. Hub neurons from LRGQ are displayed based on whether each match with one of the top hub neurons in the estimated graph from the full observations.}
    \label{fig:jg2}
\end{figure}

Figure \ref{fig:jan_stuff} shows the results for $K = 2$ patches with patch sizes $o = 1100, 1150, 1200, 1250$, and $1300$, as well as for $K = 2, 3, 4, 5$ and $6$ with $o = 1200, 800, 600, 480$, and $400$, respectively. From these results, the NNgq methods appear to do best for imputing the full covariance matrix and recovering the same edges as are found when the data set is fully observed, with the spiked covariance model assumption doing particularly well for covariance imputation accuracy. On the other hand, the other low-rank graph quilting methods and the MAD\textsubscript{gq} method are less accurate, occasionally performing worse than the zero imputation method; this possibly indicates that the eigenvalue decay structure from nuclear norm regularization may be more appropriate for the calcium imaging application. Across the varying observation block parameters, we find that the accuracy of the LRGQ methods for covariance imputation and graph recovery generally increases with increasing block sizes and decreases with an increasing number of blocks, which matches what we would expect.

We then show in Figure \ref{fig:jg2} a subset of the estimated functional neuronal connectivity graphs estimated from the low-rank Graph Quilting methods with an exact low-rank assumption, alongside the graph estimate with all joint observations. We see that the topologies and hub neurons of estimated graphs from low-rank Graph Quilting are similar to the one estimated using the full data. The most well-connected neurons in each of the graph estimates are also marked; these are known as hub neurons, and are of interest in the study of the functional neuronal architecture of the brain as potential drivers of distinct neuronal units \citep{liska2015}. We compare the top 25 hub neurons from each the low-rank graph quilting estimate to those found from the full data. The low-rank graph quilting methods tend to find many of the same hub neurons as those found when fully observing the data, at a rate between 70 and 80 percent. This shows that hub neurons can reliably be identified even in the presence of missing pairwise observations by using the low-rank Graph Quilting procedures. In the Supporting Information, we perform an additional calcium imaging-based simulation study.

\section{Functional Connectivity from Calcium Imaging}\label{sec:cal}

We now investigate the efficacy of the low-rank Graph Quilting methods for estimating functional neuronal connectivity on calcium imaging data, in which subsets of neurons are observed across temporal replications of the experiment. For this case study, we use functional neural activity from the MICrONS data set containing recorded neural activity for a single mouse V1 visual cortex \citep{microns2021functional}. Here, we analyze neural activity from two separate recording sessions containing 1018 and 1123 neurons with 275 neurons overlapping, and we consider a 15 minute spontaneous activity period from both sessions. Before applying our analysis, we first detrend the raw fluorescence traces by first differencing, then centering and scaling each column individually. From this pre-processed data, we calculate the partially observed empirical covariance matrix that we use as input to the different LRGQ methods the MAD\textsubscript{gq} algorithm, and the zero procedure described in Section \ref{sec:sim} in order to obtain graph estimates. For this example, we set the sparsity of graph estimates to be the same amongst all quilting methods in order to create an equal comparison. The rank of the BSVDgq and LRFgq methods and the regularization parameters of the NNgq method are selected via the cross validation procedure described in Section \ref{sub:pms}. For the MAD\textsubscript{gq} algorithm, we set the minimum threshold hyperparameter to be 0 when selecting graph sparsity.

\begin{figure}[t]
\begin{center}
    \begin{subfigure}[t]{0.35\linewidth}
    \centering
        \includegraphics[width=\linewidth]{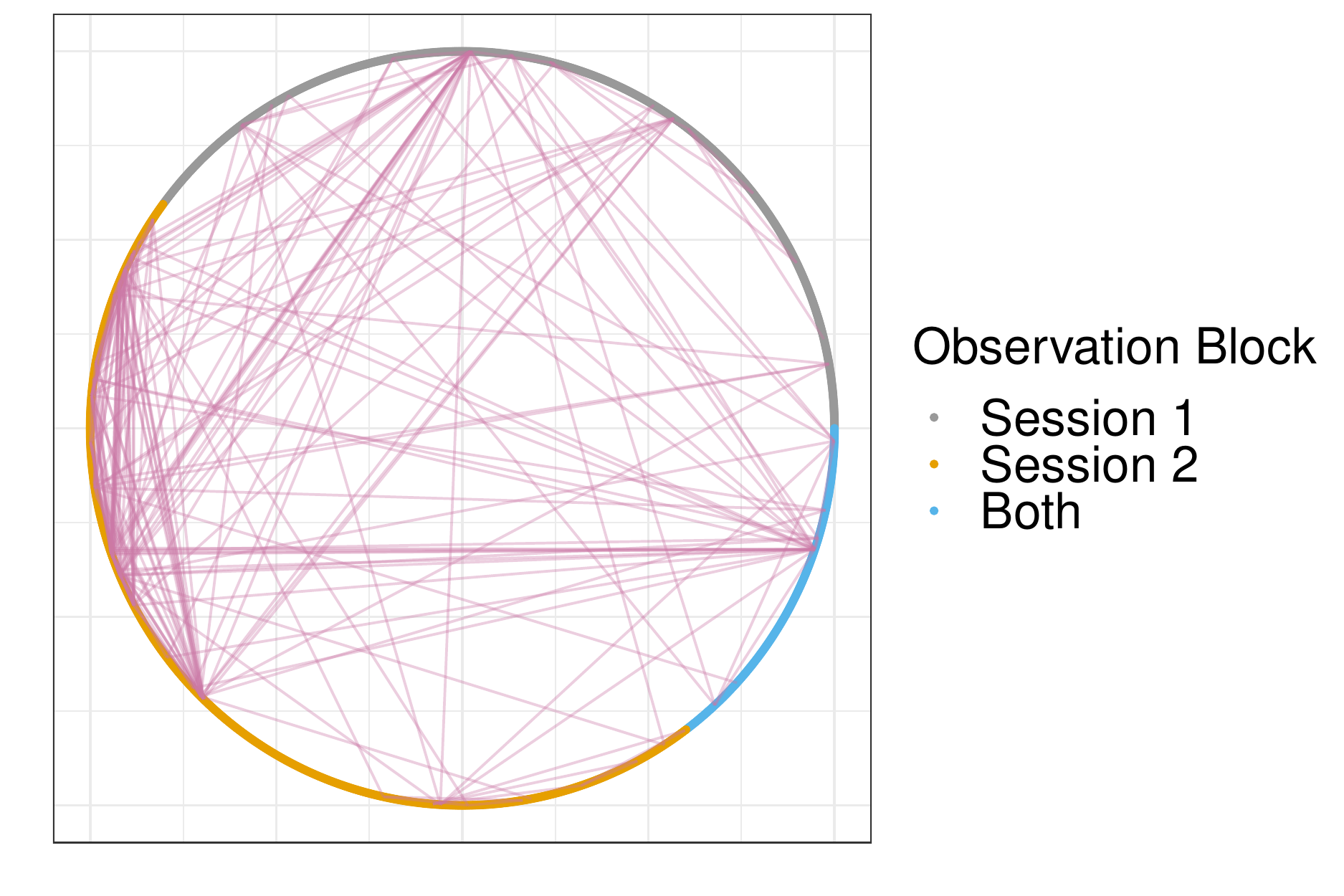}
        \caption{Approximate BSVDgq.}
    \label{fig:nsg4}
    \end{subfigure}%
    \begin{subfigure}[t]{0.35\linewidth}
    \centering
        \includegraphics[width=\linewidth]{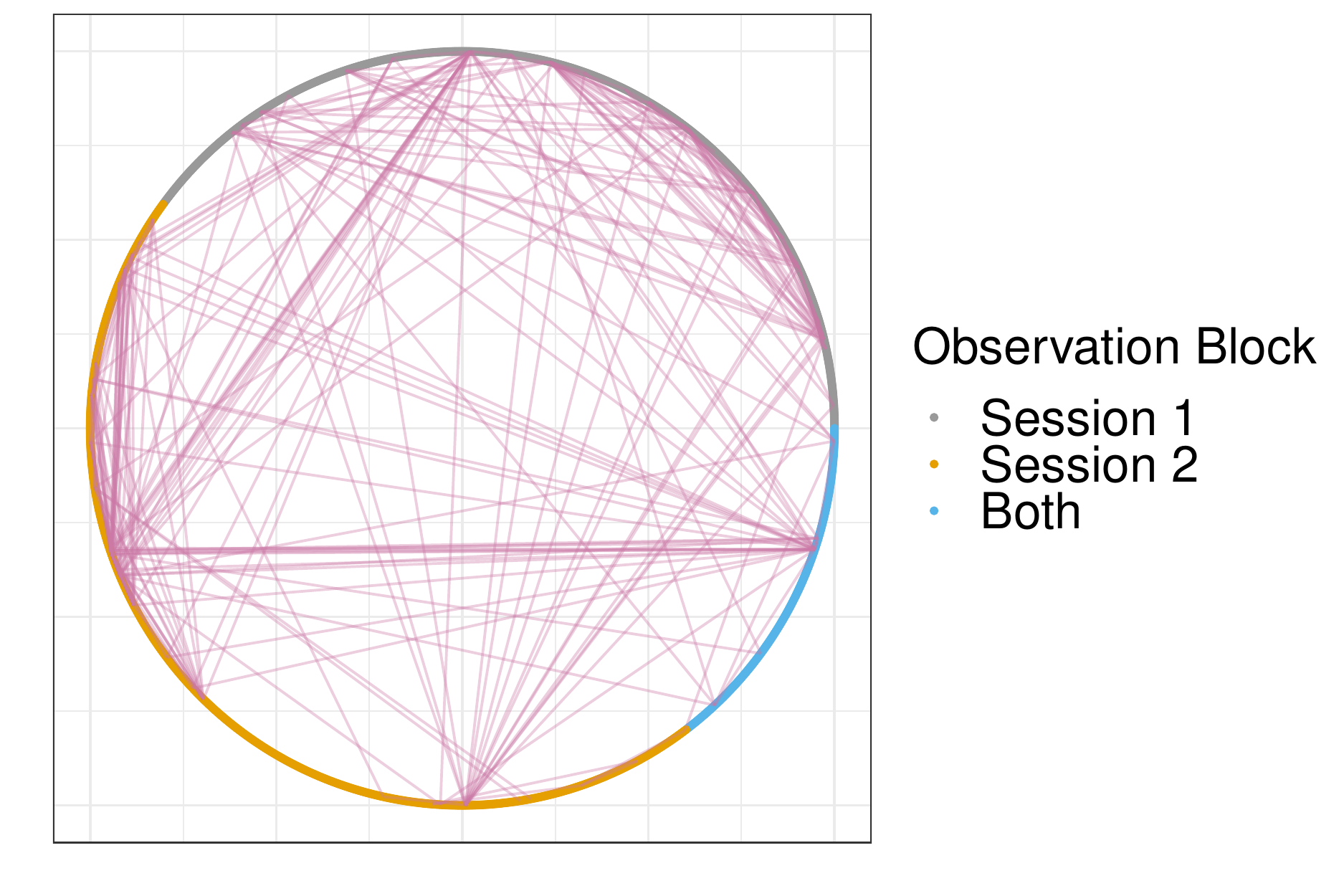}
        \caption{Approximate LRFgq.}
    \label{fig:nsg3}
    \end{subfigure}
    \begin{subfigure}[t]{0.35\linewidth}
    \centering
        \includegraphics[width=\linewidth]{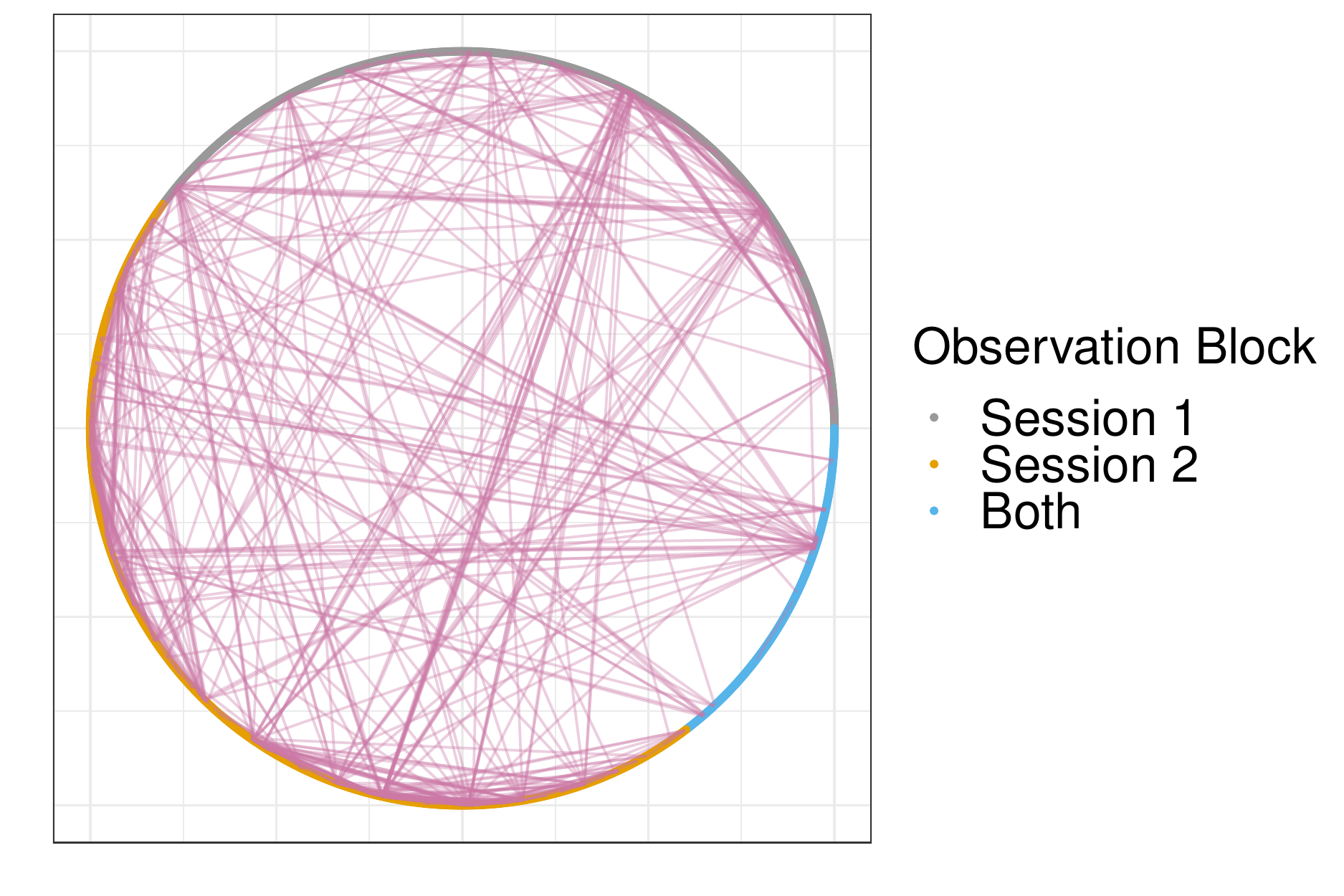}
        \caption{Approximate NNgq.}
    \label{fig:nsg5}
    \end{subfigure}%
    \begin{subfigure}[t]{0.35\linewidth}
    \centering
        \includegraphics[width=\linewidth]{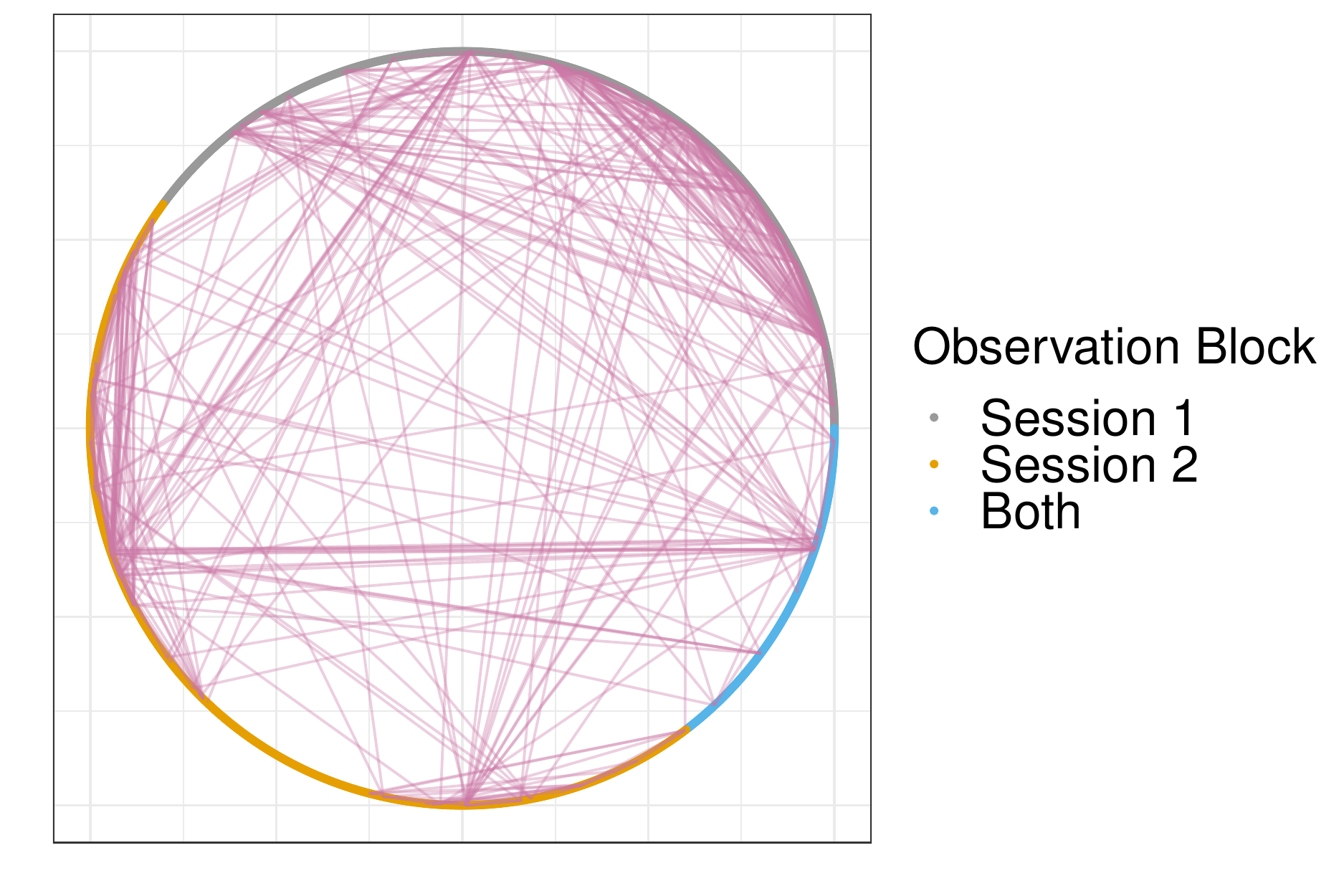}
        \caption{MAD\textsubscript{gq}.}
    \label{fig:nsg6}
    \end{subfigure}
\end{center}
    \caption{Estimated graphs for each Graph Quilting method. Neurons are ordered by observation block membership.}
    \label{fig:microns1}
\end{figure}

In Figure \ref{fig:microns1}, we show estimated functional neuronal connectivity graphs for the approximate low-rank versions of each of the low-rank Graph Quilting methods as well as that estimated by MAD\textsubscript{gq}. While the edge set from each graph estimate varies by method, we do see in general that the estimated graphs all exhibit a small-world structure; this matches what has been previously proposed about functional connectivity in the brain \citep{Sporns2007, Pandarinath2018}. Additionally, the graphs estimated by different methods share some of the same hub neurons, which indicates that some community structure can be identified by with the low-rank assumption across all different estimation procedures. One large difference we see between the functional connectivity graph estimates is the proportion of edges that are estimated between pairs of neurons in distinct observation blocks; in particular, the NNgq method finds many more connections between neurons recorded in two separate sessions, while the other methods tend to find communities that occur within a single observation block.

\spacingset{1.1}
\begin{table}[t]

\centering
\begin{footnotesize}
\begin{tabular}{|l|l|}
\hline
Method  & \textbf{Mean Directional Tuning Correlation} \\
\hline
Exact BSVDgq  & 0.1123 \\
Approx BSVDgq  & 0.1355 \\
Exact LRFgq  & 0.1132 \\
Approx LRFgq & 0.1264 \\
Exact NNgq  & 0.2147 \\
Approx NNgq  & 0.2337 \\
MAD\textsubscript{gq}  & 0.0435 \\
Zero  & 0.0154 \\
\hline
\end{tabular}
\end{footnotesize}
\caption{\small Mean directional tuning correlations of pairs of connected neurons for edges estimated in graphs from each Graph Quilting method.}
\label{tab:mdtc}
\end{table}
\spacingset{1.9}

We then validate the estimated graphs obtained by each method by comparing the mean of the correlation of directional tuning between the pairs of neurons with edges between each other in the graph. 
The preferred directional tuning is a functional property of neurons, defined in the visual cortex to be the particular direction of visual stimulus that causes the greatest rate of activity. It has been posited that neural tuning is related to functional neuronal connectivity \citep{tune}; thus, we expect the estimated edges in the functional neuronal connectivity graph to exhibit some interrelationship to the directional tuning of the neurons in the data. The mean directional tuning correlation of selected edges are shown in Table \ref{tab:mdtc}. We find that the NNgq methods substantially outperform all other methods. However, the other low-rank Graph Quilting methods still considerably outperform the MAD\textsubscript{gq} algorithm and zero imputation, which estimate functional connectivity graphs with edges that are uncorrelated with directional tuning. This indicates that the LRGQ methods, in particular NNgq, may find functional connections that align more closely with what we expect from the scientific literature relative to the comparison methods. Overall, this shows that the low-rank Graph Quilting methods can be used to estimate functional connectivity when neurons that are not simultaneously observed and thus allow for the study of functional connectivity across larger brain volumes.

\section{Discussion}

In this paper, motivated by the approximately low-rank structure in real neuronal functional data sets, we have studied three methods for the Graph Quilting problem based on the assumption that the full covariance matrix with respect to all features is low-rank. These methods are based upon a two-step procedure of low-rank covariance imputation followed by graph estimation. We have shown in both simulated and real data studies that the low-rank graph quilting methods perform better than other existing graph quilting approaches for recovering the edge structure of the graph of the complete data set in the case where the covariance matrix exhibits an approximately low-rank structure. We have also shown that the LRGQ methods are applicable for the analysis of functional neuronal connectivity in calcium imaging.

There are several possible methodological extensions to the low-rank Graph Quilting problem that can be explored in the future. For the calcium imaging application, latent variables and covariates are commonly assumed to have an effect on recorded neural activity and thus may need to be accounted for. To adjust for latent variables, we can apply a sparse plus low-rank decomposition \citep{chang2019graphical} on the imputed covariance matrices from the first step of Graph Quilting. In the presence of covariates, we can use supervised learning methods to condition on covariate effects in the raw data, then fit a graphical model to the covariance structure of the residuals. With regards to the spiked covariance model, we have use the simplest method for estimating $\sigma^{*2}$ in the literature, but other methods could potentially produce better graph estimates. Additionally, our current spiked covariance model assumes that we have a single constant $\sigma^{*2}$ that is applied to all features; however, in some applications, a more flexible model with different values for each diagonal entry may be more appropriate, and further empirical investigations will be necessary to assess the efficacy of this approach; it is also of future interest to leverage recent literature \citep{zhang2022heteroskedastic,yan2024inference} on heteroskedastic PCA to address this problem under our graph quilting setting. From a theoretical standpoint, while we have shown general consistency results of the two-step low-rank graph quilting procedure assuming a consistent low-rank covariance imputation procedure as well as specific results for the BSVDgq method, the theoretical guarantees for the other low-rank Graph Quilting procedures and with unknown $\sigma^{*2}$ may be useful to study in the future.

\section*{Acknowledgements}

The authors gratefully acknowledge support by NSF NeuroNex-1707400, NIH 1R01GM140468, and NSF DMS-2210837.

\bibliographystyle{apalike}
\bibliography{reference}

\newpage



\section*{Supplementary material}

# Supporting Information for "Low-Rank Covariance Completion for Graph Quilting with Applications to Functional Connectivity"

August 1, 2024

## A  Low Rank Graph Quilting Algorithms

In this section, we show how estimates for each of the low-rank covariance completion methods in Section 2 of the main text can be obtained computationally.

## A.1 Block Singular Value Decomposition

For the block singular value decomposition method, we directly follow the algorithm outlined in (Bishop and Yu, 2014).

---
**Algorithm A.1:** Block Singular Value Decomposition (BSVD$_{\text{gq}}$)

---
**Input:** Observation block memberships $\{V_k, k \in 1, \ldots K\}$, observed covariance matrix $\widehat{\boldsymbol{\Sigma}}_O \in \mathbb{R}^{p \times p}$, desired rank $r > 0$
**Initialize:** $\widetilde{\mathbf{H}} = \mathbf{0}_{p \times r}$.
**Find:** low-rank solution for first patch:
    (1) Calculate SVD of $\widehat{\boldsymbol{\Sigma}}_{V_1, V_1} = \mathbf{U}\boldsymbol{\Lambda}\mathbf{U}^\top$.
    (2) Set $\mathbf{h}$ as indices of the largest $r$ diagonal elements of $\boldsymbol{\Lambda}$.
    (3) Set $\widetilde{\mathbf{H}}_{V_1, :} = \mathbf{U}_{:, \mathbf{h}} \boldsymbol{\Lambda}_{\mathbf{h}, \mathbf{h}}^{1/2}$.
**for** $s = 2, \ldots K$ **do**
    (1) Find low-rank solution for $s$-th patch:
      (a) Calculate SVD of $\widehat{\boldsymbol{\Sigma}}_{V_s, V_s} = \mathbf{U}\boldsymbol{\Lambda}\mathbf{U}^\top$.
      (b) Set $\mathbf{h}$ as indices of the largest $r$ diagonal elements of $\boldsymbol{\Lambda}$.
      (c) Calculate $\widehat{\mathbf{H}}_s = \mathbf{U}_{:, \mathbf{h}} \boldsymbol{\Lambda}_{\mathbf{h}, \mathbf{h}}^{1/2}$.
    (2) Merge with previous patches:
      (a) Find overlaps $J^{(1)} = (\bigcup_{k=1}^{s-1} V_k) \cap V_s$, $J^{(2)} = \{j : V_s[j] \in J^{(1)}\}$.
      (b) Calculate SVD of $(\widehat{\mathbf{H}}_s)_{J^{(2)}, :}^\top \widetilde{\mathbf{H}}_{J^{(1)}, :} = \mathbf{W}\boldsymbol{\Lambda}\mathbf{U}^\top$.
      (c) Set $\widetilde{\mathbf{H}}_{V_s \setminus J^{(1)}, :} = (\widehat{\mathbf{H}}_s)_{\setminus J^{(2)}, :} \mathbf{W}\mathbf{U}^\top$
**end**
**return** $\widetilde{\boldsymbol{\Sigma}} = \widetilde{\mathbf{H}}\widetilde{\mathbf{H}}^\top$.

---

## A.2 Nuclear Norm Penalization

For the nuclear norm penalization method, we minimize the objective function $\mathcal{L}(\boldsymbol{\Sigma}_O) = \frac{1}{2}\|\boldsymbol{\Sigma}_O - \widehat{\boldsymbol{\Sigma}}_O\|_F^2 + \nu\|\boldsymbol{\Sigma}\|_*$; this is a composite function of two convex functions, which are the loss function $g(\boldsymbol{\Sigma}) = \frac{1}{2}\|\boldsymbol{\Sigma}_O - \widehat{\boldsymbol{\Sigma}}_O\|_F^2$ and the penalty term $h(\boldsymbol{\Sigma}) = \nu\|\boldsymbol{\Sigma}\|_*$. We use a proximal gradient descent algorithm to find the objective solution, which utilizes the singular value threshold function (Cai et al., 2010) for the proximal step.

---

**Algorithm A.2:** Nuclear Norm Penalization ($\text{NN}_\text{gq}$)

---

**Input:** Set of observed entries $O = \bigcup_{k=1}^{K} V_k \times V_k$, observed covariance matrix $\widehat{\boldsymbol{\Sigma}}_O \in \mathbb{R}^{p \times p}$, nuclear-norm regularization hyperparameters $\lambda > 0$ and $\nu > 0$, iteration step size $\alpha > 0$, error tolerance $\delta > 0$.
**Initialize:** $\boldsymbol{\Sigma}^{(0)} = I_{p \times p}$, $\eta_1 = 1$.
**while** $\frac{1}{\eta_1 \|\widehat{\boldsymbol{\Sigma}}_O\|_2} \|\boldsymbol{\Sigma}^{(r)} - \boldsymbol{\Sigma}^{(r-1)}\|_F \geq \delta$ **do**

    (1) Find gradient $\nabla g(\boldsymbol{\Sigma}^{(r)})$ and optimal step size $\eta_1$ via backtracking:

        (a) Calculate $\nabla g(\boldsymbol{\Sigma}^{(r)})_{ij} = \begin{cases} \boldsymbol{\Sigma}^{(r)}_{ij} - \widehat{\boldsymbol{\Sigma}}_{O_{ij}} & (i,j) \in O \\ 0 & (i,j) \notin O \end{cases}$.

        (b) Set $\eta_1 = \frac{\|\boldsymbol{\Sigma}^{(r)} - \boldsymbol{\Sigma}^{(r-1)}\|_F^2}{\sum_{i=1}^{p}\sum_{j=1}^{p}(\boldsymbol{\Sigma}^{(r)}_{ij} - \boldsymbol{\Sigma}^{(r-1)}_{ij})(\nabla g(\boldsymbol{\Sigma}^{(r)})_{ij} - \nabla g(\boldsymbol{\Sigma}^{(r-1)})_{ij})}$.

        (c) **Repeat:**
            (i) $\mathbf{Z} = \text{SingularValueThreshold}_{\lambda\eta_1}(\boldsymbol{\Sigma}^{(r)} - \eta_1 \nabla g(\boldsymbol{\Sigma}^{(r)}))$
            (ii) $\eta_1 = \alpha \eta_1$
        **until** $\|\mathbf{Z} - \widehat{\boldsymbol{\Sigma}}_O\|_2^2 \leq \|\boldsymbol{\Sigma}^{(r-1)} - \widehat{\boldsymbol{\Sigma}}_O\|_2^2 + 2\lambda(\|\boldsymbol{\Sigma}^{(r-1)}\|_* - \|\mathbf{Z}\|_*)$

    (2) Update $\boldsymbol{\Sigma}^{(r+1)} = \mathbf{Z}$.

    (3) Update $r = r + 1$.

**end**
**return** $\widetilde{\boldsymbol{\Sigma}} = \boldsymbol{\Sigma}^{(r)}$.

## A.3 Low-Rank Matrix Factorization

In the low-rank matrix factorization method, estimates are attained by minimizing the likelihood $\mathcal{L}(\boldsymbol{\Sigma}_O) = \frac{1}{2}\|\boldsymbol{\Sigma}_O - \widehat{\boldsymbol{\Sigma}}_O\|_F^2$. For the initialization of $\boldsymbol{\Sigma}^{(0)}$ below, we can use low-rank solution of the block singular value decomposition method, i.e. matrix $\mathbf{C}$, or the low-rank decomposition of $\widetilde{\boldsymbol{\Sigma}}$ from the nuclear norm penalization method.

---

**Algorithm A.3:** Low-Rank Matrix Factorization (LRF$_{\text{gq}}$)

---

**Input:** Set of observed entries $O = \bigcup_{k=1}^{K} V_k \times V_k$, observed covariance matrix $\widehat{\boldsymbol{\Sigma}}_O \in \mathbb{R}^{p \times p}$, initial low-rank matrix $\mathbf{U}^{(0)} \in \mathbb{R}^{p \times r}$ error tolerance $\delta > 0$.
**Initialize:** $\boldsymbol{\Sigma}^{(0)} = \mathbf{U}^{(0)}\mathbf{U}^{(0)T}$, $\eta_1 = 1$.
**while** $\frac{1}{\eta_1 \|\widehat{\boldsymbol{\Sigma}}_O\|_2}\|\boldsymbol{\Sigma}^{(r)} - \boldsymbol{\Sigma}^{(r-1)}\|_F \geq \delta$ **do**

  (1) Find gradient $\nabla_{\mathbf{U}}\mathcal{L}(\boldsymbol{\Sigma}^{(r)})$ and optimal step size $\eta_1$ via backtracking:
    (a) Calculate $\nabla_{\mathbf{U}}\mathcal{L}(\boldsymbol{\Sigma}^{(r)})_{ij} = ((\boldsymbol{\Sigma}^{(r)} - \widehat{\boldsymbol{\Sigma}}_{O^*})\mathbf{U}^{(r)})_{ij}$
      where $\widehat{\boldsymbol{\Sigma}}_{O^*} = \begin{cases} \widehat{\boldsymbol{\Sigma}}_O & (i,j) \in O \\ 0 & (i,j) \notin O \end{cases}$
    (b) Set $\eta_1 = \frac{\|\mathbf{U}^{(r)} - \mathbf{U}^{(r-1)}\|_F^2}{\sum_{i=1}^{p}\sum_{j=1}^{r}(\mathbf{U}_{ij}^{(r)} - \mathbf{U}_{ij}^{(r-1)})(\nabla_{\mathbf{U}}\mathcal{L}(\boldsymbol{\Sigma}^{(r)})_{ij} - \nabla_{\mathbf{U}}\mathcal{L}(\boldsymbol{\Sigma}^{(r-1)})_{ij})}$.
    (c) **Repeat:**
      (i) $\mathbf{Z} = \mathbf{U}^{(r)} - \eta_1 \nabla_{\mathbf{U}}\mathcal{L}(\boldsymbol{\Sigma}^{(r)})$
      (ii) $\eta_1 = \alpha\eta_1$
    **until** $\|(\mathbf{ZZ}^T) - \widehat{\boldsymbol{\Sigma}}_O\|_2^2 \leq \|\boldsymbol{\Sigma}^{(r-1)} - \widehat{\boldsymbol{\Sigma}}_O\|_2^2$.
  (2) Update:
    (a) $\mathbf{U}^{(r+1)} = \mathbf{Z}$
    (b) $\boldsymbol{\Sigma}^{(r+1)} = \mathbf{U}^{(r+1)}\mathbf{U}^{(r+1)T}$
  (3) Update $r = r + 1$.
**end**
**return** $\widetilde{\boldsymbol{\Sigma}} = \boldsymbol{\Sigma}^{(r)}$.

# B  Approximate Low Rank Graph Quilting Algorithms

In this section, we outline the low-rank graph quilting methods as applied to the spiked covariance matrix model; this requires the inclusion of an extra parameter for the diagonal entries of the covariance matrix.

## B.1  Spiked Block Singular Value Decomposition

**Algorithm B.1:** Spiked Block Singular Value Decomposition (Spiked BSVD$_{\text{gq}}$)

**Input:** Observation block memberships $\{V_k, k \in 1, \ldots K\}$, observed covariance matrix $\widehat{\mathbf{\Sigma}}_O \in \mathbb{R}^{p \times p}$, desired rank $r > 0$
**Initialize:** $\widetilde{\mathbf{H}} = \mathbf{0}_{p \times r}$, $A = V_1$, $\widehat{q} = \text{median}(\{\widehat{\mathbf{\Sigma}}_{ii}, 1 \leq i \leq p\})$.
**Find:** low-rank solution for first patch:
  (1) Calculate SVD of $(\widehat{\mathbf{\Sigma}} - \widehat{q}\mathbf{I})_{V_1, V_1} = \mathbf{W}\mathbf{\Lambda}\mathbf{U}^T$.
  (2) Set $\mathbf{h}$ as indices of the largest $r$ diagonal elements of $\mathbf{\Lambda}$.
  (3) Set $\mathbf{C}_{V_1,:} = \mathbf{U}_{:,\mathbf{h}}\mathbf{\Lambda}_{\mathbf{h},\mathbf{h}}^{1/2}$.
**for** $s \in 2, \ldots K$ **do**
  (1) Find low-rank solution for $s$-th patch:
    (a) Calculate SVD of $(\widehat{\mathbf{\Sigma}} - \widehat{q}\mathbf{I})_{V_s, V_s} = \mathbf{W}\mathbf{\Lambda}\mathbf{U}^T$.
    (b) Set $\mathbf{h}$ as indices of the largest $r$ diagonal elements of $\mathbf{\Lambda}$.
    (c) Calculate $\mathbf{D} = \mathbf{U}_{:,\mathbf{h}}\mathbf{\Lambda}_{\mathbf{h},\mathbf{h}}^{1/2}$.
  (2) Merge with previous patches:
    (a) Find overlaps $E = \{A : a \in V_s\}$, $J = \{j : V_s[j] \in A\}$.
    (b) Calculate SVD of $(\mathbf{D}_J^T \mathbf{C}_{E,:}) = \mathbf{W}\mathbf{\Lambda}\mathbf{U}^T$.
    (c) Set $\mathbf{M} = \mathbf{C}_{E,:}$, $\mathbf{C}_{V_s,:} = \mathbf{D}\mathbf{W}\mathbf{U}^T$
    (d) Set $\mathbf{C}_{E,:} = \mathbf{M}$
  (3) Update $A = \bigcup_{k=1}^{s} V_s$
**end**
**return** $\widetilde{\mathbf{\Sigma}} = \mathbf{C}\mathbf{C}^T$.

## B.2 Spiked Nuclear Norm Penalization

---

**Algorithm B.2:** Spiked Nuclear Norm Penalization (Spiked $\text{NN}_{\text{gq}}$)

---

**Input:** Set of observed entries $O = \bigcup_{k=1}^{K} V_k \times V_k$, observed covariance matrix $\widehat{\boldsymbol{\Sigma}}_O \in \mathbb{R}^{p \times p}$, nuclear-norm regularization hyperparameters $\lambda > 0$ and $\nu > 0$, iteration step size $\alpha > 0$, error tolerance $\delta > 0$, initial spiked covariance constant $c^{(0)} > 0$.

**Initialize:** $\boldsymbol{L}^{(0)} = \boldsymbol{0}_{p \times p}$, $\boldsymbol{\Sigma}^{(0)} = \boldsymbol{L}^{(0)} + c^{(0)}\boldsymbol{I}$, $\eta_1 = 1$, $\eta_2 = 1$.

**while** $\frac{1}{\eta_1 \|\widehat{\boldsymbol{\Sigma}}_O\|_2} \|\boldsymbol{\Sigma}^{(r)} - \boldsymbol{\Sigma}^{(r-1)}\|_F \geq \delta$ **do**

(1) Find gradient $\nabla g(\boldsymbol{\Sigma}^{(r)})$ and optimal step size $\eta_1$ via backtracking:

  (a) Calculate components of $\nabla g(\boldsymbol{\Sigma}^{(r)})$:

  (i) $\nabla_{\boldsymbol{L}} g(\boldsymbol{\Sigma}^{(r)})_{ij} = \begin{cases} \boldsymbol{\Sigma}^{(r)}_{ij} - \widehat{\boldsymbol{\Sigma}}_{O_{ij}} & (i,j) \in O \\ 0 & (i,j) \notin O \end{cases}.$

  (ii) $\nabla_c g(\boldsymbol{\Sigma}^{(r)}) = \text{Tr}(\boldsymbol{\Sigma}^{(r)} - \widehat{\boldsymbol{\Sigma}}_O)$

  (b) Set initial gradient step parameters:

  (i) $\eta_1 = \frac{\|\boldsymbol{L}^{(r)} - \boldsymbol{L}^{(r-1)}\|_F^2}{\sum_{i=1}^{p} \sum_{j=1}^{p} (\boldsymbol{L}^{(r)}_{ij} - \boldsymbol{L}^{(r-1)}_{ij})\{(\nabla_{\boldsymbol{L}} g(\boldsymbol{\Sigma}^{(r)}))_{ij} - (\nabla_{\boldsymbol{L}} g(\boldsymbol{\Sigma}^{(r-1)}))_{ij}\}}.$

  (ii) $\eta_2 = \frac{(c^{(r)} - c^{(r-1)})^2}{(c^{(r)} - c^{(r-1)})(\nabla_c g(\boldsymbol{\Sigma}^{(r)}) - \nabla_c g(\boldsymbol{\Sigma}^{(r-1)}))}.$

  (c) **Repeat:**

  (i) $\boldsymbol{Z} = \text{SingularValueThreshold}_{\lambda \eta_1}(\boldsymbol{L}^{(r)} - \eta_1 \nabla_{\boldsymbol{L}} g(\boldsymbol{\Sigma}^{(r)}))$

  (ii) Set $\eta_1 = \alpha \eta_1$, $\eta_2 = \alpha \eta_2$

  **until** $\|(\boldsymbol{Z} + c^{(r)}\boldsymbol{I}) - \widehat{\boldsymbol{\Sigma}}_O\|_2^2 \leq \|\boldsymbol{\Sigma}^{(r-1)} - \widehat{\boldsymbol{\Sigma}}_O\|_2^2 + 2\lambda(\|\boldsymbol{L}^{(r-1)}\|_* - \|\boldsymbol{Z}\|_*)$

(2) Update:

  (a) $\boldsymbol{L}^{(r+1)} = \boldsymbol{Z}$

  (b) $c^{(r+1)} = c^{(r)} - \eta_2 \nabla_c g(\boldsymbol{\Sigma}^{(r)})$

  (c) $\boldsymbol{\Sigma}^{(r+1)} = \boldsymbol{L}^{(r+1)} + c^{(r+1)} \boldsymbol{I}$

(3) Update $r = r + 1$.

**end**

**return** $\widetilde{\boldsymbol{\Sigma}} = \boldsymbol{\Sigma}^{(r)}$.

---

## B.3 Spiked Low-Rank Matrix Factorization

---

**Algorithm B.3:** Spiked Low-Rank Matrix Factorization (Spiked LRF$_{\text{gq}}$)

**Input:** Set of observed entries $O = \bigcup_{k=1}^{K} V_k \times V_k$, observed covariance matrix $\widehat{\boldsymbol{\Sigma}}_O \in \mathbb{R}^{p \times p}$, initial low-rank matrix $\mathbf{U}^{(0)} \in \mathbb{R}^{p \times r}$, initial spiked covariance constant $c^{(0)} > 0$, error tolerance $\delta > 0$.

**Initialize:** $\boldsymbol{\Sigma}^{(0)} = \mathbf{U}^{(0)}\mathbf{U}^{(0)T} + c^{(0)}\boldsymbol{I}$, $\eta_1 = 1$, $\eta_2 = 1$.

**while** $\frac{1}{\eta_1 \|\widehat{\boldsymbol{\Sigma}}_O\|_2} \|\boldsymbol{\Sigma}^{(r)} - \boldsymbol{\Sigma}^{(r-1)}\|_F \geq \delta$ **do**

    (1) Find gradient $\nabla \mathcal{L}(\boldsymbol{\Sigma}^{(r)})$ and optimal step size $\eta_1$ via backtracking:

      (a) Calculate components of $\nabla \mathcal{L}(\boldsymbol{\Sigma}^{(r)})$:

        (i) $\nabla_{\mathbf{U}}\mathcal{L}(\boldsymbol{\Sigma}^{(r)})_{ij} = ((\boldsymbol{\Sigma}^{(r)} - \widehat{\boldsymbol{\Sigma}}_{O^*})\mathbf{U}^{(r)})_{ij}$

          where $\widehat{\boldsymbol{\Sigma}}_{O^*} = \begin{cases} \widehat{\boldsymbol{\Sigma}}_O & (i,j) \in O \\ 0 & (i,j) \notin O \end{cases}$

        (ii) $\nabla_c \mathcal{L}(\boldsymbol{\Sigma}^{(r)}) = \text{Tr}(\boldsymbol{\Sigma}^{(r)} - \widehat{\boldsymbol{\Sigma}}_O)$

      (b) Set initial gradient step parameters:

        (i) $\eta_1 = \frac{\|\boldsymbol{U}^{(r)} - \boldsymbol{U}^{(r-1)}\|_F^2}{\sum_{i=1}^{p}\sum_{j=1}^{r}[\boldsymbol{U}_{ij}^{(r)} - \boldsymbol{U}_{ij}^{(r-1)}][\nabla_{\boldsymbol{U}}\mathcal{L}(\boldsymbol{\Sigma}^{(r)})_{ij} - \nabla_{\boldsymbol{U}}\mathcal{L}(\boldsymbol{\Sigma}^{(r-1)})_{ij}]}$.

        (ii) $\eta_2 = \frac{(c^{(r)} - c^{(r-1)})^2}{(c^{(r)} - c^{(r-1)})(\nabla_c \mathcal{L}(\boldsymbol{\Sigma}^{(r)}) - \nabla_c \mathcal{L}(\boldsymbol{\Sigma}^{(r-1)}))}$.

      (c) **Repeat:**

        (i) $\mathbf{Z} = \mathbf{U}^{(r)} - \eta_1 \nabla_{\boldsymbol{U}}\mathcal{L}(\boldsymbol{\Sigma}^{(r)})$

        (ii) Set $\eta_1 = \alpha \eta_1$, $\eta_2 = \alpha \eta_2$

      **until** $\|(\mathbf{Z}\mathbf{Z}^T + c^{(r)}\boldsymbol{I}) - \widehat{\boldsymbol{\Sigma}}_O\|_2^2 \leq \|\boldsymbol{\Sigma}^{(r-1)} - \widehat{\boldsymbol{\Sigma}}_O\|_2^2$.

    (2) Update:

      (a) $\mathbf{U}^{(r+1)} = \mathbf{Z}$

      (b) $c^{(r+1)} = c^{(r)} - \eta_2 \nabla_c \mathcal{L}(\boldsymbol{\Sigma}^{(r)})$

      (c) $\boldsymbol{\Sigma}^{(r+1)} = \mathbf{U}^{(r+1)}\mathbf{U}^{(r+1)T} + c^{(r+1)}\boldsymbol{I}$

    (3) Update $r = r + 1$.

**end**

**return** $\widetilde{\boldsymbol{\Sigma}} = \boldsymbol{\Sigma}^{(r)}$.

---

# C  Detailed Theoretical Results

**Notations:** For any vector $u \in \mathbb{R}^p$, we define its norms as follows: $\|u\|_2 = (\sum_i u_i^2)^{\frac{1}{2}}$; $\|u\|_\infty = \max_i |u_i|$; $\|u\|_1 = \sum_i |u_i|$. For any $p > 0$, we denote the $p-1$-dimensional sphere in $\mathbb{R}^p$ by $\mathbb{S}^{p-1} = \{u \in \mathbb{R}^p : \|u\|_2 = 1\}$. For any matrix $\mathbf{A} \in \mathbb{R}^{p \times q}$, we denote its norms as follows: the spectral norm $\|\mathbf{A}\|_2 = \sup_{u \in \mathbb{S}^{q-1}} \|\mathbf{A}u\|_2$, the entry-wise infinity error bound $\|\mathbf{A}\|_{\max} = \max_{i,j} |\mathbf{A}_{i,j}|$, the Frobenius norm $\|\mathbf{A}\|_F = \left(\sum_{i,j} \mathbf{A}_{i,j}^2\right)^{\frac{1}{2}}$, the two-to-infinity norm $\|\mathbf{A}\|_{2 \to \infty} = \sup_{u \in \mathbb{S}^{q-1}} \|\mathbf{A}u\|_\infty$, the matrix operator norm w.r.t. $\ell_\infty$-norm: $\|\mathbf{A}\|_\infty = \max_{i=1,\ldots,p} \sum_{j=1}^q |\mathbf{A}_{ij}|$, and the nuclear norm $\|\mathbf{A}\|_* = \sum_{i=1} \sigma_i(\mathbf{A})$, where $\sigma_i(\mathbf{A})$'s are the singular values of $\mathbf{A}$. For any quantities $\alpha, \beta > 0$, we say that $\alpha \asymp \beta$ if there exists universal constants $0 < c < C$ such that $\alpha \leq C\beta$ and $c\beta \leq \alpha$.

## C.1  Meta-theorem for Graph Selection Consistency of Algorithm 2.1

In this section, we show theoretical guarantees for edge selection consistency of Algorithm 2.1 with respect the true underlying graph, which will apply regardless of the low-rank covariance completion method used in the first step of the algorithm as long as the imputed covariance matrix $\widetilde{\mathbf{\Sigma}}$ is a good estimator for the full sample covariance $\widehat{\mathbf{\Sigma}}$. We follow the notation in Ravikumar et al. (2011): let $\mathbf{\Gamma}^* = \mathbf{\Sigma}^* \otimes \mathbf{\Sigma}^*$, $S = \{(i,j) \in [p] \times [p] : i \neq j, \mathbf{\Theta}^*_{ij} \neq 0\}$, $d = \max_i |\{j \neq i : \mathbf{\Theta}_{ij} \neq 0\}|$. Also define

$$\kappa_{\mathbf{\Sigma}^*} = \|\mathbf{\Sigma}^*\|_\infty, \quad \kappa_{\mathbf{\Gamma}^*} = \|(\mathbf{\Gamma}^*_{S,S})^{-1}\|_\infty.$$

We require the following incoherence assumption:

**Assumption C.1** (Incoherence condition). *There exists some constant $\alpha \in (0,1]$ such that*

$$\max_{e \in S^c} \|\mathbf{\Gamma}^*_{e,S}(\mathbf{\Gamma}^*_{S,S})^{-1}\|_1 \leq 1 - \alpha.$$

8

With these, we can now state the following result for model selection consistency of Algorithm 2.1:

**Proposition C.1** (Graph Estimation Consistency of Algorithm 2.1). *Consider Algorithm 2.1 and its output $\widehat{\boldsymbol{\Theta}}_G$. If Assumption 2.1 holds,*

$$\|\widetilde{\boldsymbol{\Sigma}} - \boldsymbol{\Sigma}^*\|_{\max} = O\left(\min\left\{\frac{\alpha^2}{\kappa_{\boldsymbol{\Sigma}^*}\kappa_{\boldsymbol{\Gamma}^*}d}, \frac{\alpha^2}{\kappa_{\boldsymbol{\Sigma}^*}^3\kappa_{\boldsymbol{\Gamma}^*}^2 d}, \frac{\alpha\theta_{\min}}{\kappa_{\boldsymbol{\Gamma}}^*}\right\}\right),$$

*where $\theta_{\min} = \min_{\boldsymbol{\Theta}_{ij}^* \neq 0} |\boldsymbol{\Theta}_{ij}^*|$, and $\lambda \asymp \frac{1}{\alpha}\|\widetilde{\boldsymbol{\Sigma}} - \boldsymbol{\Sigma}^*\|_{\max}$ (i.e. $\lambda$ is on the order of $\frac{1}{\alpha}\|\widetilde{\boldsymbol{\Sigma}} - \boldsymbol{\Sigma}^*\|_{\max}$, then*

$$\{(i,j) : i \neq j, (\widehat{\boldsymbol{\Theta}}_G)_{ij} \neq 0\} = \{(i,j) : i \neq j, \boldsymbol{\Theta}_{ij}^* \neq 0\}.$$

The proof for this proposition follows directly from the Theorem 1 in Ravikumar et al. (2011). In the above result, the influence of the missing entries is reflected in the term $\|\widetilde{\boldsymbol{\Sigma}} - \boldsymbol{\Sigma}^*\|_{\max}$, i.e. the error of the imputation step.

## C.2 Detailed Theoretical Results for the BSVDgq Algorithm

To establish the entrywise error bound for the imputed covariance using the BSVD algorithm, let us first introduce the following notations and assumptions. Suppose the true covariance matrix $\boldsymbol{\Sigma}^* = \boldsymbol{L}^* + \sigma^{*2}\mathbf{I}$, where $\boldsymbol{L}^*$ is positive semidefinite and of rank $r$. For $1 \leq k \leq K$, let $p_k = |V_k| \leq p$, and define $J_k = V_k \cap (\cup_{j=1}^{k-1} V_j)$ as the joint of the $k$th node set with prior sets. Also define the following quantities for the covariance $\boldsymbol{\Sigma}_{V_k,V_k}^*$ or $\boldsymbol{L}_{J_k,J_k}^*$ corresponding to the $k$th node set $V_k$:

(i) The effective rank $\tau_k = \frac{\text{tr}(\boldsymbol{\Sigma}_{V_k,V_k}^*)}{\lambda_1(\boldsymbol{\Sigma}_{V_k,V_k}^*)}$;

(ii) The incoherence parameter $\mu_k = \frac{p_k}{r}\|\mathbf{U}_k^*\|_{2\to\infty}^2$ where $\mathbf{U}_k^* \in \mathbb{O}^{p_k \times r}$ is defined by the SVD $\boldsymbol{L}_{V_k,V_k}^* = \mathbf{U}_k^*\Lambda_k^*\mathbf{U}_k^{*\top}$;

9

(iii) $\xi_k$ that quantifies both the signal strength in $V_k$ compared to $J_k$, and also the condition number corresponding to $J_k$: $\xi_k = 2\sqrt{\frac{|J_k|\|\mathbf{L}^*_{V_k,V_k}\|_{\max}}{\lambda_r(\mathbf{L}^*_{J_k,J_k})}}\sqrt{\frac{\lambda_1(\mathbf{L}^*_{J_k,J_k})}{\lambda_r(\mathbf{L}^*_{J_k,J_k})}}$. A smaller condition number $\frac{\lambda_1(\mathbf{L}^*_{J_k,J_k})}{\lambda_r(\mathbf{L}^*_{J_k,J_k})}$ and a stronger signal strength $\frac{\lambda_r(\mathbf{L}^*_{J_k,J_k})}{|J_k|\|\mathbf{L}^*_{V_k,V_k}\|_{\max}}$ in the joint observational set $J_k$ would lead to a smaller $\xi_k$. We let $\xi = \max_k \xi_k$.

**Assumption C.2** (Approximate low-rankness and condition number of each block). *For all $1 \leq k \leq K$, $\max\{\lambda_1(\mathbf{L}^*_{V_k,V_k}), \sigma^{*2}\} \leq C\lambda_r(\mathbf{L}^*_{V_k,V_k})$ for some constant $C > 0$.*

**Assumption C.3** (Blocks not too different).

$$\mu_1 \asymp \mu_2 \asymp \cdots \asymp \mu_K, \quad p_1 \asymp p_2 \asymp \cdots \asymp p_K, \quad \lambda_1(\mathbf{L}^*_{V_1,V_1}) \asymp \lambda_1(\mathbf{L}^*_{V_2,V_2}) \asymp \cdots \asymp \lambda_1(\mathbf{L}^*_{V_K,V_K}),$$

*i.e. all elements of $\boldsymbol{\mu}, \boldsymbol{p}$, and $\boldsymbol{\lambda}$ are on the same order of magnitude.*

**Assumption C.4** (Sample size). *For all $1 \leq k \leq K$,*

$$n_k \geq C(r + \frac{\tau_k}{\mu_k})(\tau_k \vee \log p_k)(\xi - 1)^{-2}\xi^{2K},$$

*i.e. the required sample size increases with increases in the rank and effective rank of the graph, an increase the size of the covariance matrix, and a decrease in signal strength of the overlapping portions of the sequential blocks.*

The following proposition characterizes the key quantity $\xi$ when the low-rank component $\mathbf{L}^*$ in the covariance matrix is randomly generated.

**Proposition C.2** (Scaling of $\xi$ for random $\mathbf{L}^*$). *Suppose that $\mathbf{L}^* = \mathbf{H}^*\mathbf{H}^{*\top}$ where all entries $\mathbf{H}^*_{j,k}$ are i.i.d. mean zero Gaussian random variables. If $|J_k| \geq 2r$, then with probability at least $1 - C\sum_{k=1}^K \exp\{-c\min\{|J_k|^2, r\log p_k\}\}$,*

$$\xi_k \leq C\sqrt{r\log p_k}, \quad k = 1, \ldots, K.$$

The theoretical guarantees for the BSVDgq method is presented as follows:

**Theorem C.1** (Guarantees for BSVDgq). *Under Assumptions C.2-C.4, with probability at least $1 - C\sum_{k=1}^{K} p_k^{-c}$, the output $\widetilde{\Sigma}$ of the BSVDgq algorithm with the input $\widehat{\Sigma}_O - \sigma^{*2}\mathbf{I}$ satisfies*

$$\|\widetilde{\Sigma} - \Sigma^*\|_{\max} \leq \frac{C\|\mathbf{L}^*\|_{\max}\xi^K}{\xi - 1} \max_k \sqrt{\frac{(r + \frac{\tau_k}{\mu_k})(\tau_k \vee \log p_k)}{n_k}}, \quad (1)$$

*where $c, C > 0$ are universal constants.*

Assumptions C.2-C.4 and the full proof of Theorem 2 can be found in Section D.

**Remark 1.** *Theorem 2 proves that the block singular value decomposition method can lead to entry-wise consistent estimates for the covariance matrix $\Sigma^*$, if the sample size for each block is larger than a polynomial of the rank $r$, effective rank $\tau_k$, and $\log p_k$ where $p_k$ is the number of nodes in the kth block. Here, Theorem 2 assumes $\sigma^{*2}$ to be known only for simplicity. Otherwise, we expect the error bound to also depend on $|\widehat{\sigma}^2 - \sigma^{*2}|$ linearly.*

**Remark 2** (Technical novelty). *Although the estimation procedure for the BSVDgq method, as outlined in Algorithm A.1 in the Supporting Information, has been theoretically studied in Bishop and Yu (2014), their result hinges on a Frobenius norm error bound for $\widehat{\Sigma}_O - \Sigma^*$, which would be too large for our purpose of bounding the entry-wise error bound $\|\widetilde{\Sigma} - \Sigma^*\|_{\infty}$. Hence we developed some new proofs, borrowing tools and ideas from spectral norm error bounds for sample covariances (Koltchinskii and Lounici, 2017a) and $\ell_{2\to\infty}$-norm error bounds for spectral methods with perturbed low-rank matrices (Cape et al., 2019). In addition, the error bound for sample covariances with high probability help us get rid of the conditions imposed upon the random quantity $\widehat{\Sigma}_O$ in Bishop and Yu (2014).*

**Remark 3** (Dependence on the number of blocks $K$). *One might notice that our error bound and sample size requirement in Assumption C.4 has an exponential dependence upon the number of blocks $K$. This is due to that the BSVDgq algorithm employs a sequential*

*matching step to find the best rotation matrix for each block that aligns with previous blocks, and the estimation errors for each block accumulates in an exponential manner. Such dependence on K also appears in Bishop and Yu (2014), where a Frobenius error bound for the BSVDgq algorithm is provided.*

**Corollary C.1.** *Suppose we apply the block SVD algorithm with input $\widehat{\boldsymbol{\Sigma}}_O - \sigma^{*2}\mathbf{I}$ as the first step of Algorithm 2.1 in the main paper. If Assumptions C.2-C.4 hold, $\lambda \asymp \frac{1}{\alpha}\|\widetilde{\boldsymbol{\Sigma}} - \boldsymbol{\Sigma}^*\|_{\max}$, and for $1 \leq k \leq K$,*

$$n_k \geq C\|\boldsymbol{\Sigma}^*\|_{\max}^2 \left(\frac{\kappa_{\boldsymbol{\Sigma}^*}^2 \kappa_{\boldsymbol{\Gamma}^*}^2 + \kappa_{\boldsymbol{\Sigma}^*}^6 \kappa_{\boldsymbol{\Gamma}^*}^4}{\alpha^4}d^2 + \frac{\kappa_{\boldsymbol{\Gamma}^*}^2}{\alpha^2 \theta_{\min}^2}\right) \frac{\xi^{2K}}{(\xi-1)^2}\left(r + \frac{\tau_k}{\mu_k}\right)(\tau_k \vee \log p_k),$$

*where $\theta_{\min} = \min_{\boldsymbol{\Theta}_{ij}^* \neq 0} |\boldsymbol{\Theta}_{ij}^*|$, $\kappa_{\boldsymbol{\Sigma}^*} = \|\boldsymbol{\Sigma}^*\|_\infty$, $\kappa_{\boldsymbol{\Gamma}}^* = \|(\boldsymbol{\Gamma}_{S,S}^*)^{-1}\|_\infty$, $0 < \alpha < 1$ is the incoherence parameter defined in Assumption C.1, then with probability at least $1 - C\sum_{k=1}^K p_k^{-c}$, we achieve exact edge recovery of the graph:*

$$\{(i,j) : i \neq j, (\widehat{\boldsymbol{\Theta}}_G)_{ij} \neq 0\} = \{(i,j) : i \neq j, \boldsymbol{\Theta}_{ij}^* \neq 0\}.$$

Under appropriate conditions and with known $\sigma^{*2}$, Corollary 1 establishes that graph consistency can be achieved with high probability if $n_k \geq C(d^2 + \theta_{\min}^{-2})\frac{\xi^{2K}}{(\xi-1)^2}\left(r + \frac{\tau_k}{\mu_k}\right)(\tau_k \vee \log p_k)$ for $1 \leq k \leq K$. Compared to the sample size requirement ($n \geq C(d^2 + \theta_{\min}^{-2})\log p$) in prior literature (Ravikumar et al., 2011), the additional cost due to the block observational pattern is reflected in the effective rank $\tau_k$ of each $\boldsymbol{\Sigma}_{V_k, V_k}^*$, rank $r$, incoherence parameter $\mu_k$, and the factor $\xi^K$ depending on number of blocks $K$.

12

# D Proofs for Theoretical Guarantees

## D.1 Proofs of Proposition 1, Theorem 1, and Theorem 2

*Proof of Proposition 1.* Suppose we have the SVD $\frac{1}{c}\mathbf{L}_0 = \mathbf{U}\boldsymbol{\Lambda}\mathbf{U}^\top$ for some $\mathbf{U} \in \mathbb{O}^{p \times r}$ and an $r \times r$ diagonal matrix $\boldsymbol{\Lambda}$. Since $\lambda_r(\mathbf{L}_0) > \frac{c}{2}$ and $\boldsymbol{\Theta} = c(\mathbf{I} - \mathbf{U}\boldsymbol{\Lambda}\mathbf{U}^\top) \succ 0$, we have $\frac{1}{2} < \lambda_r(\boldsymbol{\Lambda}) \leq \lambda_1(\boldsymbol{\Lambda}) < 1$. We then note that

$$\boldsymbol{\Sigma} = \boldsymbol{\Theta}^{-1} = \frac{1}{c}(\mathbf{I} - \mathbf{U}\boldsymbol{\Lambda}\mathbf{U}^\top)^{-1} = \frac{1}{c}(\mathbf{I} + \mathbf{U}\boldsymbol{\Lambda}(\mathbf{I} - \boldsymbol{\Lambda})^{-1}\mathbf{U}^\top),$$

which can be verified by simply multiplying $\mathbf{I} + \mathbf{U}\boldsymbol{\Lambda}(\mathbf{I} - \boldsymbol{\Lambda})^{-1}\mathbf{U}^\top$ and $(\mathbf{I} - \mathbf{U}\boldsymbol{\Lambda}\mathbf{U}^\top$ to obtain the identity matrix. Therefore, we can let $\mathbf{L} = \frac{1}{c}\mathbf{U}\boldsymbol{\Lambda}(\mathbf{I} - \boldsymbol{\Lambda})^{-1}\mathbf{U}^\top$, which satisfies $\lambda_r(\mathbf{L}) = \frac{\lambda_r(\boldsymbol{\Lambda})}{c(1-\lambda_r(\boldsymbol{\Lambda}))} > \frac{1}{c}$. □

*Proof of Theorem 1.* Let $\mathbf{L}^* = \mathbf{U}^*\boldsymbol{\Lambda}^*\mathbf{U}^{*\top}$ with $\mathbf{U}^* \in \mathbb{O}^{p \times r}$ and $\boldsymbol{\Lambda}^* \in \mathbb{R}^{r \times r}$ being a diagonal matrix. Define $\mathbf{H}^* = \mathbf{U}^*\boldsymbol{\Lambda}^{*\frac{1}{2}}$, $\mathbf{H}_k^* = \mathbf{H}_{V_k,:}^*$. To fix the notation for the proofs, here we recall the definition $J_k = (\cup_{j=1}^{k-1} V_j) \cap V_k$ in Section 2.1; also, we define $J_k^V = \{j : V_k[j] \in J_k\}$ and $S_k = \cup_{j=1}^k V_j$. Let $\widehat{\mathbf{H}}_k = \widehat{\mathbf{U}}_k \widehat{\boldsymbol{\Lambda}}_k^{\frac{1}{2}}$, where $\widehat{\mathbf{U}}_k \in \mathbb{O}^{p_k \times r}$ is the matrix containing the top $r$ eigenvectors of $\widehat{\boldsymbol{\Sigma}}_{V_k,V_k} - \sigma^{*2}\mathbf{I}$ and $\widehat{\boldsymbol{\Lambda}}_k \in \mathbb{R}^{r \times r}$ is a diagonal matrix consisting of the largest $r$ eigenvalues of $\widehat{\boldsymbol{\Sigma}}_{V_k,V_k} - \sigma^{*2}\mathbf{I}$. Then the steps in Algorithm A.1 for computing $\widetilde{\boldsymbol{\Sigma}}$ can also be written as follows:

- $\widetilde{\boldsymbol{\Sigma}} = \widetilde{\mathbf{H}}\widetilde{\mathbf{H}}^\top$ where $\widetilde{\mathbf{H}} \in \mathbb{R}^{p \times r}$ is defined sequentially:

  - $\widetilde{\mathbf{H}}_{V_1,:} = \widehat{\mathbf{H}}_{V_1,:}$;
  - For $k = 2, \ldots, K$, let the SVD of $(\widehat{\mathbf{H}}_k)_{J_k^V,:}^\top \widetilde{\mathbf{H}}_{J_k,:}$ be $\widetilde{\mathbf{W}}_k^{(1)} \widetilde{\boldsymbol{\Lambda}} \widetilde{\mathbf{W}}_k^{(2)\top}$ and $\widetilde{\mathbf{W}}_k = \widetilde{\mathbf{W}}_k^{(1)} \widetilde{\mathbf{W}}_k^{(2)\top}$, $\widetilde{\mathbf{H}}_{V_k \setminus J_k,:} = (\widehat{\mathbf{H}}_k)_{\setminus J_k^V,:} \widetilde{\mathbf{W}}_k$.

We also require the following notion of sub-Gaussian random variables.

**Definition 1.** *For any random variable $X$, if $(\mathbb{E}|X|^p)^{\frac{1}{p}} \leq C\sqrt{p}$ for some $C > 0$ and all $p \geq 1$, it is sub-Gaussian with norm $\|X\|_{\psi_2} = \sup_{p \geq 1} p^{-\frac{1}{2}}(\mathbb{E}|X|^p)^{\frac{1}{p}}$.*

Our proof can be summarized as two major steps: (i) bounding $\min_{\mathbf{W} \in \mathbb{O}^{p_k \times r}} \|\widehat{\mathbf{H}}_k \mathbf{W} - \mathbf{H}_k^*\|_2$ for each block $1 \leq k \leq K$, which requires tools from matrix perturbation theory and spectral analysis; (ii) controlling the error induced by the merging step and showing the final bound for $\|\widetilde{\mathbf{H}}\widetilde{\mathbf{H}} - \mathbf{H}\mathbf{H}^*\|_{\max}$ by induction.

We first restate the necessary assumptions:

**Assumption D.3** (Approximate low-rankness and condition number of each block). *For all $1 \leq k \leq K$, $\max\{\lambda_1(\mathbf{L}_{V_k,V_k}^*), \sigma^{*2}\} \leq C\lambda_r(\mathbf{L}_{V_k,V_k}^*)$ for some constant $C > 0$.*

Requiring $\sigma^{*2} \leq C\lambda_r(\mathbf{L}_{V_k,V_k}^*)$ is to ensure the approximate low-rankness of each block; while $\lambda_1(\mathbf{L}_{V_k,V_k}^*) \leq C\lambda_r(\mathbf{L}_{V_k,V_k}^*)$ means constant condition number.

**Assumption D.4** (Blocks not too different).

$$\mu_1 \asymp \mu_2 \asymp \cdots \asymp \mu_K, \quad p_1 \asymp p_2 \asymp \cdots \asymp p_K, \quad \lambda_1(\mathbf{L}_{V_1,V_1}^*) \asymp \lambda_1(\mathbf{L}_{V_2,V_2}^*) \asymp \cdots \asymp \lambda_1(\mathbf{L}_{V_K,V_K}^*),$$

*i.e. all elements of $\boldsymbol{\mu}, \boldsymbol{p}$, and $\boldsymbol{\lambda}$ are on the same order of magnitude.*

**Assumption D.5** (Sample size). *For all $1 \leq k \leq K$,*

$$n_k \geq C\left(r + \frac{\tau_k}{\mu_k}\right)(\tau_k \vee \log p_k)(\xi - 1)^{-2}\xi^{2K},$$

*i.e. the required sample size increases with increases in the rank and effective rank of the graph, an increase the size of the covariance matrix, and a decrease in signal strength of the overlapping portions of the sequential blocks.*

**Controlling error for each block:** The following lemma completes the first step:

**Lemma D.1** (Error bound for each block). *Assume that $x_1, \ldots, x_n \overset{i.i.d}{\sim} \mathcal{N}(0, \mathbf{\Sigma}^*)$ and $\widehat{\mathbf{\Sigma}} = \frac{1}{n}\sum_{i=1}^n x_i x_i^\top \in \mathbb{R}^{p \times p}$ is the sample covariance. Consider the eigendecomposition $\mathbf{\Sigma}^* - \sigma^{*2}\mathbf{I} = \mathbf{L}^* = \mathbf{U}^*\mathbf{\Lambda}^*\mathbf{U}^{*\top} \in \mathbb{R}^{p \times p}$ for $\mathbf{U}^* \in \mathbb{O}^{p \times r}$ and $\mathbf{\Lambda}^* \in \mathbb{R}^{r \times r}$; $\widehat{\mathbf{\Sigma}} - \sigma^{*2}\mathbf{I} = \widehat{\mathbf{U}}\widehat{\mathbf{\Lambda}}\widehat{\mathbf{U}}^\top + \widehat{\mathbf{U}}_\perp \widehat{\mathbf{\Lambda}}_\perp \widehat{\mathbf{U}}_\perp^\top$ where $\widehat{\mathbf{\Lambda}}$ consists the top $r$ eigenvalues of $\widehat{\mathbf{\Sigma}}$. Let $\tau = \frac{\operatorname{tr}(\mathbf{\Sigma}^*)}{\lambda_1(\mathbf{\Sigma}^*)}$, then as long as $\sqrt{\frac{\tau \vee \log p}{n}} \leq \frac{C\lambda_r^2(\mathbf{L}^*)}{\lambda_1^2(\mathbf{\Sigma}^*)}$, with probability at least $1 - Cp^{-c}$, there exists $\widehat{\mathbf{W}} \in \mathbb{O}^{r \times r}$ such that*

$$\|\widehat{\mathbf{U}}\widehat{\mathbf{\Lambda}}^{\frac{1}{2}}\widehat{\mathbf{W}} - \mathbf{U}^*\mathbf{\Lambda}^{*\frac{1}{2}}\|_{2\to\infty} \leq C\left(\frac{\lambda_1^2(\mathbf{\Sigma}^*)}{\lambda_r^{3/2}(\mathbf{L}^*)}\sqrt{\frac{\mu r^2}{p}} + \frac{\lambda_1^{3/2}(\mathbf{\Sigma}^*)}{\lambda_r(\mathbf{L}^*)}\sqrt{\frac{\tau r}{p}}\right)\sqrt{\frac{\tau \vee \log p}{n}}.$$

*where $\mu = \frac{p}{r}\|\mathbf{U}^*\|_{2\to\infty}^2$ is the incoherence parameter.*

Applying Lemma D.1 to each block $1 \leq k \leq K$ and taking a union bound, we have that with probability at least $1 - C\sum_{k=1}^K p_k^{-c}$, there exist $\widehat{\mathbf{W}}_1, \ldots, \widehat{\mathbf{W}}_K \in \mathbb{O}^{r \times r}$ such that

$$\begin{aligned}\|\widehat{\mathbf{H}}_k\widehat{\mathbf{W}}_k - \mathbf{H}_k^*\|_{2\to\infty} &\leq C\left(\frac{\lambda_1^2(\mathbf{\Sigma}^*_{V_k,V_k})}{\lambda_r^{3/2}(\mathbf{L}^*_{V_k,V_k})}\sqrt{\frac{\mu_k r^2}{p_k}} + \frac{\lambda_1^{3/2}(\mathbf{\Sigma}^*_{V_k,V_k})}{\lambda_r(\mathbf{L}^*_{V_k,V_k})}\sqrt{\frac{\tau_k r}{p_k}}\right)\sqrt{\frac{\tau_k \vee \log p_k}{n_k}} \\ &\leq C\sqrt{\frac{(\mu_k r^2 + \tau_k r)(\tau_k \vee \log p_k)}{p_k n_k}}\lambda_1^{\frac{1}{2}}(\mathbf{\Sigma}^*_{V_k,V_k}),\end{aligned} \quad (2)$$

holds for all $1 \leq k \leq K$. Here we have applied the fact that $\lambda_1(\mathbf{\Sigma}^*_{V_k,V_k}) \geq C\lambda_r(\mathbf{L}^*_{V_k,V_k})$ in the last line. In the following we denote $\max_k \|\widehat{\mathbf{H}}_k\widehat{\mathbf{W}}_k - \mathbf{H}_k^*\|_{2\to\infty}$ by $\varepsilon$.

**Merging step:** The following calculation shows that one can upper bound $\|\widetilde{\mathbf{H}} - \mathbf{H}^*\|_{\max}$ via $\|\widetilde{\mathbf{H}}\widehat{\mathbf{W}}_1 - \mathbf{H}^*\|_{2\to\infty}$:

$$\begin{aligned}&\|\widetilde{\mathbf{H}}\widetilde{\mathbf{H}}^\top - \mathbf{H}^*\mathbf{H}^{*\top}\|_{\max} \\ &\leq \|(\widetilde{\mathbf{H}}\widehat{\mathbf{W}}_1 - \mathbf{H}^*)(\widetilde{\mathbf{H}}\widehat{\mathbf{W}}_1 - \mathbf{H}^*)^\top\|_{\max} + 2\|\mathbf{H}^*(\widetilde{\mathbf{H}}\widehat{\mathbf{W}}_1 - \mathbf{H}^*)^\top\|_{\max} \\ &\leq \|\widetilde{\mathbf{H}}\widehat{\mathbf{W}}_1 - \mathbf{H}^*\|_{2\to\infty}^2 + 2\|\mathbf{H}^*\|_{2\to\infty}\|\widetilde{\mathbf{H}}\widehat{\mathbf{W}}_1 - \mathbf{H}^*\|_{2\to\infty} \\ &= \|\widetilde{\mathbf{H}}\widehat{\mathbf{W}}_1 - \mathbf{H}^*\|_{2\to\infty}^2 + 2\|\mathbf{L}^*\|_{\max}^{\frac{1}{2}}\|\widetilde{\mathbf{H}}\widehat{\mathbf{W}}_1 - \mathbf{H}^*\|_{2\to\infty}.\end{aligned} \quad (3)$$

Recall the definition of $\xi$ and $\xi_k, k = 1, \ldots, K$ in Section 2.1:

$$\xi_k = 2\sqrt{\frac{|J_k|\|\mathbf{L}^*_{V_k,V_k}\|_{\max}}{\lambda_r(\mathbf{L}^*_{J_k,J_k})}}\sqrt{\frac{\lambda_1(\mathbf{L}^*_{J_k,J_k})}{\lambda_r(\mathbf{L}^*_{J_k,J_k})}}, \quad \xi = \max_k \xi_k. \quad (4)$$

Now we define $\epsilon_k = \frac{3\xi^k + \xi^{k-1} - 2\xi - 2}{\xi - 1}\varepsilon$, where $\xi = \max_k \xi_k$ and and we will show by induction that when (2) holds, for $1 \leq k \leq K$,

$$\|\widetilde{\mathbf{H}}_{S_k,:}\widehat{\mathbf{W}}_1 - \mathbf{H}^*_{S_k,:}\|_{2\to\infty} \leq \epsilon_k. \tag{5}$$

When $k = 1$, $\epsilon_k = \varepsilon$ and $\widetilde{\mathbf{H}}_{S_1,:} = \widehat{\mathbf{H}}_1$, $\mathbf{H}^*_{S_1,:} = \mathbf{H}^*_1$, and (5) is immediately implied by the definition of $\varepsilon$. If (5) holds for $k \leq l-1$ and $l \leq K-1$, then

$$\begin{aligned}
&\|\widetilde{\mathbf{H}}_{S_l,:}\widehat{\mathbf{W}}_1 - \mathbf{H}^*_{S_l,:}\|_{2\to\infty} \\
&= \max\{\epsilon_{l-1}, \|(\widehat{\mathbf{H}}_l)_{\setminus J_l^V,:}\widetilde{\mathbf{W}}_l\widehat{\mathbf{W}}_1 - (\mathbf{H}^*_l)_{\setminus J_l^V,:}\|_{2\to\infty}\} \\
&\leq \max\{\epsilon_{l-1}, \varepsilon + \|(\widehat{\mathbf{H}}_l)_{\setminus J_l^V,:}(\widetilde{\mathbf{W}}_l\widehat{\mathbf{W}}_1 - \widehat{\mathbf{W}}_l)\|_{2\to\infty}\} \\
&\leq \max\{\epsilon_{l-1}, \varepsilon + \|(\widehat{\mathbf{H}}_l)_{\setminus J_l^V,:}\|_{2\to\infty}\|\widetilde{\mathbf{W}}_l\widehat{\mathbf{W}}_1 - \widehat{\mathbf{W}}_l\|_2\} \\
&\leq \max\{\epsilon_{l-1}, \varepsilon + (\varepsilon + \|\mathbf{L}^*_{V_l,V_l}\|^{\frac{1}{2}}_{\max})\|\widetilde{\mathbf{W}}_l\widehat{\mathbf{W}}_1 - \widehat{\mathbf{W}}_l\|_2\}
\end{aligned} \tag{6}$$

where we have applied the definition of $\varepsilon$ in the 3rd line, and the 5th line is due to

$$\|(\widehat{\mathbf{H}}_l)_{\setminus J_l^V,:}\|_{2\to\infty} = \|(\widehat{\mathbf{H}}_l)_{\setminus J_l^V,:}\widehat{\mathbf{W}}_l\|_{2\to\infty},$$

and $\|\mathbf{H}^*_l\|_{2\to\infty} = \|\mathbf{L}^*_{V_l,V_l}\|^{\frac{1}{2}}_{\max}$. On the other hand, by the definition of $\widetilde{\mathbf{W}}_l$, $\widetilde{\mathbf{W}}_l$ is the unitary polar factor (Li, 1995) of $(\widehat{\mathbf{H}}_l)^\top_{J_l^V,:}\widetilde{\mathbf{H}}_{J_l,:}$. Meanwhile, since $\mathbf{H}^*_{J_l,:}$ is of rank $r$, we can write $\mathbf{H}^{*\top}_{J_l,:}\mathbf{H}^*_{J_l,:} = \mathbf{P}^\top \boldsymbol{\Lambda} \mathbf{P}$ where $\mathbf{P}$ is orthonormal and $\boldsymbol{\Lambda}$ is a diagonal matrix with positive diagonal entries. Hence $\widehat{\mathbf{W}}_l\mathbf{P}^\top\mathbf{P}\widehat{\mathbf{W}}_1^\top = \widehat{\mathbf{W}}_l\widehat{\mathbf{W}}_1^\top$ is the unitary factor of

$$\widehat{\mathbf{W}}_l(\mathbf{H}^*_l)^\top_{J_l^V,:}\mathbf{H}^*_{J_l,:}\widehat{\mathbf{W}}_1^\top = \widehat{\mathbf{W}}_l\mathbf{H}^{*\top}_{J_l,:}\mathbf{H}^*_{J_l,:}\widehat{\mathbf{W}}_1^\top = \widehat{\mathbf{W}}_l\mathbf{P}^\top\boldsymbol{\Lambda}\mathbf{P}\widehat{\mathbf{W}}_1^\top.$$

By the perturbation bounds of unitary polar factors (Theorem 1 in Li (1995)),

$$\begin{aligned}
\|\widetilde{\mathbf{W}}_l\widehat{\mathbf{W}}_1 - \widehat{\mathbf{W}}_l\|_2 &= \|\widetilde{\mathbf{W}}_l - \widehat{\mathbf{W}}_l\widehat{\mathbf{W}}_1^\top\|_2 \\
&\leq \frac{2\|(\widehat{\mathbf{H}}_l)^\top_{J_l^V,:}\widetilde{\mathbf{H}}_{J_l,:} - \widehat{\mathbf{W}}_l(\mathbf{H}^*_l)^\top_{J_l^V,:}\mathbf{H}^*_{J_l,:}\widehat{\mathbf{W}}_1^\top\|_2}{\sigma_{\min}((\widehat{\mathbf{H}}_l)^\top_{J_l^V,:}\widetilde{\mathbf{H}}_{J_l,:}) + \sigma_{\min}(\widehat{\mathbf{W}}_l(\mathbf{H}^*_l)^\top_{J_l^V,:}\mathbf{H}^*_{J_l,:}\widehat{\mathbf{W}}_1^\top)} \\
&\leq \frac{2\|(\widehat{\mathbf{H}}_l)^\top_{J_l^V,:}\widetilde{\mathbf{H}}_{J_l,:} - \widehat{\mathbf{W}}_l(\mathbf{H}^*_l)^\top_{J_l^V,:}\mathbf{H}^*_{J_l,:}\widehat{\mathbf{W}}_1^\top\|_2}{2\lambda_r(\mathbf{L}^*_{J_l,J_l}) - \|(\widehat{\mathbf{H}}_l)^\top_{J_l^V,:}\widetilde{\mathbf{H}}_{J_l,:} - \widehat{\mathbf{W}}_l(\mathbf{H}^*_l)^\top_{J_l^V,:}\mathbf{H}^*_{J_l,:}\widehat{\mathbf{W}}_1^\top\|_2}.
\end{aligned} \tag{7}$$

16

Meanwhile,

$$\|(\widehat{\mathbf{H}}_l)_{J_l^V,:}^\top \widetilde{\mathbf{H}}_{J_l,:} - \widehat{\mathbf{W}}_l (\mathbf{H}_l^*)_{J_l^V,:}^\top \mathbf{H}_{J_l,:}^* \widehat{\mathbf{W}}_1^\top\|_2$$

$$=\|\widehat{\mathbf{W}}_l^\top (\widehat{\mathbf{H}}_l)_{J_l^V,:}^\top \widetilde{\mathbf{H}}_{J_l,:} \widehat{\mathbf{W}}_1 - (\mathbf{H}_l^*)_{J_l^V,:}^\top \mathbf{H}_{J_l,:}^*\|_2$$

$$\leq \|(\widehat{\mathbf{H}}_l)_{J_l^V,:}\widehat{\mathbf{W}}_l - (\mathbf{H}_l^*)_{J_l^V,:}\|_2 \|\mathbf{H}_{J_l,:}^*\|_2 + \|\widetilde{\mathbf{H}}_{J_l,:}\widehat{\mathbf{W}}_1 - \mathbf{H}_{J_l,:}^*\|_2 \|\mathbf{H}_{J_l,:}^*\|_2$$

$$+ \|(\widehat{\mathbf{H}}_l)_{J_l^V,:}\widehat{\mathbf{W}}_l - (\mathbf{H}_l^*)_{J_l^V,:}\|_2 \|\widetilde{\mathbf{H}}_{J_l,:}\widehat{\mathbf{W}}_1 - \mathbf{H}_{J_l,:}^*\|_2$$

$$\leq \sqrt{|J_l| \lambda_1(\mathbf{L}_{J_l,J_l}^*)} (\varepsilon + \epsilon_{l-1}) + |J_l| \varepsilon \epsilon_{l-1}.$$

Here, the last line above is due to the definition of $\varepsilon$, the induction assumption that (5) holds for $k \leq l-1$, and the fact that for any matrix $\mathbf{A}$ with $m$ rows, $\|\mathbf{A}\|_2 \leq \sqrt{m} \|\mathbf{A}\|_{2\to\infty}$. Recall the definition of $\xi_k$ and $\xi$ in (4), one can show that $\xi_k \geq 2\sqrt{\frac{\|\mathbf{L}_{J_k,J_k}^*\|_F}{\lambda_r(\mathbf{L}_{J_k,J_k}^*)}} > 1$, and hence the definition of $\epsilon_{l-1}$ implies that $\epsilon_{l-1} \leq \frac{4\xi^{l-1}}{\xi-1}\varepsilon$. Furthermore, when (2) holds, Assumption C.2-C.4 implies that

$$\begin{aligned}
\epsilon_{l-1} &\leq C \frac{\xi^{l-1}}{\xi-1} \max_k \sqrt{\frac{(\mu_k r^2 + \tau_k r)(\tau_k \vee \log p_k)}{p_k n_k}} \lambda_1^{\frac{1}{2}}(\mathbf{\Sigma}_{V_k,V_k}^*) \\
&\leq C \frac{\xi^{l-1}}{\xi-1} \min_k \sqrt{\frac{\lambda_r(\mathbf{L}_{V_k,V_k}^*)\mu_k r}{p_k}} \max_k \sqrt{\frac{(r+\frac{\tau_k}{\mu_k})(\tau_k \vee \log p_k)}{n_k}} \\
&\leq C \frac{\sqrt{\min_k \|\mathbf{L}_{V_k,V_k}^*\|_{\max}}}{\max_k \xi_k} \\
&\leq C \sqrt{\frac{\lambda_1(\mathbf{L}_{J_l,J_l}^*)}{|J_l|}},
\end{aligned} \quad (8)$$

where the third line is due to the fact that $\|\mathbf{L}_{V_k,V_k}^*\|_{\max} \geq \frac{\lambda_r(\mathbf{L}_{V_k,V_k}^*)\mu_k r}{p_k}$. This further implies

17

that $|J_l|\varepsilon\epsilon_{l-1} \leq \sqrt{|J_l|\lambda_1(\mathbf{L}^*_{J_l,J_l})}\varepsilon$ and

$$
\begin{aligned}
&\|(\widehat{\mathbf{H}}_l)^\top_{J_l^V,:}\widetilde{\mathbf{H}}_{J_l,:} - \widehat{\mathbf{W}}_l(\mathbf{H}^*_l)^\top_{J_l^V,:}\mathbf{H}^*_{J_l,:}\widehat{\mathbf{W}}_1^\top\|_2 \\
&\leq \sqrt{|J_l|\lambda_1(\mathbf{L}^*_{J_l,J_l})}(2\varepsilon + \epsilon_{l-1}) \\
&\leq \sqrt{|J_l|\lambda_1(\mathbf{L}^*_{J_l,J_l})}\frac{3\xi^{l-1} + \xi^{l-2} - 4}{\xi - 1}\varepsilon \\
&\leq \sqrt{|J_l|\lambda_1(\mathbf{L}^*_{J_l,J_l})}\frac{4\xi^{K-1}}{\xi - 1}\varepsilon \\
&\leq C\sqrt{|J_l|\lambda_1(\mathbf{L}^*_{J_l,J_l})}\frac{\sqrt{\min_k \|\mathbf{L}^*_{V_k,V_k}\|_{\max}}}{\max_k \xi_k} \\
&\leq \lambda_r(\mathbf{L}^*_{J_l,J_l}),
\end{aligned} \tag{9}
$$

where the 5th line follows similar arguments to (8), and the last line is due to the definition of $\xi_k$. Therefore, (9) and (7) together imply $\|\widetilde{\mathbf{W}}_l\widehat{\mathbf{W}}_1 - \widehat{\mathbf{W}}_l\|_2 \leq \frac{\sqrt{|J_l|\lambda_1(\mathbf{L}^*_{J_l,J_l})}}{\lambda_r(\mathbf{L}^*_{J_l,J_l})}\frac{6\xi^{l-1}+2\xi^{l-2}-8}{\xi-1}\varepsilon \leq 1$. Plugging this bound for $\|\widetilde{\mathbf{W}}_l\widehat{\mathbf{W}}_1 - \widehat{\mathbf{W}}_l\|_2$ into (6), we have

$$
\begin{aligned}
&\|\widetilde{\mathbf{H}}_{S_l,:}\widehat{\mathbf{W}}_1 - \mathbf{H}^*_{S_l,:}\|_{2\to\infty} \\
&\leq \max\left\{\epsilon_{l-1}, 2\varepsilon + \frac{\sqrt{|J_l|\lambda_1(\mathbf{L}^*_{J_l,J_l})\|\mathbf{L}^*_{V_l,V_l}\|_{\max}}}{\lambda_r(\mathbf{L}^*_{J_l,J_l})}\frac{6\xi^{l-1}+2\xi^{l-2}-8}{\xi-1}\varepsilon\right\} \\
&\leq \max\left\{\epsilon_{l-1}, \frac{3\xi^l + \xi^{l-1} - 2\xi - 2}{\xi - 1}\varepsilon\right\} = \epsilon_l.
\end{aligned}
$$

Therefore, with probability at least $1 - C\sum_{k=1}^K p_k^{-c}$,

$$\|\widetilde{\mathbf{H}}\widehat{\mathbf{W}}_1 - \mathbf{H}^*\|_{2\to\infty} \leq \epsilon_K \leq \frac{4\xi^K}{\xi-1}\varepsilon \leq \min_k \frac{\lambda_r(\mathbf{L}^*_{J_k,J_k})}{\sqrt{|J_k|\lambda_1(\mathbf{L}^*_{J_k,J_k})}} \leq \min_k \sqrt{\frac{\|\mathbf{L}^*_{J_k,J_k}\|_F}{|J_k|}} \leq \|\mathbf{L}^*\|_{\max}^{\frac{1}{2}},$$

where the 3rd inequality follows similar arguments to (8), and the 4th and 5th inequalities are caused by the facts that $\lambda_1(\mathbf{L}^*_{J_k,J_k}) \leq \|\mathbf{L}^*_{J_k,J_k}\|_F \leq \|\mathbf{L}^*_{J_k,J_k}\|_{\max}|J_k|$. Plugging this bound into (3), we have

$$\|\widetilde{\mathbf{H}}\widetilde{\mathbf{H}}^\top - \mathbf{H}^*\mathbf{H}^{*\top}\|_{\max} \leq \frac{12\|\mathbf{L}^*\|_{\max}^{\frac{1}{2}}\xi^K}{\xi - 1}\max_k\sqrt{\frac{(\mu_k r^2 + \tau_k r)(\tau_k \vee \log p_k)}{p_k n_k}}\lambda_1^{\frac{1}{2}}(\mathbf{\Sigma}^*_{V_k,V_k}).$$

18

Also noting the fact that $\frac{\mu_k r}{p_k}\lambda_1(\boldsymbol{\Sigma}^*_{V_k,V_k}) \leq C\|\mathbf{L}^*_{V_k,V_k}\|_{\max} \leq \|\mathbf{L}^*\|_{\max}$, $\|\widetilde{\mathbf{H}}\widetilde{\mathbf{H}}^\top - \mathbf{H}^*\mathbf{H}^{*\top}\|_{\max} = \|\widetilde{\boldsymbol{\Sigma}} - \boldsymbol{\Sigma}^*\|_{\max}$ and we can further give the following bound:

$$\|\widetilde{\boldsymbol{\Sigma}} - \boldsymbol{\Sigma}*\|_{\max} \leq \frac{C\|\mathbf{L}^*\|_{\max}\xi^K}{\xi - 1}\max_k \sqrt{\frac{(r + \frac{\tau_k}{\mu_k})(\tau_k \vee \log p_k)}{n_k}}.$$

$\square$

*Proof of Theorem 2.* Recall that we assumed $\mathbf{L}^* = \mathbf{H}^*\mathbf{H}^{*\top}$ where all entries $\mathbf{H}^*_{j,k}$ are i.i.d. mean zero Gaussian random variables. Since $\xi_k = 2\sqrt{\frac{|J_k|\|\mathbf{L}^*_{V_k,V_k}\|_{\max}}{\lambda_r(\mathbf{L}^*_{J_k,J_k})}}\sqrt{\frac{\lambda_1(\mathbf{L}^*_{J_k,J_k})}{\lambda_r(\mathbf{L}^*_{J_k,J_k})}}$ does not change when $\mathbf{L}^*$ is multiplied by a factor, we can simply assume $\mathbf{H}^*_{j,k}$ are i.i.d. $\mathcal{N}(0,1)$ without loss if generality.

We first provide upper and lower bounds for $\lambda_1(\mathbf{L}^*_{J_k,J_k})$, $\lambda_r(\mathbf{L}^*_{J_k,J_k})$ with high probability. Note that $\lambda_1(\mathbf{L}^*_{J_k,J_k}) = s^2_{\max}(\mathbf{H}^*_{J_k,:})$, and $\lambda_r(\mathbf{L}^*_{J_k,J_k}) = s^2_{\min}(\mathbf{H}^*_{J_k,:})$, where $s_{\max}(\cdot)$ and $s_{\min}(\cdot)$ are maximum and minimum singular values. Applying the singular value bounds for random matrices (see Vershynin, 2010, Corollary 5.35), we have

$$\sqrt{|J_k|} - \sqrt{r} - t_k \leq s_{\min}(\mathbf{H}^*_{J_k,:}) \leq s_{\max}(\mathbf{H}^*_{J_k,:}) \leq \sqrt{|J_k|} + \sqrt{r} + t_k,$$

with probability at least $1 - 2\exp\{-t_k^2/2\}$. Let $t_k = \frac{3-\sqrt{2}}{4}\sqrt{|J_k|}$, then with probability at least $1 - C\sum_{k=1}^K \exp\{-c|J_k|\}$,

$$c|J_k| \leq \lambda_1(\mathbf{L}^*_{J_k,J_k}) \leq \lambda_r(\mathbf{L}^*_{J_k,J_k}) \leq C|J_k|,$$

for some constants $c, C > 0$. Furthermore, one can show that $\|\mathbf{L}^*_{V_k,V_k}\|_{\max} = \max_{j \in V_k}\|\mathbf{H}^*_{j,:}\|_2^2$, and $\{\|\mathbf{H}^*_{j,:}\|_2^2\}_{j \in V_k}$ are i.i.d. random variables, each being the sum of $r$ sub-exponential random variables with mean 1 and constant sub-exponential norm. Therefore, we can apply the Bernstein-type inequality for sub-exponential sums (see Vershynin, 2010, Proposition 5.16) and obtain the following:

$$\mathbb{P}(|\|\mathbf{H}^*_{i,:}\|_2^2 - r| > Cr\log p_k) \leq 2\exp\{-cr\log p_k\}.$$

Taking a union bound over $i \in V_k$, we have

$$\mathbb{P}(\max_{i \in V_k} \|\mathbf{H}^*_{i,:}\|_2^2 > Cr \log p_k) \leq 2\exp\{-cr \log p_k\}.$$

Therefore, $\xi_k \leq C\sqrt{r \log p_k}, k = 1, \ldots, K$ with probability at least $1 - C \sum_{k=1}^{K} \exp\{-c \min\{|J_k|^2, r \log p_k\}\}$.

$\square$

## D.2 Proofs of Supporting Lemmas

*Proof of Lemma D.1.* Consider the SVD of $\widehat{\mathbf{U}}^\top \mathbf{U}^* = \mathbf{W}^{(1)} \mathbf{\Lambda}_{\mathbf{U}, \widehat{\mathbf{U}}} \mathbf{W}^{(2)\top}$, and define $\widehat{\mathbf{W}} = \mathbf{W}^{(1)} \mathbf{W}^{(2)\top} \in \mathbb{R}^{r \times r}$. First note that

$$\begin{aligned}
&\|\widehat{\mathbf{U}} \widehat{\mathbf{\Lambda}}^{\frac{1}{2}} \widehat{\mathbf{W}} - \mathbf{U}^* \mathbf{\Lambda}^{*\frac{1}{2}}\|_{2 \to \infty} \\
\leq & \|\widehat{\mathbf{U}} \widehat{\mathbf{\Lambda}}^{\frac{1}{2}} \widehat{\mathbf{W}} - \widehat{\mathbf{U}} \widehat{\mathbf{W}} \mathbf{\Lambda}^{*\frac{1}{2}}\|_{2 \to \infty} + \|\widehat{\mathbf{U}} \widehat{\mathbf{W}} \mathbf{\Lambda}^{*\frac{1}{2}} - \mathbf{U}^* \mathbf{\Lambda}^{*\frac{1}{2}}\|_{2 \to \infty} \\
\leq & \|\widehat{\mathbf{U}}\|_{2 \to \infty} \|\widehat{\mathbf{\Lambda}}^{\frac{1}{2}} - \widehat{\mathbf{W}} \mathbf{\Lambda}^{*\frac{1}{2}} \widehat{\mathbf{W}}^\top\|_2 + \|\widehat{\mathbf{U}} \widehat{\mathbf{W}} - \mathbf{U}^*\|_{2 \to \infty} \|\mathbf{\Lambda}^{*\frac{1}{2}}\|_2 \\
\leq & (\|\widehat{\mathbf{U}} \widehat{\mathbf{W}} - \mathbf{U}^*\|_{2 \to \infty} + \|\mathbf{U}^*\|_{2 \to \infty}) \|\widehat{\mathbf{\Lambda}}^{\frac{1}{2}} - \widehat{\mathbf{W}} \mathbf{\Lambda}^{*\frac{1}{2}} \widehat{\mathbf{W}}^\top\|_2 + \|\widehat{\mathbf{U}} \widehat{\mathbf{W}} - \mathbf{U}^*\|_{2 \to \infty} \|\mathbf{\Sigma}^*\|_2^{\frac{1}{2}},
\end{aligned} \quad (10)$$

where we have applied the fact that $\|\widehat{\mathbf{U}}\|_{2 \to \infty} \leq \|\widehat{\mathbf{U}} \widehat{\mathbf{W}}\|_{2 \to \infty} \|\widehat{\mathbf{W}}^\top\|_2 = \|\widehat{\mathbf{U}} \widehat{\mathbf{W}}\|_{2 \to \infty}$, and $\|\mathbf{\Lambda}^{*\frac{1}{2}}\|_2 = \|\mathbf{L}^*\|_2^{\frac{1}{2}} \leq \|\mathbf{\Sigma}^*\|_2^{\frac{1}{2}}$. In the following, we provide bounds for $\|\widehat{\mathbf{U}} \widehat{\mathbf{W}} - \mathbf{U}^*\|_{2 \to \infty}$ and $\|\widehat{\mathbf{\Lambda}}^{\frac{1}{2}} - \widehat{\mathbf{W}} \mathbf{\Lambda}^{*\frac{1}{2}} \widehat{\mathbf{W}}\|_2$ separately.

**Bounding** $\|\widehat{\mathbf{U}} \widehat{\mathbf{W}} - \mathbf{U}^*\|_{2 \to \infty}$**:** Our proof for this step closely follows the proof of Theorem 1.1 in Cape et al. (2019), but with a few modifications that leads to a tighter bound w.r.t. rank $r$. First note that $\|\widehat{\mathbf{U}} \widehat{\mathbf{W}} - \mathbf{U}^*\|_{2 \to \infty} = \|(\widehat{\mathbf{U}} - \mathbf{U}^* \widehat{\mathbf{W}}^\top) \widehat{\mathbf{W}}\|_{2 \to \infty} = \|\widehat{\mathbf{U}} - \mathbf{U}^* \widehat{\mathbf{W}}^\top\|_{2 \to \infty}$, and we can decompose $\widehat{\mathbf{U}} - \mathbf{U}^* \widehat{\mathbf{W}}^\top$ as follows:

$$\begin{aligned}
\widehat{\mathbf{U}} - \mathbf{U}^* \widehat{\mathbf{W}}^\top = & \mathbf{U}^* (\mathbf{U}^{*\top} \widehat{\mathbf{U}} - \widehat{\mathbf{W}}^\top) + (\mathbf{I} - \mathbf{U}^* \mathbf{U}^{*\top})(\widehat{\mathbf{\Sigma}} - \mathbf{\Sigma}^*) \widehat{\mathbf{U}} \widehat{\mathbf{\Lambda}}^{-1} + (\mathbf{I} - \mathbf{U}^* \mathbf{U}^{*\top}) \mathbf{L}^* \widehat{\mathbf{U}} \widehat{\mathbf{\Lambda}}^{-1} \\
= & \mathbf{U}^* (\mathbf{U}^{*\top} \widehat{\mathbf{U}} - \widehat{\mathbf{W}}^\top) + (\mathbf{I} - \mathbf{U}^* \mathbf{U}^{*\top})(\widehat{\mathbf{\Sigma}} - \mathbf{\Sigma}^*) \mathbf{U}^* \mathbf{U}^{*\top} \widehat{\mathbf{U}} \widehat{\mathbf{\Lambda}}^{-1} \\
& + (\mathbf{I} - \mathbf{U}^* \mathbf{U}^{*\top})(\widehat{\mathbf{\Sigma}} - \mathbf{\Sigma}^*)(\mathbf{I} - \mathbf{U}^* \mathbf{U}^{*\top}) \widehat{\mathbf{U}} \widehat{\mathbf{\Lambda}}^{-1},
\end{aligned}$$

20

where the first line is due to that $(\widehat{\boldsymbol{\Sigma}} - \sigma^{*2}\mathbf{I})\widehat{\mathbf{U}}\widehat{\boldsymbol{\Lambda}}^{-1} = \widehat{\mathbf{U}}$, and the second line utilizes the fact that $(\mathbf{I} - \mathbf{U}^*\mathbf{U}^{*\top})\mathbf{L}^* = 0$. Hence we have

$$\|\widehat{\mathbf{U}} - \mathbf{U}^*\widehat{\mathbf{W}}^\top\|_{2\to\infty} \leq \|\mathbf{U}^*\|_{2\to\infty}\|\mathbf{U}^{*\top}\widehat{\mathbf{U}} - \widehat{\mathbf{W}}^\top\|_2 + \|\mathbf{U}^*_\perp\mathbf{U}^{*\top}_\perp(\widehat{\boldsymbol{\Sigma}} - \boldsymbol{\Sigma}^*)\mathbf{U}^*\|_{2\to\infty}(\lambda_r(\widehat{\boldsymbol{\Sigma}}) - \sigma^{*2})^{-1}$$
$$+ \|\mathbf{U}^*_\perp\mathbf{U}^{*\top}_\perp(\widehat{\boldsymbol{\Sigma}} - \boldsymbol{\Sigma}^*)\mathbf{U}^*_\perp\mathbf{U}^{*\top}_\perp\|_{2\to\infty}(\lambda_r(\widehat{\boldsymbol{\Sigma}}) - \sigma^{*2})^{-1}.$$

Further note that

$$\|\widehat{\mathbf{W}} - \widehat{\mathbf{U}}^\top\mathbf{U}^*\|_2 = \|\mathbf{W}^{(1)}(\mathbf{I} - \boldsymbol{\Lambda}_{\mathbf{U},\widehat{\mathbf{U}}})\mathbf{W}^{(2)\top}\|_2$$
$$= 1 - \sigma_r(\widehat{\mathbf{U}}^\top\mathbf{U}^*) \leq 1 - \sigma_r^2(\widehat{\mathbf{U}}^\top\mathbf{U}^*) = \|\sin\boldsymbol{\Theta}(\widehat{\mathbf{U}}, \mathbf{U}^*)\|_2^2,$$

and the Davis-Kahan $\sin\boldsymbol{\Theta}$ theorem establishes that

$$\|\sin\boldsymbol{\Theta}(\widehat{\mathbf{U}}, \mathbf{U}^*)\|_2 \leq \frac{\|\widehat{\boldsymbol{\Sigma}} - \boldsymbol{\Sigma}^*\|_2}{\lambda_r(\boldsymbol{\Sigma}^*) - \lambda_{r+1}(\widehat{\boldsymbol{\Sigma}})} \leq \frac{\|\widehat{\boldsymbol{\Sigma}} - \boldsymbol{\Sigma}^*\|_2}{\lambda_r(\mathbf{L}^*) - \|\widehat{\boldsymbol{\Sigma}} - \boldsymbol{\Sigma}^*\|_2},$$

where the second inequality is due to Weyl's inequality: $\lambda_{r+1}(\widehat{\boldsymbol{\Sigma}}) \leq \sigma^{*2} + |\lambda_{r+1}(\widehat{\boldsymbol{\Sigma}}) - \lambda_{r+1}(\boldsymbol{\Sigma}^*)| \leq \sigma^{*2} + \|\widehat{\boldsymbol{\Sigma}} - \boldsymbol{\Sigma}^*\|_2$ and $\lambda_r(\boldsymbol{\Sigma}^*) = \sigma^{*2} + \lambda_r(\mathbf{L}^*)$. Since $\widehat{\boldsymbol{\Sigma}}$ is the sample covariance of $n$ samples from multivariate Gaussian $\mathcal{N}(0, \boldsymbol{\Sigma}^*)$, it follows from Koltchinskii and Lounici (2017a,b) that with probability at least $1 - \frac{1}{3}p^{-2}$, $\|\widehat{\boldsymbol{\Sigma}} - \boldsymbol{\Sigma}^*\|_2 \leq C\|\boldsymbol{\Sigma}^*\|_2\sqrt{\frac{\tau \vee \log p}{n}}$, which implies

$$\|\sin\boldsymbol{\Theta}(\widehat{\mathbf{U}}, \mathbf{U}^*)\|_2 \leq C\frac{\lambda_1(\boldsymbol{\Sigma}^*)}{\lambda_r(\mathbf{L}^*)}\sqrt{\frac{\tau \vee \log p}{n}}.$$

Here we have applied the condition that $\frac{\lambda_1^2(\boldsymbol{\Sigma})}{\lambda_r^2(\mathbf{L}^*)}\sqrt{\frac{\tau^* \vee \log p}{n}} \leq C$ so that $\|\widehat{\boldsymbol{\Sigma}} - \boldsymbol{\Sigma}^*\|_2 \leq \frac{1}{2}\lambda_r(\mathbf{L}^*)$.

In addition, one can show that $\lambda_r(\widehat{\boldsymbol{\Sigma}}) - \sigma^{*2} \geq \lambda_r(\mathbf{L}^*) - \|\widehat{\boldsymbol{\Sigma}} - \boldsymbol{\Sigma}^*\|_2 \geq \frac{1}{2}\lambda_r(\mathbf{L}^*)$, which implies

$$\|\widehat{\mathbf{U}} - \mathbf{U}^*\widehat{\mathbf{W}}^\top\|_{2\to\infty} \leq \frac{C\lambda_1^2(\boldsymbol{\Sigma}^*)}{\lambda_r^2(\mathbf{L}^*)}\|\mathbf{U}^*\|_{2\to\infty}\frac{\tau \vee \log p}{n} + 2\sqrt{r}\|\mathbf{U}^*_\perp\mathbf{U}^{*\top}_\perp(\widehat{\boldsymbol{\Sigma}} - \boldsymbol{\Sigma}^*)\mathbf{U}^*\|_{\max}\lambda_r^{-1}(\mathbf{L}^*)$$
$$+ 2\|\mathbf{U}^{*\top}_\perp(\widehat{\boldsymbol{\Sigma}} - \boldsymbol{\Sigma}^*)\mathbf{U}^*_\perp\|_2\lambda_r^{-1}(\mathbf{L}^*).$$

Note that $\mathbf{U}^{*\top}_\perp\widehat{\boldsymbol{\Sigma}}\mathbf{U}^*_\perp$ can be viewed as the sample covariance matrix of data $\{\mathbf{U}^{*\top}_\perp x_1, \ldots, \mathbf{U}^{*\top}_\perp x_n\}$ in $\mathbb{R}^{p-r}$, and the corresponding population covariance matrix is $\mathbf{U}^{*\top}_\perp\boldsymbol{\Sigma}^*\mathbf{U}^*_\perp = \sigma^{*2}\mathbf{I}_{p-r}$.

Therefore, one can apply the sample covariance error bound in Koltchinskii and Lounici (2017a) again to obtain

$$\|\mathbf{U}_\perp^{*\top}(\widehat{\mathbf{\Sigma}} - \mathbf{\Sigma}^*)\mathbf{U}_\perp^*\|_2 \le C\sigma^{*2}\sqrt{\frac{p-r}{n}} \le C\sqrt{\sigma^{*2}\lambda_1(\mathbf{\Sigma}^*)}\sqrt{\frac{\tau \vee \log p}{n}},$$

with probability at least $1 - \frac{1}{3}p^{-2}$, where the last inequality is due to that $\frac{\sigma^{*2}(p-r)}{\lambda_1(\mathbf{\Sigma}^*)} \le \frac{\text{tr}(\mathbf{\Sigma}^*)}{\lambda_1(\mathbf{\Sigma}^*)} = \tau$. While for $\|\mathbf{U}_\perp^*\mathbf{U}_\perp^{*\top}(\widehat{\mathbf{\Sigma}} - \mathbf{\Sigma}^*)\mathbf{U}^*\|_{\max}$, we can bound each entry $|(\mathbf{U}_\perp^*\mathbf{U}_\perp^{*\top})_{j,:}(\widehat{\mathbf{\Sigma}} - \mathbf{\Sigma}^*)\mathbf{U}^*_{:,k}|$ and then take a union bound. For any $j \in [p-r]$, $k \in [r]$, we can write $(\mathbf{U}_\perp^*\mathbf{U}_\perp^{*\top})_{j,:}(\widehat{\mathbf{\Sigma}} - \mathbf{\Sigma}^*)\mathbf{U}^*_{:,k} = \frac{1}{n}\sum_{i=1}^n (\mathbf{U}_\perp^*\mathbf{U}_\perp^{*\top})_{j,:}x_i x_i^\top \mathbf{U}^*_{:,k}$, where $(\mathbf{U}_\perp^*\mathbf{U}_\perp^{*\top})_{j,:}x_i \sim \mathcal{N}(0, \sigma^{*2}\|(\mathbf{U}_\perp^*)_{j,:}\|_2^2)$, $\mathbf{U}^{*\top}_{:,k}x_i \sim \mathcal{N}(0, \lambda_k(\mathbf{\Sigma}^*))$. Hence

$$\begin{aligned}\|(\mathbf{U}_\perp^*\mathbf{U}_\perp^{*\top})_{j,:}x_i x_i^\top \mathbf{U}^*_{:,k}\|_{\psi_1} &\le \left\|\frac{\sqrt{\lambda_1(\mathbf{\Sigma}^*)}}{2\sigma^*}[(\mathbf{U}_\perp^*\mathbf{U}_\perp^{*\top})_{j,:}x_i]^2\right\|_{\psi_1} + \left\|\frac{\sigma^*}{2\sqrt{\lambda_1(\mathbf{\Sigma}^*)}}(\mathbf{U}^{*\top}_{:,k}x_i)^2\right\|_{\psi_1}\\ &\le \frac{\sqrt{\lambda_1(\mathbf{\Sigma}^*)}}{2\sigma^*}\|(\mathbf{U}_\perp^*\mathbf{U}_\perp^{*\top})_{j,:}x_i\|_{\psi_2}^2 + \frac{\sigma^*}{2\sqrt{\lambda_1(\mathbf{\Sigma}^*)}}\|\mathbf{U}^{*\top}_{:,k}x_i\|_{\psi_2}^2\\ &\le C\sigma^*\sqrt{\lambda_1(\mathbf{\Sigma}^*)}.\end{aligned}$$

Here we have applied the Cauchy-Schwarz inequality and triangle inequality of norms in the first line; the second line is due to the relationship between $\|\cdot\|_{\psi_1}$ and $\|\cdot\|_{\psi_2}$ (see e.g., Lemma 5.14 in Vershynin, 2010); the third line is due to that standard Gaussian random variables have constant $\|\cdot\|_{\psi_2}$ norm and $\|(\mathbf{U}_\perp^*)_{j,:}\|_2^2 \le 1$. Therefore, by applying the Bernstein type inequality for sum of sub-exponential random variables (see Proposition 5.16 in Vershynin, 2010), we have

$$\left|(\mathbf{U}_\perp^*\mathbf{U}_\perp^{*\top})_{j,:}(\widehat{\mathbf{\Sigma}} - \mathbf{\Sigma}^*)\mathbf{U}^*_{:,k}\right| \le C\sigma^*\sqrt{\lambda_1(\mathbf{\Sigma}^*)}\sqrt{\frac{\log p}{n}}$$

holds for all $1 \le j \le p-r$, $1 \le k \le r$, with probability at least $1 - Cp^{-c}$. Combining the

bounds discussed above, we have

$$\|\widehat{\mathbf{U}} - \mathbf{U}^*\widehat{\mathbf{W}}^\top\|_{2\to\infty} \leq \frac{C\lambda_1^2(\mathbf{\Sigma}^*)}{\lambda_r^2(\mathbf{L}^*)}\|\mathbf{U}^*\|_{2\to\infty}\frac{\tau \vee \log p}{n} + \frac{C\sigma^*\sqrt{\lambda_1(\mathbf{\Sigma}^*)}}{\lambda_r(\mathbf{L}^*)}\sqrt{\frac{\tau \vee (r\log p)}{n}}$$

$$\leq C\left(\|\mathbf{U}^*\|_{2\to\infty} + \frac{\sigma^*\sqrt{\lambda_1(\mathbf{\Sigma}^*)}}{\lambda_r(\mathbf{L}^*)}\right)\sqrt{\frac{\tau \vee (r\log p)}{n}}$$

$$\leq C\left(\sqrt{\frac{\mu r^2}{p}} + \frac{\lambda_1(\mathbf{\Sigma}^*)}{\lambda_r(\mathbf{L}^*)}\sqrt{\frac{\tau r}{p}}\right)\sqrt{\frac{\tau \vee \log p}{n}},$$

where the last line is due to that $\frac{\lambda_1(\mathbf{\Sigma}^*)}{\sigma^{*2}} \geq \frac{\lambda_1(\mathbf{\Sigma}^*)}{\text{tr}(\mathbf{\Sigma}^*)/p} = \frac{p}{\tau}$, and $\|\mathbf{U}^*\|_{2\to\infty} \leq \sqrt{\frac{\mu r}{p}}$. Therefore,

**Bounding** $\|\widehat{\mathbf{\Lambda}}^{\frac{1}{2}} - \widehat{\mathbf{W}}\mathbf{\Lambda}^{*\frac{1}{2}}\widehat{\mathbf{W}}\|_2$: While for $\|\widehat{\mathbf{\Lambda}}^{\frac{1}{2}} - \widehat{\mathbf{W}}\mathbf{\Lambda}^{*\frac{1}{2}}\widehat{\mathbf{W}}\|_2$, we first apply the Taylor's theorem and bounds on the derivative of square root matrices, showing that $\|\widehat{\mathbf{\Lambda}}^{\frac{1}{2}} - \widehat{\mathbf{W}}\mathbf{\Lambda}^{*\frac{1}{2}}\widehat{\mathbf{W}}\|_2$ can be controlled by $\|\widehat{\mathbf{\Lambda}} - \widehat{\mathbf{W}}\mathbf{\Lambda}^*\widehat{\mathbf{W}}\|_2$, and then bounding the latter via $\|\sin\mathbf{\Theta}(\widehat{\mathbf{U}}, \mathbf{U}^*)\|_2$. Our proof idea for the first step is similar to Mathias (1997) while we adopt a non-asymptotic analysis that suits our needs.

Denote $\widehat{\mathbf{W}}\mathbf{\Lambda}^*\widehat{\mathbf{W}}^\top - \widehat{\mathbf{\Lambda}}$ by $\mathbf{\Delta}$. Define matrix-valued function $g(\eta) = (\widehat{\mathbf{\Lambda}} + \eta\mathbf{\Delta})^{\frac{1}{2}}$, then $g(0) = \widehat{\mathbf{\Lambda}}^{\frac{1}{2}}$ and $g(1) = \widehat{\mathbf{W}}\mathbf{\Lambda}^{*\frac{1}{2}}\widehat{\mathbf{W}}^\top$. By the Taylor's theorem, there exists $\eta_0 \in [0,1]$ such that

$$g(1) - g(0) = \frac{dg(\eta)}{d\eta}\bigg|_{\eta=\eta_0} = \mathbf{U}(\eta_0)\left\{\left(\frac{\sqrt{\mathbf{\Lambda}_{i,i}(\eta_0)} - \sqrt{\mathbf{\Lambda}_{j,j}(\eta_0)}}{\mathbf{\Lambda}_{i,i}(\eta_0) - \mathbf{\Lambda}_{j,j}(\eta_0)}\right)_{i,j} \circ \mathbf{U}(\eta_0)\mathbf{\Delta}\mathbf{U}(\eta_0)\right\}\mathbf{U}(\eta_0)^\top,$$

where the second equality is due to the matrix derivative formula (see Theorem 6.6.30 in Horn and Johnson (1991)), and $\mathbf{U}(\eta_0) \in \mathbb{O}^{p\times r_0}, \mathbf{\Lambda}(\eta_0) \in \mathbb{R}^{r\times r}$ are defined by the eigendecomposition $\widehat{\mathbf{\Lambda}} + \eta_0\mathbf{\Delta} = \mathbf{U}(\eta_0)\mathbf{\Lambda}(\eta_0)\mathbf{U}(\eta_0)^\top$. Denote $\left(\frac{\sqrt{\mathbf{\Lambda}_{i,i}(\eta_0)} - \sqrt{\mathbf{\Lambda}_{j,j}(\eta_0)}}{\mathbf{\Lambda}_{i,i}(\eta_0) - \mathbf{\Lambda}_{j,j}(\eta_0)}\right)_{i,j}$ by $\mathbf{Z}(\eta_0)$, then

$$\|\widehat{\mathbf{W}}\mathbf{\Lambda}^{*\frac{1}{2}}\widehat{\mathbf{W}}^\top - \widehat{\mathbf{\Lambda}}^{\frac{1}{2}}\|_2 = \|g(1) - g(0)\|_2 \leq \|\mathbf{Z}(\eta_0) \circ \mathbf{U}(\eta_0)^\top\mathbf{\Delta}\mathbf{U}(\eta_0)\|_2.$$

The following lemma shows that $\mathbf{Z}(\eta_0)$ is positive semidefinite:

**Lemma D.2.** *For any $u_1, \ldots, u_p \in \mathbb{R}$, $v_1, \ldots, v_p > 0$, $\sum_{i,j}\frac{u_i u_j}{v_i + v_j} \geq 0$.*

23

Thus we can apply Theorem 5.5.18 in Horn and Johnson (1991) on the spectral norm of Hadamard product matrix and obtain

$$\|\mathbf{Z}(\eta_0) \circ \mathbf{U}(\eta_0)^\top \mathbf{\Delta}\mathbf{U}(\eta_0)\|_2 \leq \|\mathbf{U}(\eta_0)^\top \mathbf{\Delta}\mathbf{U}(\eta_0)\|_2 \max_i |\mathbf{Z}_{i,i}(\eta_0)| \leq \frac{1}{2}\|\mathbf{\Delta}\|_2 (\min_i \mathbf{\Lambda}_{i,i}(\eta_0))^{-\frac{1}{2}}.$$

Furthermore, by Weyl's inequality, $\mathbf{\Lambda}_{i,i}(\eta_0) = \lambda_i(\widehat{\mathbf{W}}\mathbf{\Lambda}^*\widehat{\mathbf{W}}^\top - (1-\eta_0)\mathbf{\Delta}) \geq \lambda_i(\mathbf{\Lambda}^*) - \|\mathbf{\Delta}\|$, which implies

$$\|\widehat{\mathbf{W}}\mathbf{\Lambda}^{*\frac{1}{2}}\widehat{\mathbf{W}}^\top - \widehat{\mathbf{\Lambda}}^{\frac{1}{2}}\|_2 \leq \frac{\|\widehat{\mathbf{W}}\mathbf{\Lambda}^*\widehat{\mathbf{W}}^\top - \widehat{\mathbf{\Lambda}}\|_2}{2(\lambda_r(\mathbf{L}^*) - \|\widehat{\mathbf{W}}\mathbf{\Lambda}^*\widehat{\mathbf{W}}^\top - \widehat{\mathbf{\Lambda}}\|_2)^{\frac{1}{2}}}. \tag{11}$$

Now we focus on bounding $\|\widehat{\mathbf{W}}\mathbf{\Lambda}^*\widehat{\mathbf{W}}^\top - \widehat{\mathbf{\Lambda}}\|_2$:

$$\|\widehat{\mathbf{W}}\mathbf{\Lambda}^*\widehat{\mathbf{W}}^\top - \widehat{\mathbf{\Lambda}}\|_2$$
$$\leq \|\widehat{\mathbf{U}}^\top \mathbf{U}^* \mathbf{\Lambda}^* \mathbf{U}^{*\top}\widehat{\mathbf{U}} - \widehat{\mathbf{\Lambda}}\|_2 + 2\|(\widehat{\mathbf{W}} - \widehat{\mathbf{U}}^\top \mathbf{U}^*)\mathbf{\Lambda}^* \mathbf{U}^{*\top}\widehat{\mathbf{U}}\|_2 + \|(\widehat{\mathbf{W}} - \widehat{\mathbf{U}}^\top \mathbf{U}^*)\mathbf{\Lambda}^*(\widehat{\mathbf{W}} - \widehat{\mathbf{U}}^\top \mathbf{U}^*)^\top\|_2$$
$$\leq (\|\widehat{\mathbf{W}} - \widehat{\mathbf{U}}^\top \mathbf{U}^*\|_2^2 + 2\|\widehat{\mathbf{W}} - \widehat{\mathbf{U}}^\top \mathbf{U}^*\|_2)\|\mathbf{L}^*\|_2 + \|\widehat{\mathbf{U}}^\top \mathbf{L}^* \widehat{\mathbf{U}} - \widehat{\mathbf{\Lambda}}\|_2.$$

Further note that

$$\|\widehat{\mathbf{W}} - \widehat{\mathbf{U}}^\top \mathbf{U}^*\|_2 = \|\mathbf{W}^{(1)}(\mathbf{I} - \mathbf{\Lambda}_{\mathbf{U},\widehat{\mathbf{U}}})\mathbf{W}^{(2)\top}\|_2$$
$$= 1 - \sigma_r(\widehat{\mathbf{U}}^\top \mathbf{U}^*) \leq 1 - \sigma_r^2(\widehat{\mathbf{U}}^\top \mathbf{U}^*) = \|\sin\mathbf{\Theta}(\widehat{\mathbf{U}},\mathbf{U}^*)\|_2^2 \leq \frac{C\lambda_1^2(\mathbf{\Sigma}^*)}{\lambda_r^2(\mathbf{L}^*)}\frac{\tau \vee \log p}{n},$$

and

$$\|\widehat{\mathbf{U}}^\top \mathbf{L}^* \widehat{\mathbf{U}} - \widehat{\mathbf{\Lambda}}\|_2 \leq \|\widehat{\mathbf{U}}^\top(\widehat{\mathbf{\Sigma}} - \mathbf{\Sigma}^*)\widehat{\mathbf{U}}\|_2 \leq \|\widehat{\mathbf{\Sigma}} - \mathbf{\Sigma}^*\|_2 \leq C\|\mathbf{\Sigma}^*\|_2\sqrt{\frac{\tau \vee \log p}{n}}.$$

Combine the bounds above and (11), then we have

$$\|\widehat{\mathbf{W}}\mathbf{\Lambda}^{*\frac{1}{2}}\widehat{\mathbf{W}}^\top - \widehat{\mathbf{\Lambda}}^{\frac{1}{2}}\|_2 \leq C\frac{\lambda_1^2(\mathbf{\Sigma}^*)}{\lambda_r^{3/2}(\mathbf{L}^*)}\sqrt{\frac{\tau \vee \log p}{n}}. \tag{12}$$

Now we can plug the upper bounds for $\|\widehat{\mathbf{U}}\widehat{\mathbf{W}} - \mathbf{U}^*\|_2$, $\|\widehat{\mathbf{U}}\widehat{\mathbf{W}} - \mathbf{U}^*\|_{2\to\infty}$, and $\|\widehat{\mathbf{\Lambda}}^{\frac{1}{2}} - \widehat{\mathbf{W}}\mathbf{\Lambda}^{*\frac{1}{2}}\mathbf{W}^\top\|_2$ into (10) and the proof is complete. $\square$

*Proof of Lemma D.2.* Define function $h(x) = \sum_{i,j} \frac{u_i u_j}{v_i + v_j} x^{v_i + v_j}$, then some calculation shows that

$$\frac{dh(x)}{dx} = \sum_{i,j} u_i u_j x^{v_i + v_j - 1} = \frac{1}{x}(\sum_i u_i x^{v_i})^2,$$

which implies $\frac{dh(x)}{dx} \geq 0$ for $x > 0$. Since $h(x)$ is continuous on $[0, \infty)$, and differentiable on $(0, \infty)$, Taylor's theorem thus implies that that $\sum_{i,j} \frac{u_i u_j}{v_i + v_j} = h(1) \geq h(0) = 0$.

□

# E   Additional Empirical Studies

We now present additional simulation studies.

## E.1   Details on the Illustrative Simulations

In our illustrative simulations, we generate a sparse inverse covariance matrix with 100 features for each graph. The non-zero pattern is as indicated in the illustrative figures (Figure 3), and the specific value of each non-zero off-diagonal entry is randomly selected from uniform distribution on $[0.5, 0.8]$. We set the diagonal entries as a constant value such that the minimum eigenvalue of the precision matrix is 0.2, in order to avoid degeneracy. For the population illustration with low-rankness, we either project the population covariance onto a nuclear norm ball or find its best low-rank approximation; we then apply the neighborhood Lasso on the low-rank covariance.

For the finite sample illustration, the 100 features are split into two observation blocks with 75 features observed in each block. The observational block is as illustrated in Figure 4. From the true underlying covariance matrix, we then draw 5000 observations from a multivariate Gaussian distribution, which we then use to construct a stochastic empirical covariance matrix that we use as input to the different Graph Quilting procedures. We use

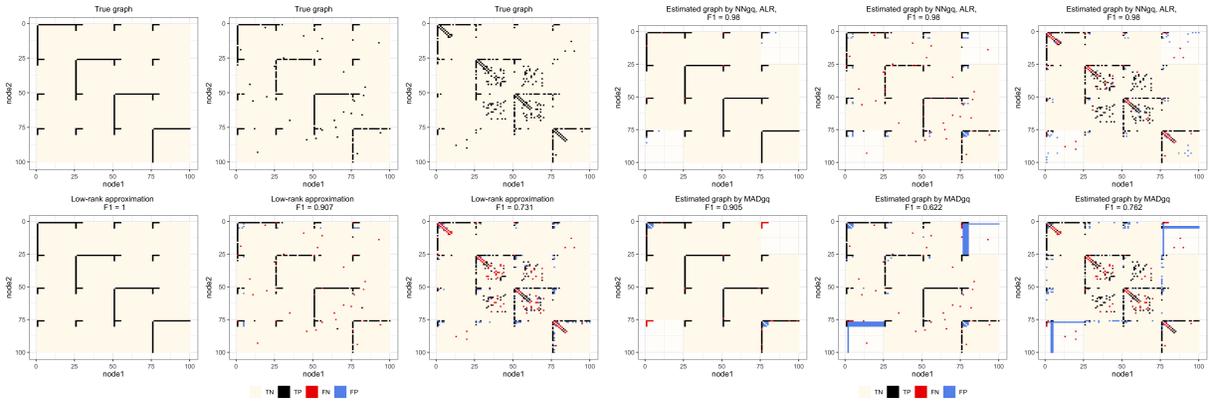

(a) Figure 3 from Section 2 of the main text.     (b) Figure 4 from Section 3 of the main text.

Figure 1: Illustrative low-rank graph structures and estimation results from the main text.

oracle sparsity tuning for the graph sparsity. The rank parameter and nuclear norm penalty parameter are selected to yield the highest graph F1 score in the final graph estimate. From these simulation studies, we find that the low-rank Graph Quilting procedures can better recover edges in the unobserved portion of the covariance matrix that are affiliated with the low-rank structure as compared to the $\text{MAD}_{\text{gq}}$ algorithm.

## E.2    Data-Driven Hyperparameter Tuning

Below, we replicate the analysis performed in Section 3.2.1 of the main text using the same generated Gaussian graphical model data, but now with hyperparameters chosen by data-driven tuning. For the low-rank Graph Quilting methods, we use the methods described in Section 2.5 of the main text, while for the $\text{MAD}_{\text{gq}}$ algorithm we use a minimum threshold of 0 and a maximum threshold selected by an a priori stability score threshold of 0.1 (Liu et al., 2010). The results are shown in Figure 2. The relative comparative accuracy amongst each of the methods is fairly consistent with the results of the main text. Compared to the oracle hyperparameter tuning results, all methods tend to perform slightly worse with regards to accurate edge selection and covariance imputation (expect for zero-imputation

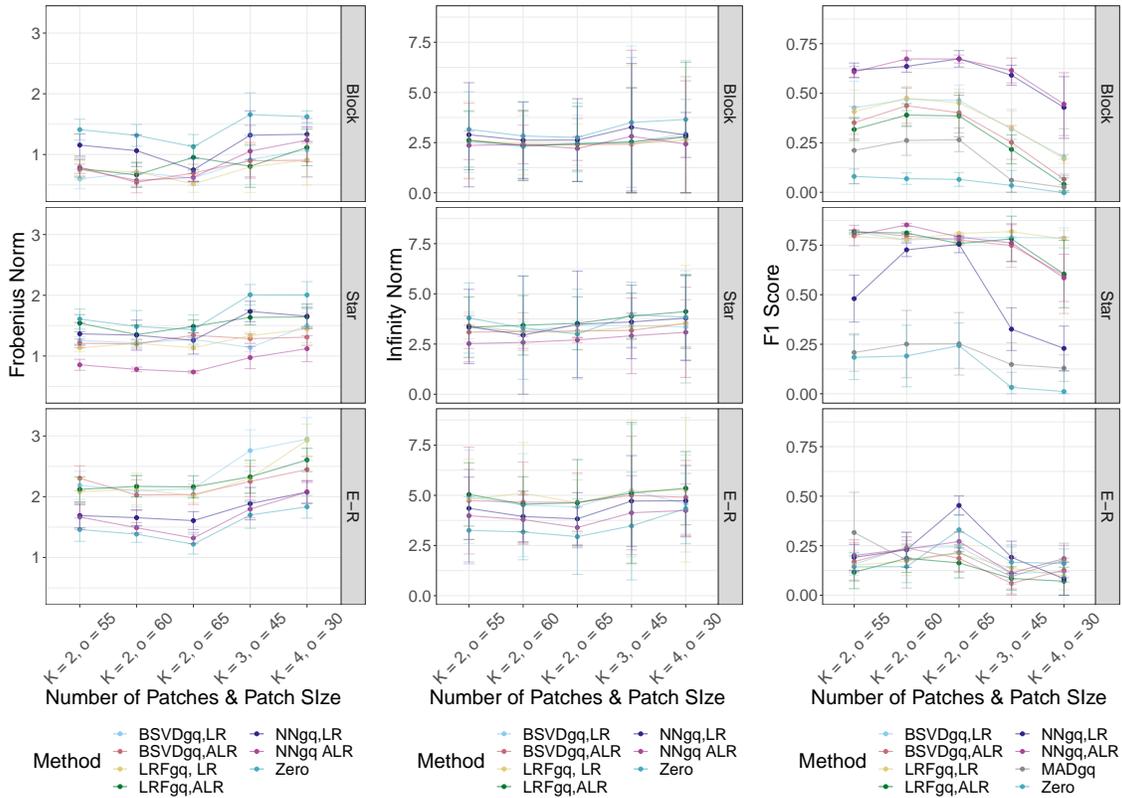

(a) Frobenius norm.   (b) Infinity norm.   (c) F1 score.

Figure 2: Performance of LRGQ, MAD$_{gq}$, and zero imputation methods for covariance imputation and graph recovery on Gaussian graphical model data simulation studies using data-driven hyperparameter tuning.

for the latter task, as there is no hyperparameter selection involved.) We find that the degradation in performance is substantially larger for the data from a Erdös Rényi graph compared to the other two; this is likely because the block diagonal and multi-star graphs exhibit an approximate low-rank structure, leading to more accurate covariance imputation and selection of the optimal rank. However, in the Erdös Rényi case, the optimal rank (i.e. a full-rank matrix) is never selected.

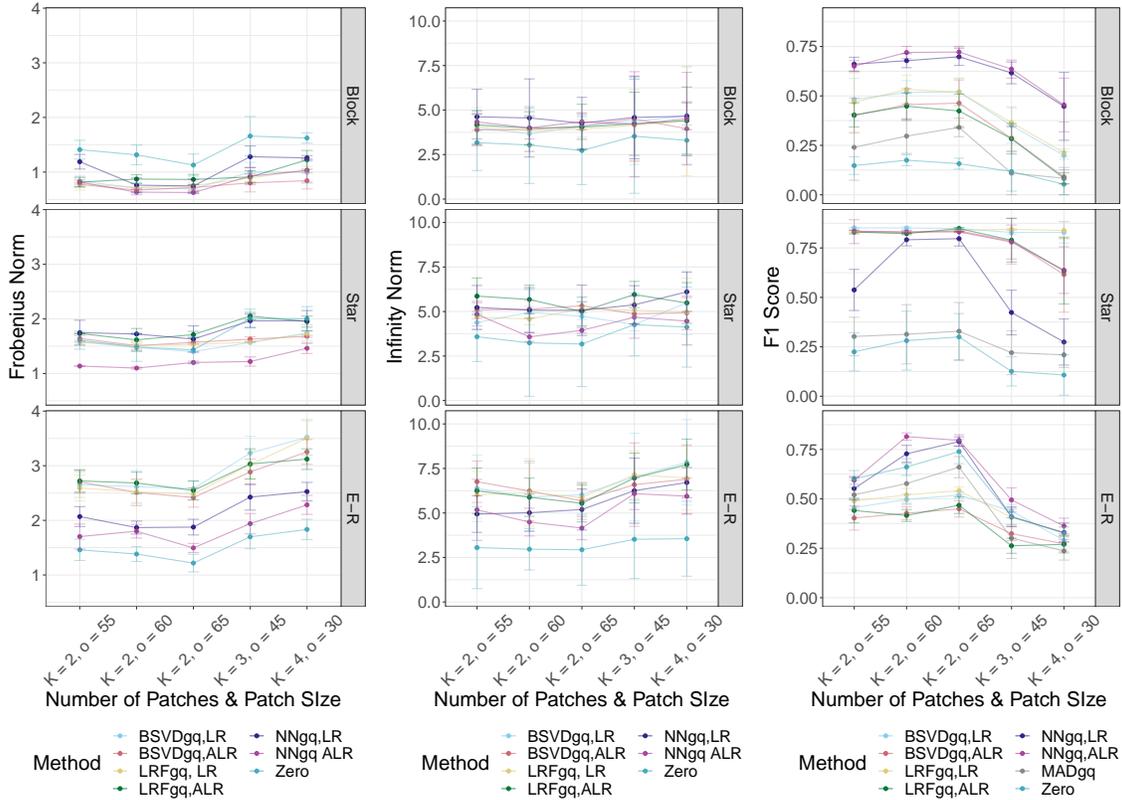

(a) Frobenius norm.     (b) Infinity norm.     (c) F1 score.

Figure 3: Performance of LRGQ, $\text{MAD}_{gq}$, and zero imputation methods for covariance imputation and graph recovery on T-distribution copula data simulation studies.

## E.3   t-distributed Copula Graphical Model

Here, we replicate the analysis performed in Section 3.2.1 of the main text, but with the marginal distribution of each feature following a t-distribution with 3 degrees of freedom rather than a Gaussian distribution. To create the data for this study, we take the simulated data sets used in the simulations in Section 3.2.1 of the main text and apply a univariate copula transform to each column. Hyperparameters for the simulations in this section are selected via oracle tuning. In Figure 3, we see that the relative performance between the different Graph Quilting methods is fairly consistent. Also, the F1 scores for graph recovery and Frobenius norm between the true and imputed covariance matrix only worsen slightly. However, in this case, we see that the infinity norm increases substantially compared to

the Gaussian case.

## E.4 Additional Calcium Imaging Inspired Simulation Studies

We perform an additional calcium-imaging inspired simulation below, using the same process used in Section . Here, we analyze from the Allen Institute contains 227 neurons over 9000 measured time points for one single mouse and recording session. In Figure 4, we show the results for different sizes of observed blocks $o$, as well as different numbers of patches $K$ while keeping $o \times K$ constant. Specifically, we have $K = 2$ patches with patch sizes $o = 130, 140, 150, 160$, and 170, and $K = 2, 3, 4, 5$ and 6 with $o = 150, 100, 75, 60$, and 50, respectively. For each setting, we run 50 replications and report the mean and standard error of the Frobenius norm and F-1 score metrics. We see that the approximate low-rank NNgq method performs best with respect to the Frobenius norm metric on the raw imputed covariance matrix, while NNgq with the exact low-rank is better with respect to edge selection. We also observe that BSVDgq and LRFgq methods with the approximate low-rank assumption perform relatively poorly with the larger patch sizes, with a less accurate imputed covariance matrix and graph edge selection compared to zero imputation. The $\text{MAD}_{\text{gq}}$ also performs relatively poorly in almost all scenarios with regards to edge selection even when compared to zero imputation, which matches what we saw in Section 3. As we would expect, all methods are more accurate when the patch sizes are larger, with the NNgq method exhibiting the greatest increases in performance. On the other hand, as the number of patches is increased while keeping the total number of observations the same, all methods perform worse. We see a particularly low accuracy when the number of patches reaches 5 or 6, as the amount of overlap between each pair of patches is less than 10 neurons per pair of sequential patches.

29

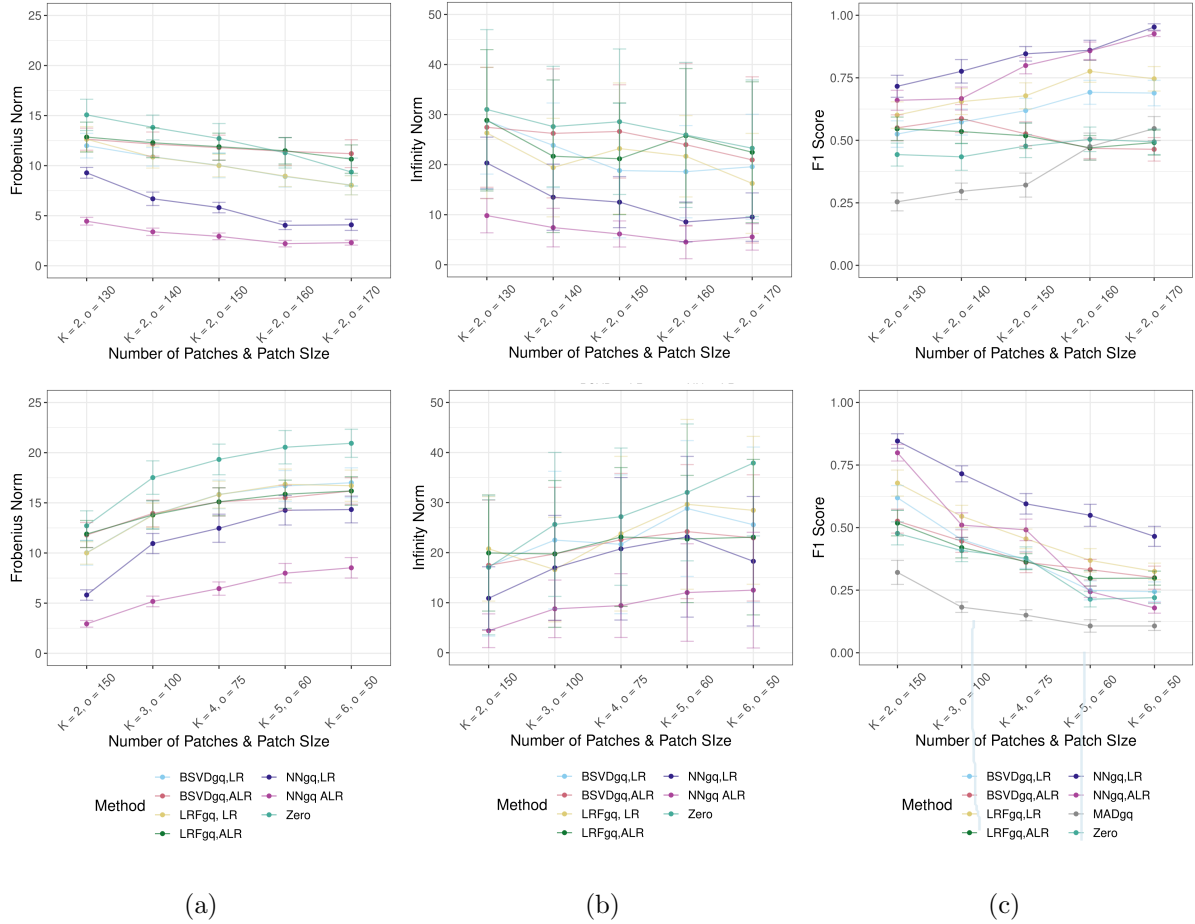

(a)           (b)           (c)

Figure 4: Allen Brain Atlas data.

We show the specific structure of one example of estimated functional connectivity graphs in Figure 5. We see that all methods appear to have few false negatives with respect to the functional connectivity network estimated with full information. However, we see that the BSVDgq and LRFgq methods tend to have more false positive selected edges, while the nuclear norm method has very few. We also validate and compare the selected functional connections using meta data on neuronal tuning, which has been posited to be related to functional connectivity (Sakia and Miyashita, 1994; Stevenson et al., 2012). In Table 1, we compare the selected functional connections by each low-rank Graph Quilting method in the unobserved portion of the covariance matrix by neuronal tuning. Here, neurons are categorized by 8 categories of angular tuning and 5 categories of frequency

30

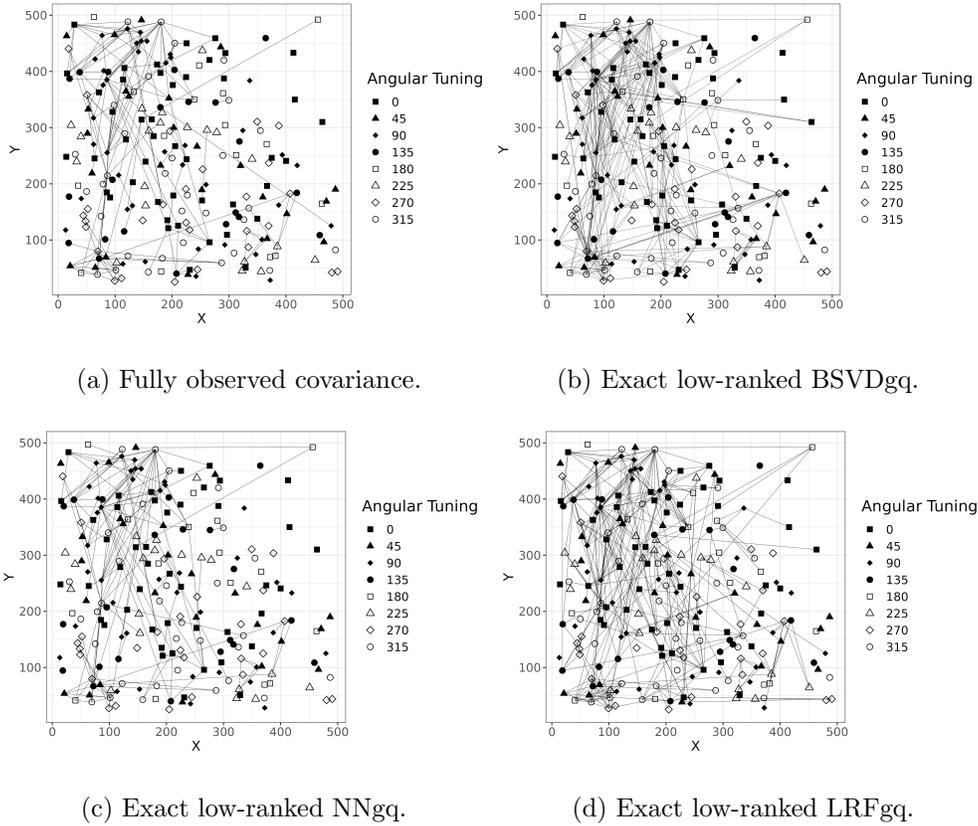

Figure 5: Comparison of one example of functional connectivity estimates from fully observed covariance and low-rank graph quilting methods on Allen Institute data with $K = 2, o = 150$. Neurons are plotted with respect to their spatial locations, and the points are shaped by their angular tuning category.

tuning. Generally, the performance of the different low-rank graph quilting methods is relatively similar for finding functional connections with matching angular tuning, while the approximate low-rank assumption appears to generally perform better for frequency tuning. We also see that all of the low-rank graph quilting methods find functional connections between neurons with the same tuning at similar rates as when the full data is observed and at a substantially higher rate compared the $\text{MAD}_{\text{gq}}$ method.

| **Model** | Angular Tuning | Frequency Tuning |
|---|---:|---:|
| (Full Data) | 0.208 | 0.357 |
| BSVDgq, LR | 0.247 (0.057) | 0.353 (0.039) |
| BSVDgq, ALR | 0.241 (0.054) | 0.413 (0.048) |
| LRFgq, LR | 0.226 (0.044) | 0.319 (0.045) |
| LRFgq, ALR | 0.251 (0.056) | 0.392 (0.051) |
| NNgq, LR | 0.284 (0.063) | 0.452 (0.061) |
| NNgq, ALR | 0.257 (0.052) | 0.377 (0.042) |
| $\mathrm{MAD_{gq}}$ | 0.145 (0.027) | 0.251 (0.040) |

Table 1: Proportion of edges from the imputed portion of the estimated functional connectivity network which link two neurons of the same tuning category for each low-rank graph quilting method. Results for $K = 2, o = 150$ in the Allen Institute data.



\end{document}